\documentclass[fleqn,usenatbib]{mnras}

\usepackage{newtxtext,newtxmath}

\usepackage[T1]{fontenc}

\DeclareRobustCommand{\VAN}[3]{#2}
\let\VANthebibliography\thebibliography
\def\thebibliography{\DeclareRobustCommand{\VAN}[3]{##3}\VANthebibliography}

\usepackage{graphicx}	
\usepackage{amsmath}	
\usepackage{adjustbox}
\usepackage{pdflscape}

\title[PPTA MSP timing models]{The Parkes pulsar timing array second data release: Timing analysis}

\author[D.~J.~Reardon et al.]{
D.~J.~Reardon,$^{1,2}$\thanks{E-mail: dreardon@swin.edu.au}
R.~M.~Shannon,$^{1,2}$
A.~D.~Cameron,$^{1,2}$
B.~Goncharov,$^{4,2}$
G.~B.~Hobbs,$^{3}$ \newauthor 
H.~Middleton,$^{1,5,2}$ 
M.~Shamohammadi,$^{1}$
N.~Thyagarajan,$^{6, 7}$
M.~Bailes,$^{1,2}$
N.~D.~R.~Bhat,$^{8}$\newauthor
S.~Dai,$^{9}$  
M.~Kerr,$^{10}$
R.~N.~Manchester,$^{3}$
C.~J.~Russell,$^{11}$
R.~Spiewak,$^{12,1,2}$
J.~B.~Wang,$^{13}$\newauthor
X.~J.~Zhu,$^{14}$
\\
$^{1}$Centre for Astrophysics and Supercomputing, Swinburne University of Technology, Hawthorn, VIC, 3122 Australia\\
$^{2}$Australia Research Council Centre for Excellence for Gravitational Wave Discovery (OzGrav)\\
$^{3}$ CSIRO Astronomy and Space Science, Australia Telescope National Facility, PO~Box~76, Epping NSW~1710, Australia \\ 
$^{4}$ School of Physics and Astronomy, Monash University, VIC 3800, Australia \\
$^{5}$ School of Physics, University of Melbourne, Parkville, VIC 3010, Australia \\
$^{6}$ National Radio Astronomy Observatory, Socorro, NM 87801, USA \\
$^{7}$ CSIRO Astronomy and Space Science (CASS), P. O. Box 1130, Bentley, WA 6102, Australia \\
$^{8}$ International Centre for Radio Astronomy Research, Curtin University, Bentley, WA 6102, Australia \\
$^{9}$ School of Science, Western Sydney University, Locked Bag 1797, Penrith NSW 2751, Australia \\
$^{10}$ Space Science Division, Naval Research Laboratory, Washington, DC 20375--5352, USA \\
$^{11}$ CSIRO Scientific Computing, Australian Technology Park, Locked Bag 9013, Alexandria, NSW 1435, Australia \\
$^{12}$ Jodrell Bank Centre for Astrophysics, Department of Physics and Astronomy, University of Manchester, Manchester M13 9PL, UK \\
$^{13}$ Xinjiang Astronomical Observatory, Chinese Academy of Science, 150 Science 1-Street, Urumqi, Xinjiang 830011, China \\
$^{14}$ Advanced Institute of Natural Sciences, Beijing Normal University at Zhuhai 519087, China
}

\date{Accepted XXX. Received YYY; in original form ZZZ}

\pubyear{2021}

\begin{document}
\label{firstpage}
\pagerange{\pageref{firstpage}--\pageref{lastpage}}
\maketitle

\begin{abstract}
The main goal of pulsar timing array experiments is to detect correlated signals such as nanohertz-frequency gravitational waves. Pulsar timing data collected in dense monitoring campaigns can also be used to study the stars themselves, their binary companions, and the intervening ionised interstellar medium.
Timing observations are extraordinarily sensitive to changes in path length between the pulsar and the Earth, enabling precise measurements of the pulsar positions, distances and velocities, and the shapes of their orbits.
Here we present a timing analysis of $25$~pulsars observed as part of the Parkes Pulsar Timing Array (PPTA) project over time spans of up to $24$~yr. The data are from the second data release of the PPTA, which we have extended by including legacy data.
We make the first detection of Shapiro delay in four Southern pulsars (PSRs J1017$-$7156, J1125$-$6014, J1545$-$4550, and J1732$-$5049), and of parallax in six pulsars. The prominent Shapiro delay of PSR~J1125$-$6014 implies a neutron star mass of $M_p = 1.5 \pm 0.2 ~M_\odot$  (68\% credibility interval). Measurements of both Shapiro delay and relativistic periastron advance in PSR~J1600$-$3053 yield a large but uncertain pulsar mass of $M_p = 2.06^{+0.44}_{-0.41}$\,M$_\odot$ (68\% credibility interval). We measure the distance to PSR~J1909$-$3744 to a precision of 10\,lyr, indicating that for gravitational wave periods over a decade, the pulsar  provides a coherent baseline for pulsar timing array experiments.

\end{abstract}

\begin{keywords}
pulsars: general -- stars: neutron -- astrometry -- parallaxes
\end{keywords}



\section{Introduction}

\begin{figure*}
\centering
\includegraphics[width=.8\textwidth]{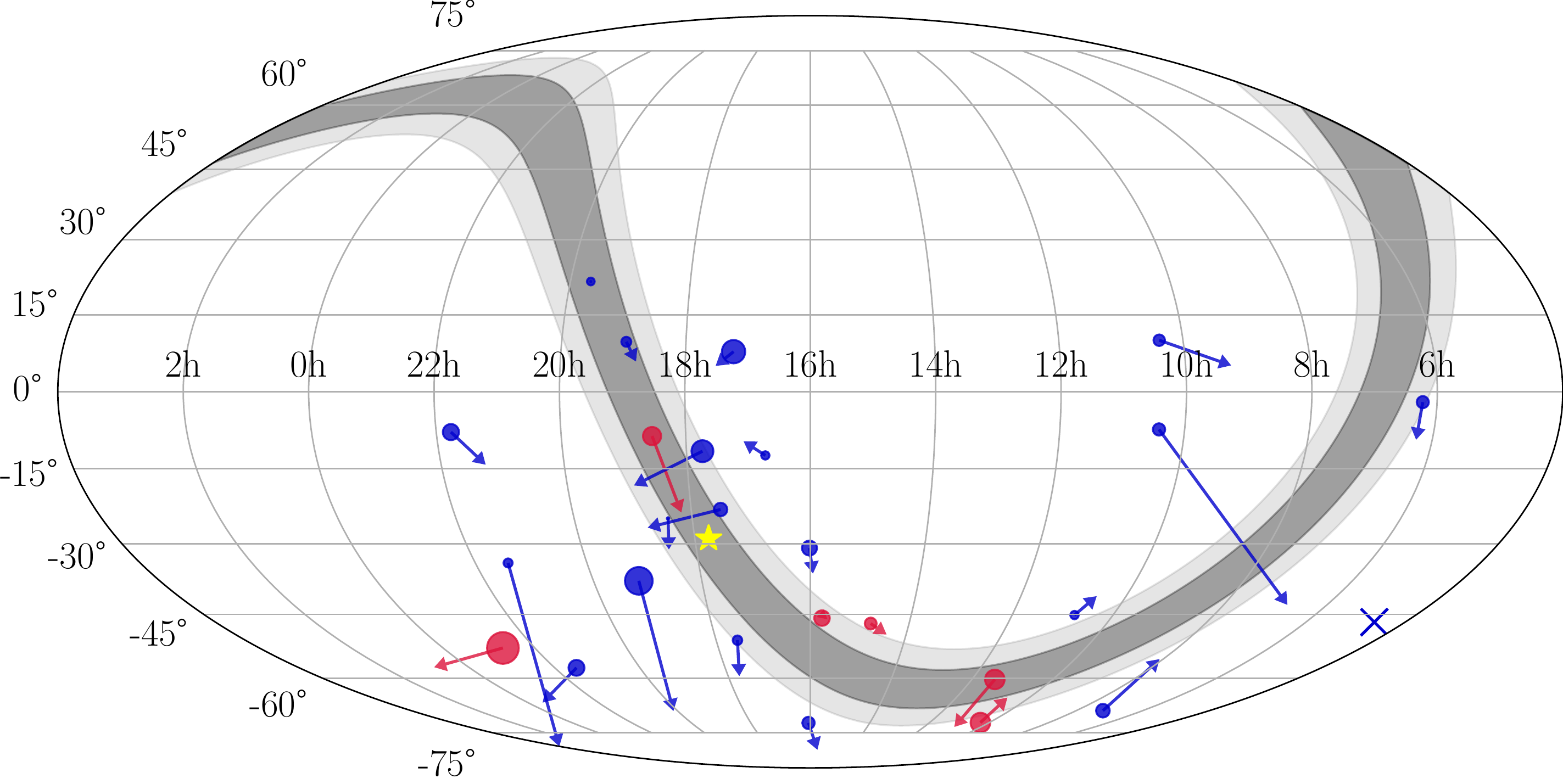}
\caption{The sky distribution in right ascension (horizontal) and declination of the PPTA pulsars. Pulsars in blue are in PPTA-DR2e, while pulsars in red have no legacy data and are only present in PPTA-DR2. The cross marks PSR~J0437$-$4715, which is in PPTA-DR2e, but not analysed in this work. The arrows indicate the direction and relative magnitude (proportional to the arrow length) of our measurement of the pulsar proper motion. The marker size is inversely proportional to the weighted rms residual after subtracting dispersion measure variations. The dark and light grey bands show the $\pm5^\circ$ and $\pm10^\circ$ regions of the Galactic plane respectively, and the yellow star marks the Galactic centre.}
\label{fig:skymap}
\end{figure*}

Utilising the 64-m Parkes radio telescope (also known as {\em Murriyang}), the Parkes Pulsar Timing Array project \citep[PPTA;][]{Manchester+13} has been monitoring an ensemble of millisecond pulsars (MSPs) since 2004.
The primary goal of the project is to search for and study correlated signals that affect the ensemble, including unmodelled contributions to terrestrial time standards \cite[][]{Hobbs+12} and the solar system ephemeris \cite[][]{Champion+10}, and light-year wavelength gravitational radiation \cite[][]{zhu+14,Shannon+15,wang+15}.

In order to search for these signals it is necessary to produce very accurate models for the arrival time variations from the pulsars. This includes properly modelling deterministic and stochastic variations in the arrival times.  Much effort has been expended in constructing robust models of the stochastic contributions, which can include achromatic spin noise \cite[][]{Shannon+10,Lam+17}, chromatic contributions attributable to the interstellar medium \cite[][]{Keith+13}, and system-dependent noise terms \cite[][]{Lentati+16}. These noise terms are often covariant with the correlated signatures of interest, including a potential stochastic gravitational wave background from binary supermassive black holes \cite[][]{Rajagopal+95}.  

It is also essential to model deterministic components to the timing model.
The fundamental deterministic terms are the pulsar spin rate and the rate at which it is slowing down, which are often expressed as the spin frequency $f$ and frequency derivative $\dot{f}$. 
The deterministic component includes geometric terms  caused by variations in the pulsar-observatory line of sight, including the motion of the observatory about the solar system barycentre, the motion of the pulsar through the Galaxy, and, if the pulsar is in  a binary system, the motion of the pulsar about its binary companion \cite[][]{Edwards+06}.   
Binary motion can include the effects of general relativity. 
There are additional relativistic effects related to the propagation of the radiation through the gravitational potential of its companion star, and the gravitational potential of the bodies of the solar system. 
Being the most precisely timed pulsars, MSPs provide the most precise measurements of many of these properties for neutron stars \citep{vanStraten+01, Fonseca+21}.

The most  massive neutron stars known in our Galaxy are the MSPs \cite[][]{Demorest+10,Antoniadis+13,Cromartie+20}.  The presence of $\gtrsim 2 M_\odot$ neutron stars provides vital constraints on equations of state of hadronic matter, which are further tightened when neutron star sizes are determined \cite[][]{Riley+21, Miller+21}.  Binary systems can also be used to test the general theory of relativity \cite[][]{Kramer+06} and more general assumptions about gravity including the equivalence principles \cite[][]{Archibald+18}. 

Here we present an analysis of the deterministic properties for pulsars that have been observed as part of the PPTA project, using the second PPTA data release \cite[PPTA-DR2][]{Kerr+20}. 
The data set provides timing baselines approximately eight years longer than those of the first PPTA data release \cite[PPTA-DR1, ][]{Manchester+13}, and includes data from six additional MSPs.  
Because data quality has improved over the course of the project, additional data improve parameter estimates beyond what would naively be expected from the scaling of parameter precision with data set length. Archival data taken prior to the commencement of the PPTA project, which were also included in the extended data release accompanying PPTA-DR1, have been included in this analysis in order to extend the timing baseline of PPTA-DR2. We refer to the new extended data release as PPTA-DR2e.


In this paper we present timing analyses for $25$ pulsars published in PPTA-DR2. 
The work extends on the analysis of PPTA-DR1, both from the longer data sets, but also through better noise characetersiation \cite[][]{PPTADR2noise}  and algorithmic improvements that have been made in the last half decade.
Another paper  (D. Reardon et al., in preparation) will describe in detail the timing properties of the millisecond pulsar J0437$-$4715, which includes improvements to the Times of Arrival (ToAs) with respect to those in PPTA-DR2, and an analysis of earlier dense observing campaigns.

In Section \ref{sec:data} we summarise the data set, including the extension with legacy data. 
In Section \ref{sec:methods} we describe the methods employed to search for and measure pulsar, binary and astrometric parameters.
In Section \ref{sec:results} we present the results of the analysis. These results are compared to those obtained using data sets produced at other radio telescopes in Section \ref{sec:discuss}. 
We summarise the key findings in Section \ref{sec:conclusions}.

\section{The data set}\label{sec:data}

The primary component of the data set used in the analysis is the PPTA second data release \citep[PPTA-DR2;][]{Kerr+20}. In brief, observations for the project span $2004-2018$, with an approximately three-week cadence for most of the pulsars. At most observing epochs the pulsars were observed in three distinct bands with centre frequencies close to $700$~MHz, $1400$~MHz, and $3100$~MHz.  Data were recorded with a set of fast sampled spectrometers that changed over the course of the data release.  The spectrometers were configured to fold the digitised data at the predicted topocentric spin period of the pulsar and average together many rotations (typically over $8-30$\,s) before writing data to disk.  These data were excised of radio frequency interference (RFI) and calibrated to produce flux-calibrated high-polarisation fidelity pulse profiles. These profiles were then cross-correlated against analytic noise-free templates to determine pulse arrival times, integrated over observations that for most pulsars at most epochs spanned $1$~h.  The analysis methods are detailed in \citet{Kerr+20}.    

\citet{Kerr+20} produced data sets containing either band-averaged or sub-banded arrival times. While the band averaged TOAs might be appropriate for some applications and reduce the size of the data sets considered, we choose to use the sub-banded version of PPTA-DR2. This allows us to account for in-band profile evolution which introduces excess noise in pulsar timing measurements \cite[][]{Demorest+13} and potentially improves measurements of dispersion-measure variations.  Data were dynamically sub-banded to maintain a minimum signal to noise ratio in each observation.  

We supplemented PPTA-DR2 with archival data. While PPTA observing formally commenced in February 2004, we included timing observations of MSPs at Parkes from 1994 that were part of earlier programs. We do not reprocess these earlier data and instead use the TOAs published in \cite{Manchester+13}. For some of the pulsars there was only limited extended data $\lesssim 20$~TOAs, which did not significantly extend the data sets.  For these pulsars we chose not produce extended data sets. We refer to the extended data set as PPTA-DR2e.

In total there are 26 pulsars in PPTA-DR2 and PPTA-DR2e.  Here we present an an analysis of $25$ of them; a detailed analysis of the timing and noise model for an updated dataset of PSR~J0437$-$4715 will be presented separately. All pulsars are continuing to be monitored, with the exception of PSR~J1732$-$5049, which was dropped from the array because of poor timing precision. The sky distribution of PPTA pulsars in this analysis, and their proper motions, are shown in Figure \ref{fig:skymap}. Pulsars new to the PPTA since the first data release are shown with red dots while the other pulsars are in blue.

\section{Methods}\label{sec:methods}

The timing models released with PPTA-DR2 in \citet{Kerr+20}, which included just an initial estimate of the noise in each pulsar, were used to determine timing offsets between observing instruments. These timing models included parameters that were the best descriptions of the pulsar prior to this data release. We have updated these timing models using a generalised least-squares (GLS) fitting routine implemented in \textsc{tempo2}~\citep{Edwards+06}, applying the more detailed noise models for each pulsar described in \cite{PPTADR2noise}. These noise models were developed simultaneously with the timing models described here, because of potential covariance between some timing model parameters and low-frequency noise. Following the inclusion of new parameters to the pulsar timing models, the noise model analysis of \citet{PPTADR2noise} was repeated. The TOAs are referred to the solar system barycentre using the JPL DE436 ephemeris, and are also referred to the TT(BIPM2018) realisation of terrestrial time. 

We searched for significant new timing model parameters using a method similar to that used with PPTA-DR1 by \citet{Reardon+16}. We use the Akaike information criterion \cite[AIC, ][]{Akaike73} to assess whether a parameter significantly improved the model.

If a pulsar's orbit is observed nearly edge-on, or is modelled with sufficiently high precision, it is possible to measure the Shapiro delay \citep{Shapiro64} incurred by pulses travelling through the gravitational field of the companion. To search for the presence of this delay at low signal to noise ratio, we first fitted using the orthometric representation of the effect \cite[][]{Freire+10}, including the amplitude of the third harmonic of orbital period, $h_3$, and adding either the fourth harmonic, $h_4$, or the ratio of harmonics, $\varsigma = h_4/h_3$ if it satisfied the AIC. The highest significance measurements of the Shapiro delay were parametrised instead by the companion mass $M_c$ and the sine of the inclination angle, $\sin{i}$.   

For near-circular binary pulsars, we use the Laplace-Lagrange parameters of the ``ELL1" model \citep{Lange+01}. This model uses a first-order Taylor series expansion of the Romer delay to approximate binary motion. Therefore, the validity of the model depends on the eccentricity of the orbit, but also the timing precision of the pulsar. We use this ELL1 model for pulsars that satisfy $x e^2 < 0.1\sigma_{\rm rms}/\sqrt{N_{\rm obs}}$, where $x$ is the projected semi-major axis (in light-seconds), $e$ is eccentricity, $\sigma_{\rm rms}$ is the rms timing residual (in seconds) across number of observations, $N_{\rm obs}$.

\subsection{Extended data set noise models}

It was necessary to extend the noise models for the PPTA-DR2e data set because the noise models in \cite{PPTADR2noise} are provided only for PPTA-DR2. To estimate white noise in the extended data set, we employed Bayesian methods. We performed our calculations using {\sc temponest} \cite[][]{Lentati+14}. As in~\cite{PPTADR2noise}, we estimated EFAC and EQUAD parameters that describe white noise in the legacy signal processors. To estimate the white noise parameters we analytically marginalised over the deterministic parameters in the timing model and numerically marginalised over the red noise.

We chose to extrapolate the red noise models from PPTA-DR2 to the earlier data, including achromatic noise, dispersion measure variations, band-specific noise, and other chromatic noise processes. This is because the improved data quality, and the existence of three observing bands in PPTA-DR2, allow these processes to be constrained more precisely than in the legacy data. We increased the number of Fourier components in the red noise processes proportionally to the fractional increase to the length of the data set with the added legacy observations, so that the red noise was modelled to the same maximum fluctuation frequency (or, equivalently, cadence). Because the early observations were conducted in bands close to $1400$~MHz, and are fully frequency-averaged and of lower precision than more recent data, we chose not to search for additional chromatic stochastic contributions to arrival times. 

We validated the use of these extrapolated noise models by analysing the whitened (noise-subtracted) residuals and by comparing the timing model results of DR2e and DR2. We found that PSR~J2145$-$0750 was the only pulsar for which the DR2 noise model did not sufficiently characterise the low-frequency noise in the legacy data of the DR2e dataset. This manifested as a 2$\sigma$ discrepancy in the measurements of proper motion between the two datasets. For this pulsar we used the noise modelling procedure of \citet{PPTADR2noise} with the DR2e dataset, and identified significant excess noise in the \textit{ftpm} data processing instrument. We then included this additional system noise in our noise model and repeated the timing model analysis. This pulsar was identified by \citet{PPTADR2noise} as having unusual red noise characteristics and perhaps profile variability, which may be why the power-law noise model does not extrapolate to the earlier data.

\subsection{Ultra-wide bandwidth dispersion measure}

Precision timing experiments emphasise the need for measurement and mitigation of variations in dispersion measure (DM).  It is necessary to have a good {\em a-priori} measurement of DM for observations in which coherent dedispersion \cite[][]{Hankins71} is employed. However, DM measurements are covariant with pulse profile evolution, which can result in large uncertainties in the DM when modelling precision timing data sets. 

Measurements of the DM of each pulsar were made using observations obtained with the Parkes Ultra-Wideband Low \cite[UWL, ][]{Hobbs+20} receiver system, which records coherently-dedispersed pulses in the band 704--4032\,MHz. For each pulsar, a representative selection of three recent UWL observations were selected for analysis. Each observation was manually inspected to excise RFI in both the time and frequency domains, before then being calibrated in both polarisation and flux. The data were then divided into six frequency sub-bands in order to account for frequency-dependent evolution of the pulse profile. The distribution of these sub-bands is given in Table~\ref{tab: DM sub-bands}.

\begin{table}
    \centering
    \begin{tabular}{lccc}
        \hline
         Name & Low Freq. & High Freq. & Bandwidth \\
         & (MHz) & (MHz) & (MHz) \\
         \hline
         RF10 & 704 & 1024 & 320\\
         RF11 & 1024 & 1344 & 320 \\
         RF20 & 1344 & 1856 & 512 \\
         RF21 & 1856 & 2368 & 512 \\
         RF30 & 2368 & 3200 & 832 \\
         RF31 & 3200 & 4032 & 832 \\
         \hline
    \end{tabular}
    \caption{Definition of the frequency sub-bands used for fitting the dispersion measure of each pulsar across the bandwidth of the Parkes UWL receiver.}
    \label{tab: DM sub-bands}
\end{table}

For each sub-band, a standard analytic profile was produced by summing all the data for each pulsar together in time, frequency and polarisation to produce a single average profile, before fitting a series of Gaussian curves using the \textsc{psrchive} program \textsc{paas}. Each observation was then individually summed in time and polarisation, before being partially summed in frequency to produce ToAs across the sub-band. We measured the DM  using \textsc{tempo2}, incorporating the TOAs from all sub-bands simultaneously, with each sub-band TOAs offset by a fitted jump parameter in order to account for the use of multiple timing standards across the band. In this way, a fit for DM was achieved which utilised the full band of the UWL receiver while still accounting for pulse-profile evolution. These DM measurements and their epochs were used in the timing model (for all pulsars except PSR~J1732$-$5049 which has no UWL data) and are treated as fixed parameters. Additional profile variations and changes in DM were modelled with FD parameters \citep{NG11} and the DM model (with stochastic and deterministic terms) of \citet{PPTADR2noise}, respectively.

\subsection{Formation of a frequency-averaged dataset}

The PPTA-DR2 dataset contains thousands of individual TOAs per pulsar, because each observation was split in frequency into several sub-bands \citep{Kerr+20}. Because of this, we are able to measure and model the effects of profile evolution within each band, and also improve the sensitivity to chromatic noise processes identified in the recent PPTA noise analysis \citep{PPTADR2noise}. However, this has greatly increased the dimensionality of the design matrix for each pulsar compared with the early frequency-averaged dataset of PPTA-DR1, which significantly increases the compute time. Further, the white noise parameter \textsc{ECORR}, which  was required for many of our pulsars, makes the GLS fitting routine computationally expensive. For this reason we have formed an alternate dataset using our new pulsar ephemerides and the white noise model of \citet{PPTADR2noise}.

The DM for each pulsar has been measured with the UWL data, and in the pulsar ephemerides we included a number of FD parameters (determined by the AIC) to produce a model of the profile evolution with frequency. For each pulsar, we corrected its sub-banded TOAs by this model and de-dispersed using the fixed value of DM. We then applied the \textsc{EFAC}s and \textsc{EQUAD}s measured in the PPTA-DR2 noise analysis to correctly weight the sub-banded TOAs. These sub-bands were then combined to produce one weighted averaged TOA, at the weighted average observing frequency, per observation. Finally, we applied the \textsc{ECORR}s to these averaged TOAs and re-introduced the dispersion delay.

The resulting dataset is a compromise between the traditional timing of frequency-averaged pulse profiles, and the detailed model of profile evolution provided by a large number of sub-bands or wide-bandwidth timing techniques \citep[e.g.][]{Pennucci+14, Liu+14}. Like the traditional approach, we have not accounted for time-dependent noise processes that may perturb individual sub-bands \citep[e.g. magnetospheric events, strong DM variations, or band- and system-dependent red noise,][]{PPTADR2noise}. This dataset is useful for accelerated compute time in applications such as spectral analysis of the residuals, and gravitational wave searches using many pulsars, but we caution against its use in general because these TOAs and their uncertainties depend on the assumed models of profile evolution and white noise. These model assumptions are not expected to affect large-scale stochastic structures.

\subsection{Parameter derivations}

Many of the parameters measured in pulsar ephemerides can be used to derive additional physical quantities. The derivations are often nonlinear, which can lead to asymmetric confidence regions for the derived parameters. To capture this, we compute the central 68\%-confidence (1$\sigma$) region (unless otherwise stated) for each derived parameter using a Monte Carlo simulation of our measurements. We assume that our GLS ephemeris solution for a given pulsar is described by a multivariate Gaussian distribution. We then draw one million samples from this distribution and compute the derived parameter for each measurement sample. In cases where the derived parameter itself depends on other derived parameters (for example the pulsar mass $M_p$, which can be constrained by multiple parameters relating to relativistic and kinematic effects), we convert the distributions of derived parameters to probability density functions, which are then multiplied to determine the probability distribution for the final derived parameter. Below we detail some of the derivations available from our ephemerides.

\subsubsection{Distances and transverse velocities}

For most pulsars we measure a significant parallax, $\pi$, which gives the pulsar distance through $D_\pi = 1/\pi$. For pulsars with distances derived in this way, we also determine the magnitude of their transverse velocity $v_t = \mu D_\pi$, where $\mu$ is the total proper motion. 

The transverse velocity of a pulsar is also associated with a small radial acceleration, which produces a time-varying Doppler shift to the observed pulsar spin period, and orbital period. This is the Shklovskii effect and is the dominant source of the measured orbital period-derivative, $\dot{P}_b$, in our sample of pulsars (with the exception of PSRs J1446$-$4701 and J2241$-$5236, which have a large $\dot{P}_b$ due to interaction with a non-degenerate companion). With a measurement of the proper motion and this $\dot{P}_b$ it is possible to determine the distance to the pulsar \cite[][]{Shklovskii70,Bell+96} with
\begin{align}
    D_{\rm shk} = & \frac{c}{\mu^2}\frac{\dot{P}_b^{\rm shk}}{P_b}
    \\ 
    \dot{P}_b^{\rm obs} = & \dot{P}_b^{\rm GR} + \dot{P}_b^{\rm shk} + \dot{P}_b^{\rm Gal},
\end{align}
    
\noindent where $P_b^{\rm shk}$, $\dot{P}_b^{\rm GR}$, and $\dot{P}_b^{\rm Gal}$ are respectively the contributions to the observed orbital period-derivative $\dot{P}_b^{\rm obs}$, from the Shklovskii effect, gravitational wave emission, and differential acceleration in the Galaxy. For each pulsar, we estimate $\dot{P}_b^{\rm Gal}$ using the \textsc{galpy} package \citep{Bovy15}, which provides the radial and vertical components of the differential acceleration. With the exception of PSR~J1909$-$3744, $\dot{P}_b^{\rm GR}$ is neglected because it is much smaller than the measurement uncertainty of $\dot{P}_b^{\rm obs}$.

\subsubsection{Kinematic Kopeikin terms}
\label{sec:kopeikin}
Among the time-derivatives of binary parameters in our ephemerides is a secular change to the projected semi-major axis, $\dot{x}$. The dominant source of this $\dot{x}$ for our pulsars is a kinematic effect, where the projected shape of the orbit varies with the transverse motion of the pulsar \citep{Kopeikin96}. 
This effect can also contribute a secular change to the longitude of periastron $\dot{\omega}$, however for our pulsars the measured values are dominated instead by the relativistic periastron advance.

A measurement of this kinematic $\dot{x}$ can be used to place a limit on the orbital inclination angle $i$, using $\tan i \leq x\mu/\dot{x}$, for total proper motion $\mu$ \citep{Sandhu+97}. 

For the most precisely timed pulsars, it is possible to also observe the annual-orbital parallax, which causes annual variations to $x$ and $\omega$ \citep{Kopeikin95}. A detection of the annual-orbital parallax provides a unique measurement for $i$ and the longitude of ascending node, $\Omega$. In the \textsc{Tempo2} binary timing model, these parameters ($i$ and $\Omega$, known as ``Kopeikin terms") include both the secular and annual-orbital parallax contributions to $\dot{x}$ and $\dot{\omega}$ \citep{Edwards+06}. In cases where the Shapiro delay is also measured, $i$ is also used to describe the ``shape" parameter $s = \sin i$. Among our sample of pulsars, only PSR~J1713+0747 is parametrised in this way.

For pulsars with highly significant measurements of $\dot{x}$, we have also tried to parameterise the timing model using the Kopeikin terms to search for a significant annual-orbital parallax. In cases of a non-detection of the annual-orbital parallax, we have explored the multiple solutions in ($i$, $\Omega$) using \textsc{Tempo2}, and its Bayesian inference extension, \textsc{TempoNest}. We present and discuss the multiple solutions in ($i$, $\Omega$) for cases where one solution is favoured over the other(s).

\subsubsection{Relativistic parameters}

Observing a Shapiro delay can directly give the mass of the companion $M_c$, but also $\sin{i}$, which is used to determine the true shape of the orbit from its precisely-measured projection. A Shapiro delay with high significance therefore provides the total system mass $M_{\rm tot}$ and also the pulsar mass $M_p$. However, for a low-significance detection, it is best to parametrise the effect using the orthometric parameters ($h_3$, and $h_4$ or $\varsigma$) of \citet{Freire+10}. 

For pulsars with Shapiro delays measured with the orthometric parameters, we compute the corresponding range of $M_c$ and $i$ using the Monte Carlo technique described above. The probability density function for $i$ is also re-weighted to reflect a prior probability distribution for isotropy, which is a uniform distribution in $\cos{i}$ \citep{Freire+10}.

The dominant source of observed $\dot{\omega}$ in our pulsars is intrinsic to the system and attributed to the relativistic precession of the orbit. Where we have detected $\dot{\omega}$, we use this measurement to derive the total system mass using \citep{Damour+86}
\begin{equation}
    M_{\rm tot} = \frac{1}{T_\odot} \left( \frac{\dot{\omega} (1-e^2)} {3 n_b^{5/3}}  \right)^{3/2},
\end{equation}
\noindent where $n_b$ is the mean angular velocity, $T_\odot$ is the mass of the Sun in time units, and $e$ is eccentricity. This derivation of $M_{\rm tot}$, paired with the $M_c$ provided by the Shapiro delay, provides a further constraint for the pulsar mass $M_p$ for systems where both relativistic effects are measured.

\begin{figure}
\centering
\includegraphics[width=.5\textwidth]{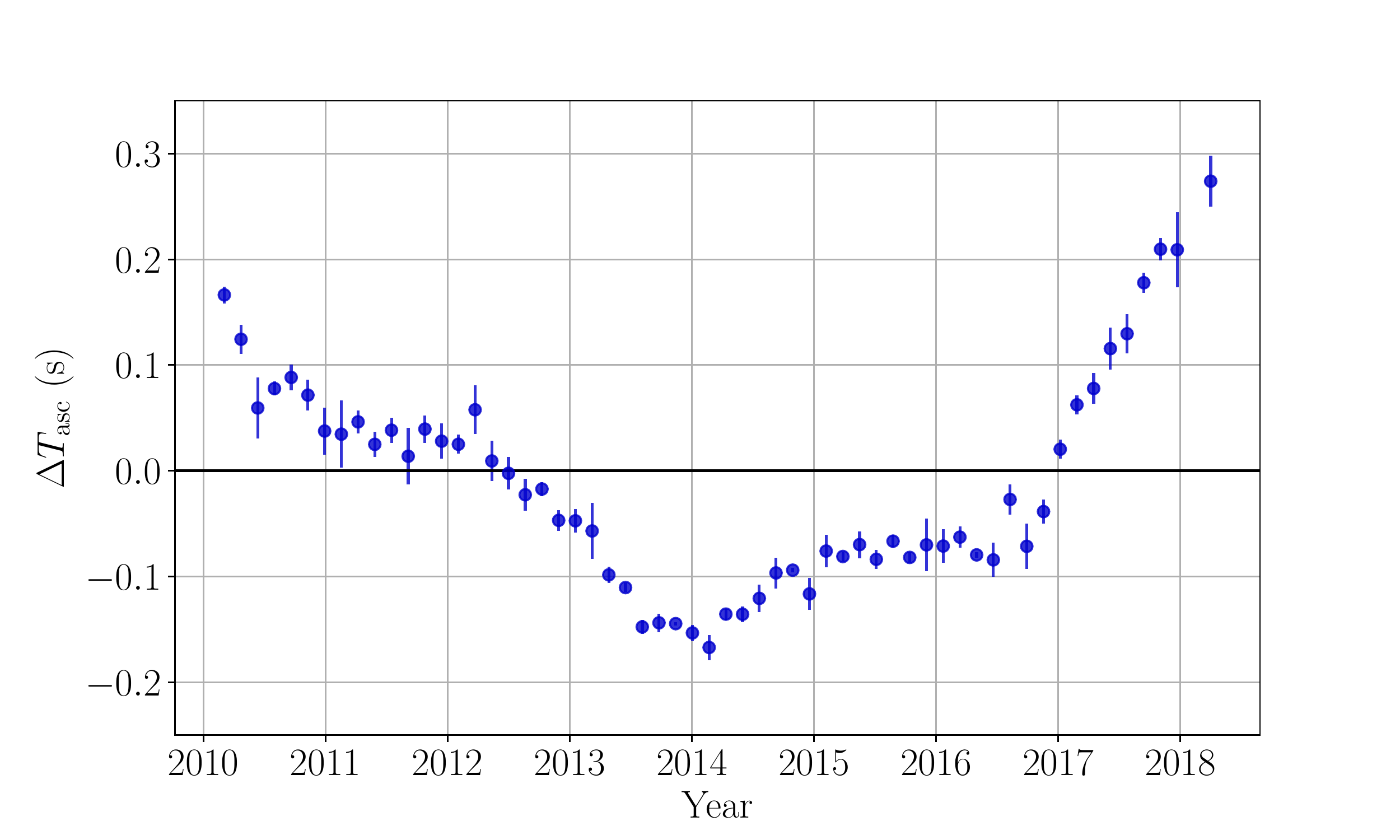}
\caption{Stochastic variability in the epoch of ascending node, $T_{\rm asc}$, for PSR J2241$-$5236. $\Delta T_{\rm asc}$ is the difference from the mean $T_{\rm asc}$, measured every 50\,days.}
\label{fig:2241}
\end{figure}



\subsection{Comparing to previous results}

We compare our updated pulsar timing models, and derived parameters, with a number of previous works to check for consistency. We also compare the results between our extended PPTA-DR2e, and the nominal PPTA-DR2 to confirm that the extrapolated noise models describe the legacy data sufficiently, and to test whether the increase in timing baseline in PPTA-DR2e adds to the sensitivity to any parameters. The comparison of measured parallaxes and derived pulsar masses (where available) are highlighted and compared for all datasets.

\begin{figure*}
\centering
\includegraphics[width=\textwidth]{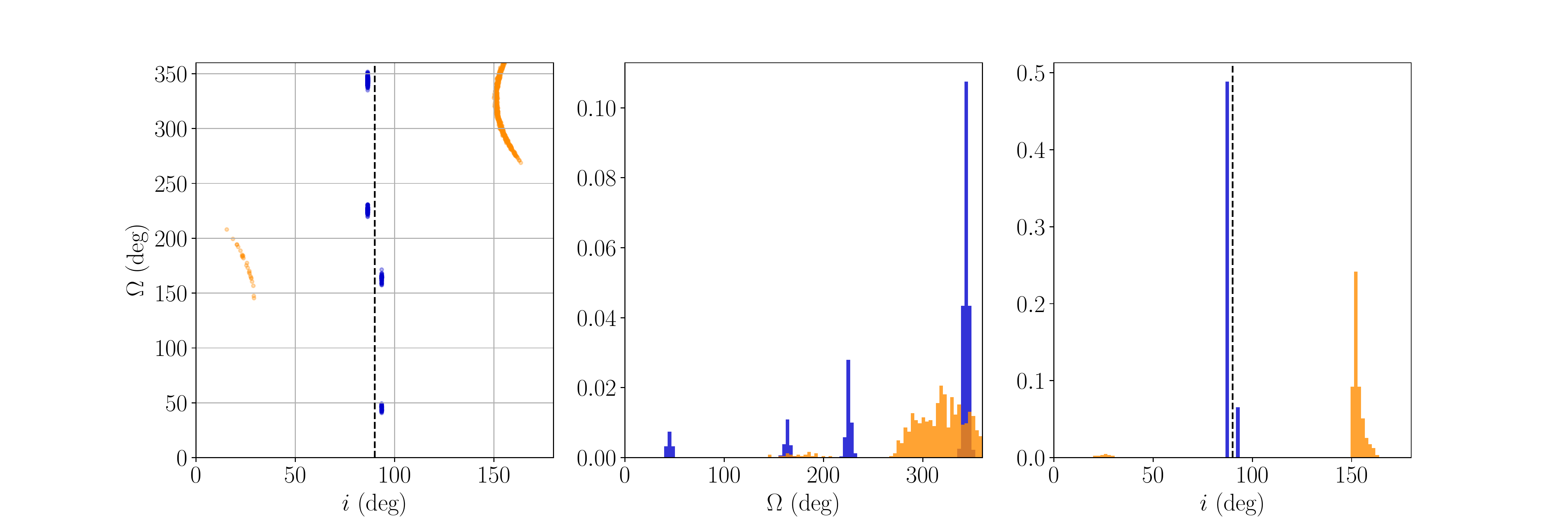}
\caption{Posterior probability samples for the Kopeikin parameters that model the effect of annual-orbital parallax for PSRs J1643$-$1224 (orange) and J1909$-$3744 (blue). Left panel shows the posterior samples, the middle and right panels show the probability density histograms for $\Omega$ and $i$, respectively. The dashed vertical line in the left and right panels indicate $i=90^\circ$.}
\label{fig:kinkom}
\end{figure*}

The previous datasets used for the comparisons are: the extended first PPTA data release \citep[PPTA-DR1;][]{Reardon+16}, the first data release of the European Pulsar Timing Array \citep[EPTA-DR1;][]{Desvignes+16} with $42$ MSPs, and the NANOGrav 12.5 year data release with $47$ MSPs. The latter was presented as two versions of the data; a ``narrowband" analysis \citep{NG12Narrowband} including sub-banded TOAs for each observation (analogous to PPTA-DR2), and a ``wideband" analysis \citep{NG12Wideband} with only one TOA per observation after accounting for profile evolution in the template profile and measuring dispersion simultaneously with the TOA. Since these NANOGrav analyses already include a discussion of the timing model consistency between the two datasets, for this work we only compare directly to the narrowband analysis (NG-12.5yr-NB).

For PSR~J1713+0747 we also compare our results with the combined NANOGrav and EPTA dataset of \citet{Zhu+19}, and for PSR~J1909$-$3744 we discuss the recent EPTA analysis using 15\,years of timing \citep{Liu+20}. \cite{wang+17} has presented a comparison of pulsar positions derived from very long baseline interferometry, to those derived from pulsar timing assuming different solar system ephemerides.

\section{Results}\label{sec:results}  

The measured and derived parameters for each pulsar are given in the tables of Appendix \ref{sec:ephemerides}, and are described in more detail in the sections below. The post-fit timing residuals are given in the figures of Appendix \ref{sec:residuals} (online only), and show, for each pulsar; the post-fit residuals, the residuals with chromatic noise sources (primarily DM variations) subtracted, the fully whitened residuals (e.g. with achromatic red noise also subtracted), and the normalised whitened residuals. The latter serve as a test for the quality of our noise model and timing solution, and demonstrate that the measured timing model parameters and uncertainties should be unbiassed.

\subsection{Spin down}
\label{sec:f2}

For every pulsar, we model the spin frequency $f$ and its time derivative $\dot{f}$. We also searched for evidence for significant second time derivatives in all of the pulsars. 
It has been previously noted that PSR~J1024$-$0719 is likely in a wide binary orbit \cite[][]{Kaplan+16,Bassa+16}, which introduces an unmodelled jerk to the radial motion of the pulsar, manifested as an apparent second time derivative of the pulsar spin frequency, $\ddot{f}$. \cite{PPTADR2noise} analysed the significance of this $\ddot{f}$ term through Bayesian model selection, and found that it is indeed preferred over a model of low-frequency spin noise alone, and we have measured $\ddot{f} = -3.89(8)\times 10^{-27}$\,s$^{-3}$. Using this same Bayesian model selection technique, we verified that no other pulsars in this data set have evidence for $\ddot{f}$.



\subsection{Binary systems}

Of the pulsars in the sample, $17$ have binary companions\footnote{This excludes PSR~J1024$-$0719, which is discussed in the previous section.}, with binary periods ranging from approximately $3$\,hr to $150$\,days and projected semimajor axes ranging from $0.026$ to $32$ lt-s.  The two pulsars with the shortest orbital periods, PSRs J1446$-$4701 and J2241$-$5236, are likely to have non-degenerate companions.  The orbital properties of PSR~J2241$-$5236 are discussed in greater detail in Section \ref{sec:2241}.

We find evidence for orbital distortions due to the Shapiro delay for ten pulsars, including four which have not been previously reported. Of these the most significant measurement is for PSR~J1125$-$6014, which is discussed in Section \ref{sec:masses}. We measure significant values of the first orthometric term $h_3$ for the first time in three other pulsars, J1017$-$7156, J1545$-$4550, and J1732$-$5049.

Strong detections of the Shapiro effect are made for four pulsars for which this effect has been previously identified. We provide the most precise measurements for PSRs J1600$-$3053 and ~J1909$-$3744.    

We detect significant binary orbital period derivatives ($\dot{P}_b$) for 10 pulsars, most of which are dominated by the Shklovskii effect. For PSR~J1446$-$4701, the large value of $\dot{P}_b$ may be attributed to tidal effects with its putatively non-degenerate companion. We detect secular changes in the projected semi-major axes $\dot{x}$ that can be attributed to kinematic effects in 11 pulsars, and we use these measurements to constrain (or measure when paired with a Shapiro delay) the orbital inclination.

Added to the PPTA project in 2010, PSR~J1017$-$7156 exhibits $\sim 410$\,ns timing precision in the 1.4\,GHz band after removing the strong DM (and scattering) variations. As well as the Shapiro delay, we detected binary derivative parameters for this pulsar, $\dot{x}$, $\dot{\omega}$, and $\dot{P}_b$. \citet{Ng+14} also measured $\dot{x}$ and $\dot{\omega}$ using $\sim$3\,yrs of observations from the High Time Resolution Universe (HTRU) pulsar survey. Our measurements are consistent with, but more precise than these earlier measurements.

\subsubsection{Black widow PSR~J2241$-$5236} \label{sec:2241}

PSR~2241$-$5236 is a non-eclipsing black widow pulsar, meaning it has a low-mass, non-degenerate companion. Tidal interactions between the pulsar and this companion may be stochastically disturbing the orbit \citep[as in other black widow pulsars, e.g.][]{Applegate+94, Arzoumanian+94, Doroshenko+01}. The pulsar itself exhibits extreme stability in its pulse profile, with $\lesssim$200\,ns weighted RMS residuals in this data set (after correcting for DM variations), and the lowest measured jitter (pulse shape instability) level of any known pulsar \citep{Bailes+20}. Because of this, the timing model demands a high number of orbital-frequency derivatives to fully characterise the orbital noise. In this data set we include the first ten of these derivatives and defer a detailed analysis to future work using the much more sensitive observations obtained with MeerKAT. 

The orbital noise can be visualised by measuring the epoch of ascending node, $T_{\rm asc}$ as a function of time, effectively timing the orbit. To do this we first removed all derivative terms relating to the binary orbit from the timing model. We then fixed all other timing model parameters and fitted only for $T_{\rm asc}$ using TOAs in each window of 50\,days across the data set. The resulting time series of $T_{\rm asc}$ is shown in Figure \ref{fig:2241}, and demonstrates the stochastic and low-frequency nature of the orbital noise. 

Fortunately the presence of orbital noise does not preclude the pulsar from being sensitive to low-frequency gravitational radiation \cite[][]{Bochenek+15}. This pulsar is already a key contributor to the sensitivity of the PPTA to gravitational waves.

\subsubsection{Annual-orbital parallax} \label{sec:aop}

Only one pulsar in this sample has a measurement of the annual-orbital parallax, PSR~J1713+0747, which uniquely gives the inclination and longitude of ascending node as $i=72.0^\circ \pm 0.5 ^\circ$ and $\Omega = 89.2^\circ \pm 1.4 ^\circ$, respectively. This is consistent with the earlier measurement of \citet{Zhu+19}, which used data from both NANOGrav and EPTA.

A number of other pulsars have a strong detection of the secular $\dot{x}$, but no significant measurement of the annual-orbital parallax. For these pulsars we substituted $\dot{x}$ for the Kopeikin terms parametrised with $i$ and $\Omega$, to explore possible solutions. The posterior probability distributions for these parameters were computed using \textsc{TempoNest} and we assessed the relative likelihoods of the multiple solutions (with the estimated Bayes factor, $\mathcal{B}$, as the ratio of likelihoods).

PSRs J1643$-$1224 and J1909-3744 showed weak evidence for one solution over others (with a log Bayes factor $2 \lesssim \log{\mathcal{B}} \lesssim 4$). The posterior samples for $i$ and $\Omega$ for these pulsars are shown in Figure \ref{fig:kinkom}. The four solutions found for PSR~J1909$-$3744 are consistent with those presented by \citet{Liu+20}, which follows from a consistent measurement of $\dot{x}$.

The most likely solutions, with weak Bayesian evidence, are $i=152.3^{+3.3 \circ}_{-1.1}$ and $\Omega =319^\circ \pm22^\circ$ for PSR~J1643$-$1224, and $i=86.47^\circ \pm0.03^\circ$ and $\Omega = 344^\circ \pm3^\circ$ for PSR~J1909$-$3744. This ascending node for PSR~J1909$-$3744 is closest to the solution of $\Omega = 352^\circ \pm5^\circ$ from \citet{Liu+20}. 


    \label{tab:kopeikin}

\subsection{Pulsar and companion masses}\label{sec:masses} 

\begin{figure}
\centering
 \includegraphics[width=.5\textwidth]{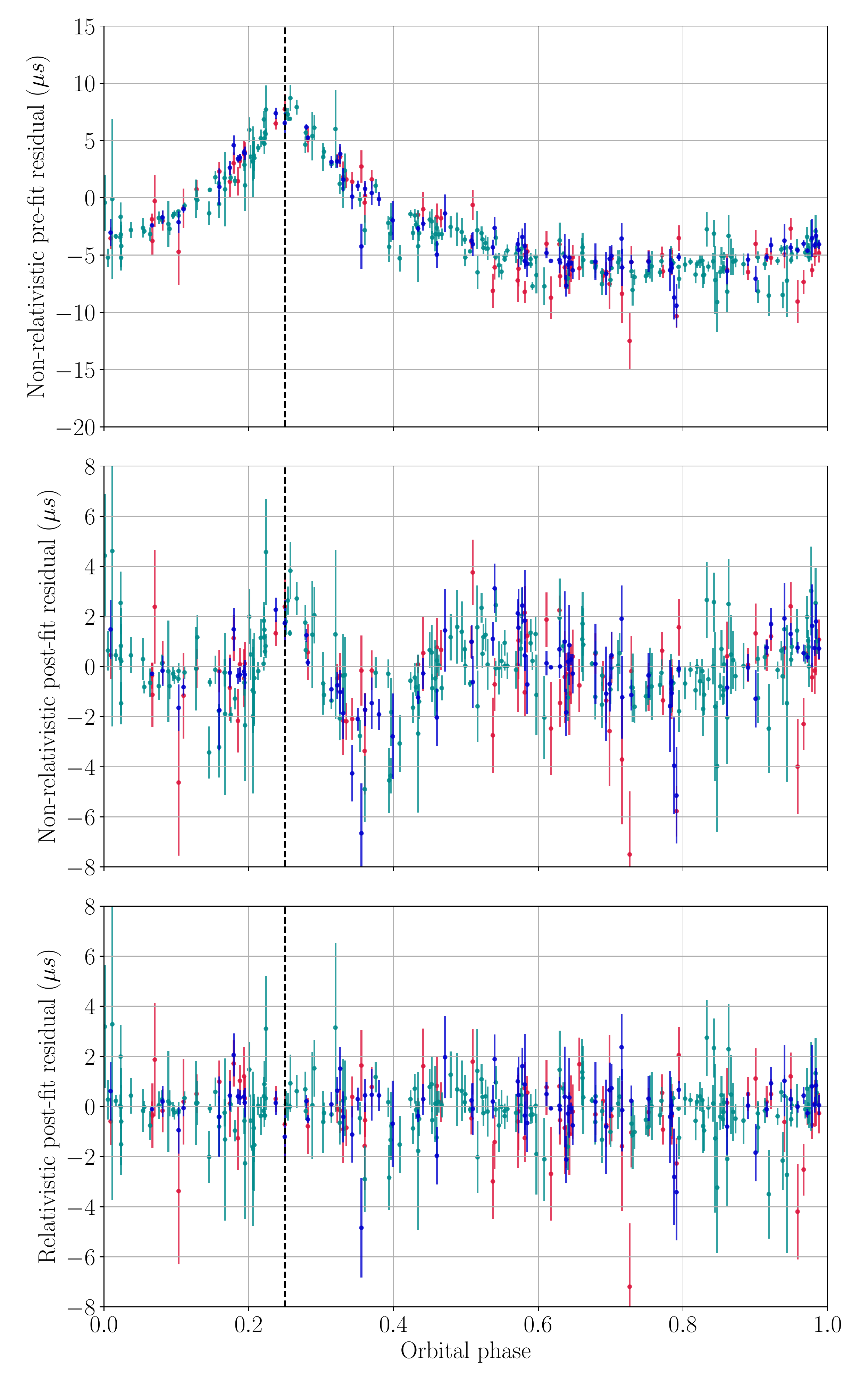}
\caption{Timing residuals as a function of binary orbital phase for PSR~J1125$-$6014, showing the Shapiro delay signal. The top panel shows the pre-fit residuals with the full Shapiro delay signature preserved, the middle panel shows the post-fit residuals after fitting for the full timing model except for the relativistic Shapiro delay parameters, and the bottom panel shows the residuals after including the Shapiro delay. The dashed vertical line indicates the orbital phase of superior conjunction, and the colours indicate the different observing bands with centre frequencies of 700MHz (red), 1400MHz (teal), and 3100MHz (blue).}
\label{fig:1125}
\end{figure}

Among our new Shapiro delay detections, PSR~J1125$-$6014 has the highest significance and we derive the pulsar mass to be $M_p = 1.5 \pm 0.2$\,M$_\odot$. The pre- and post-fit timing residual signatures of this Shapiro delay are shown in Figure \ref{fig:1125}. Being a southern pulsar, the MeerKAT radio telescope could be used for observing campaigns that target superior conjunction and other key orbital phases, to reduce the measurement uncertainty. 

\begin{figure*}
\centering
\includegraphics[width=\textwidth]{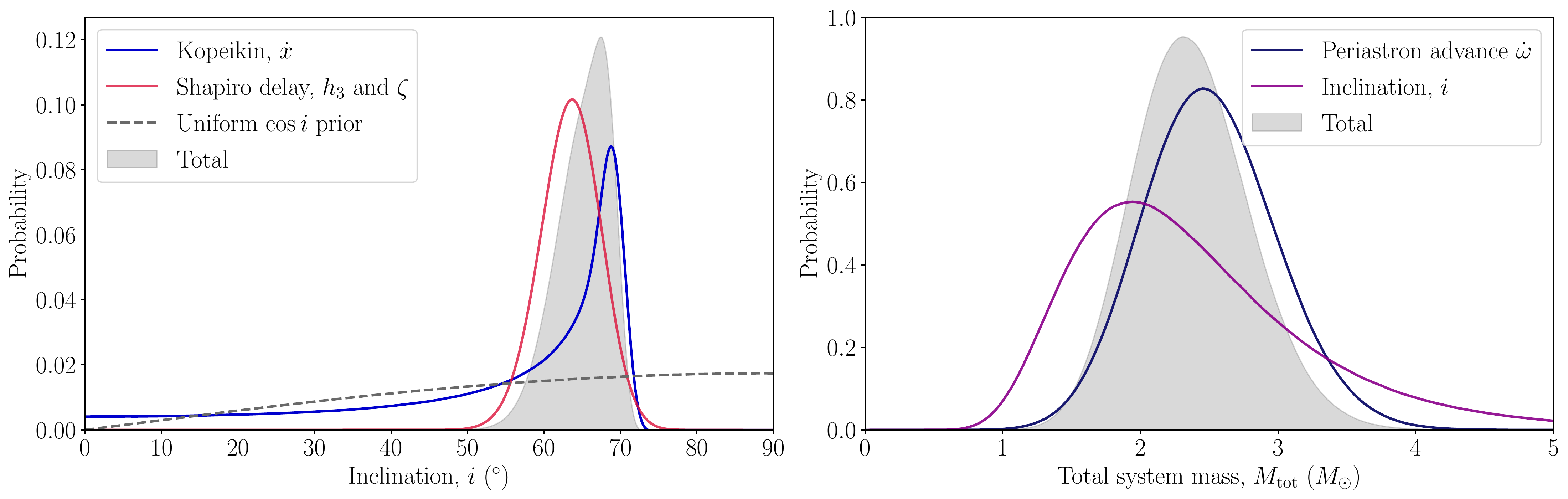}
\caption{Probability distributions (each normalised to an area of one) for orbital inclination ($i$, left panel) and total system mass ($M_{\rm tot}$, right panel) for PSR~J1600$-$3053. $i$ is constrained by our measurement of Shapiro delay (with orthometric parameters $h_3$ and $\varsigma$, red curve), and by the apparent rate-of-change of the projected semi-major axis $\dot{x}$ (blue curve). We have multiplied these distributions with a prior that is uniformly distributed in $\cos(i)$, corresponding to isotropy (grey dashed line). The total normalised probability distribution is shown as the grey, shaded region. $M_{\rm tot}$ (right panel) is constrained by the Shapiro delay parameters (providing the companion mass $M_c$) and the "total" probability distribution on $i$ shown in the left panel (purple curve), and by our measurement of the relativistic periastron advance $\dot{\omega}$ (dark blue curve). The total probability distribution for $M_{\rm tot}$ is shown in the grey shaded region. From these distributions we derive a measurement of the pulsar mass to be $M_p = 2.06^{+0.44}_{-0.41}$\,M$_\odot$.}
\label{fig:1600_mass}
\end{figure*}

For PSR~J1017$-$7156, we have detected $h_3$ with $\sim 4\sigma$ significance, but do not have a measurement of $\varsigma$. Typically in this scenario only weak constraints on $i$ and $M_c$ (and hence $M_p$) can be derived. We have computed the wide range of $i$ and $M_c$ allowed by the measured Shapiro delay parameters. By constraining the companion masses to $M_c < 1.4$\,M$_\odot$ (since it is a white dwarf), the 68\% confidence region for for these parameters are $34^\circ < i < 67^\circ$ and $0.08$\,M$_\odot$ < $M_c$ < $0.78$\,M$_\odot$. However, for this pulsar we are fortunate to have a strong detection of $\dot{x}$ which provides an independent constraint on $i$ \citep{Sandhu+97}. 

Using our $\dot{x}$ measurement alone we derive the limit $i < 47.7^\circ$ (95\% confidence), which substantially improves the above constraints derived from Shapiro delay parameters, $34.0^\circ < i < 44.2^\circ$. We also have a measurement of the relativistic periastron advance $\dot{\omega}$, which independently implies a total system mass of $M_{\rm tot} = 2.4 \pm 0.7$\,M$_\odot$. We are therefore able to bound the pulsar mass to $M_p = 2.0\pm0.8$\,M$_\odot$. 

Similarly, we determine the masses of PSRs J1022+1001 and J1600$-$3052 to be $M_p = 1.44 \pm 0.44$\,M$_\odot$ and  $M_p = 2.06^{+0.44}_{-0.41}$\,M$_\odot$, respectively. PSR~J1600$-$3053 is therefore potentially a massive neutron star. This measurement largely results from our first detection of $\dot{\omega} = 0.0043(6)$\,deg\,yr$^{-1}$. The expected kinematic contribution to this measurement \citep[Equation 12 of][]{Kopeikin96}, is orders of magnitude smaller than the measurement uncertainty. The constraints on $i$ and $M_{\rm tot}$ that are derived from our measurements of $\dot{x}$, $h_3$, $\varsigma$, and $\dot{\omega}$ are shown in Figure \ref{fig:1600_mass}.




\subsection{Astrometry}


Positions and proper motions were obtained in ecliptic coordinates (ecliptic latitude $\beta$, and longitude $\lambda$) to minimise covariances between the parameters, but we have also derived their equatorial values (right ascension $\alpha$, and declination $\delta$).

We detected significant parallax $\pi$ in all but three of the pulsars. The significance of the parallax measurements ranges from 1.3$\sigma$ for PSR~J1125$-$6014 to $64\sigma$ for J1909$-$3744.  Of the parallax measurements $18$ have  significance greater than $2\sigma$.

PSR~J0711$-$6830 is at high ecliptic latitude with $\beta\approx 82^\circ$  which makes it very insensitive to parallax. PSR~J1732$-$5049 was only observed for $7$\,yr (2004-2011), and was removed from regular observing because of its poor timing precision \cite[][]{Manchester+13}. It is now being monitored as part of the MeerTime MSP timing program.  PSR~J1824$-$2452A is associated with the globular cluster M28 (NGC 6626) which has an independent distance estimate of 5.5\,kpc \cite[][]{Harris96}.  For these three pulsars we do not include parallax terms in the pulsar ephemerides.

For all of the pulsars we detect significant proper motions. For pulsars with a significant parallax, we use these proper motions to derive the transverse space velocity $v_t$. The median magnitude from our sample of MSPs is $v_{t,50\%} = 54$\,km\,s$^{-1}$, with the lower and upper quartiles $v_{t,25\%} = 39$\,km\,s$^{-1}$ and $v_{t,75\%} = 105$\,km\,s$^{-1}$, respectively.

\subsection{Distances from orbital period-derivatives} 

\begin{figure*}
\centering
\includegraphics[width=.85\textwidth]{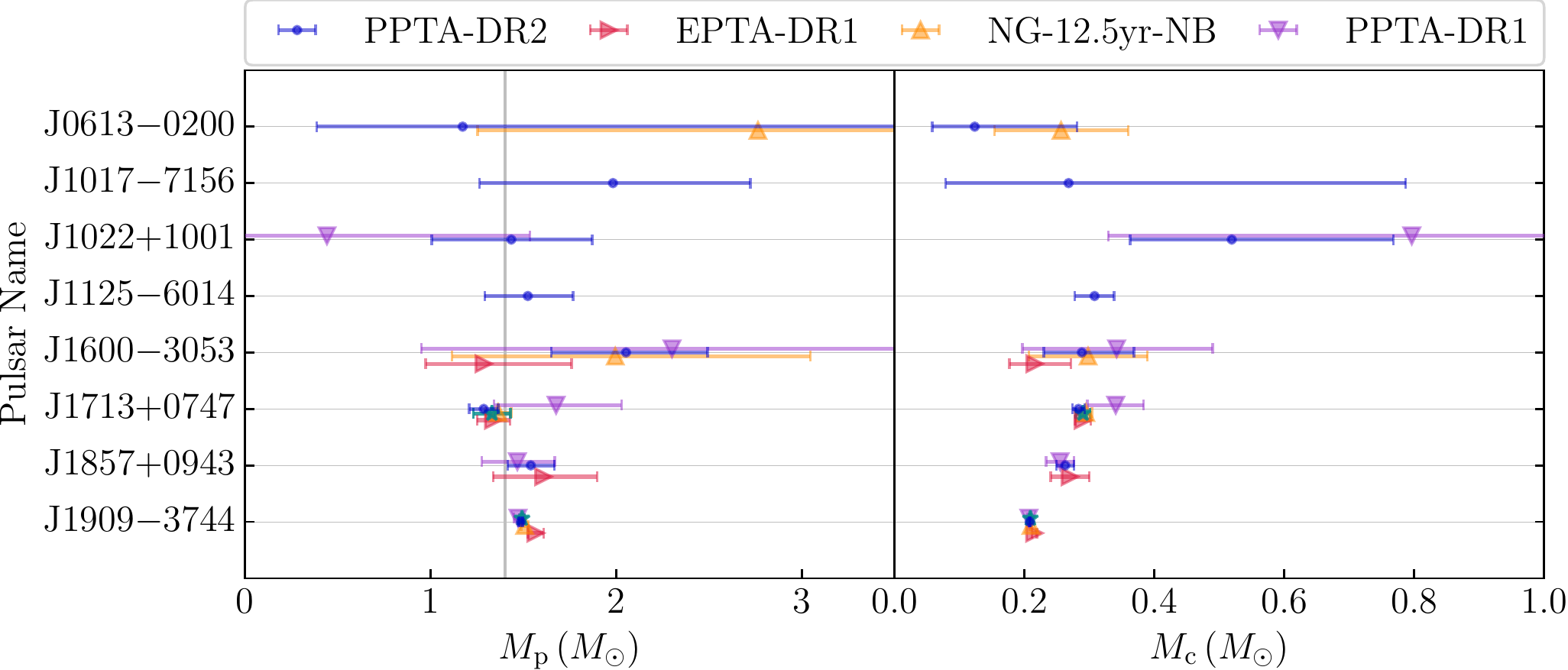}
\caption{ Pulsar and companion masses for PPTA pulsars.   The grey vertical line corresponds to the Chandrasekhar limit of $1.4~M_\odot$. PSRs~J1545$-$4550, J1732$-$5049, and J2145$-$0750 are excluded because we detect $h_3$ but cannot derive meaningful mass constraints. The cyan measurements with a star symbol for PSRs J1713+0747 and J1909$-$3744 are from the single-pulsar analyses of \citet{Zhu+19} and \citet{Liu+20}, respectively.}
\label{fig:mass_comparison}
\end{figure*}


The most precise measurement of $\dot{P}_b$ in the sample is for PSR~J1909$-$3744, for which we measure $\dot{P}_b=5.093(7)\times10^{-13}$.  For this pulsar it is necessary to both model the  Galactic contribution, which is  $\dot{P}_b^{\rm Gal} = 4(1)\times10^{-15}$, but also account for the subdominant contribution from gravitational wave emission  $\dot{P}_{\rm GR} =  -2.763(3)\times10^{-15}$.   After this correction we find  $D_{\rm shk} = 1.152(3)$\,kpc. This measurement uncertainty ($\sim 10$\,lyr), means that a search for continuous gravitational wave sources with periods greater than $\sim$10\,years can utilise both the pulsar and Earth terms for PSR~J1909$-$3744 \citep[and PSR~J0437$-$4715,][]{Reardon+16}.

The largest value of $\dot{P}_b$ is measured for PSR~J2129$-$5721, for which we find  $\dot{P}_b=1.7(2)\times10^{-12}$ with PPTA-DR2 and $\dot{P}_b=1.51(9)\times10^{-12}$ with PPTA-DR2e. The latter measurement with better precision implies a large distance of $D_{\rm shk} = 7.0\pm 0.4$\,kpc. The parallax distance is poorly constrained with $D_\pi = 3.6^{+5}_{-1.4}$\, kpc. The distance to the pulsar inferred from its dispersion measure is 1.4\,kpc using the NE2001 model \cite[][]{Cordes+02} and 6.5\,kpc according to the YMW16 model \cite[][]{Yao+17}.  The high galactic latitude $|b| \approx 44^\circ$ implies the pulsar is 4.9(3)\,kpc above the Galactic plane.   

\section{Discussion}\label{sec:discuss}


\begin{figure*}
\centering
\includegraphics[width=.85\textwidth]{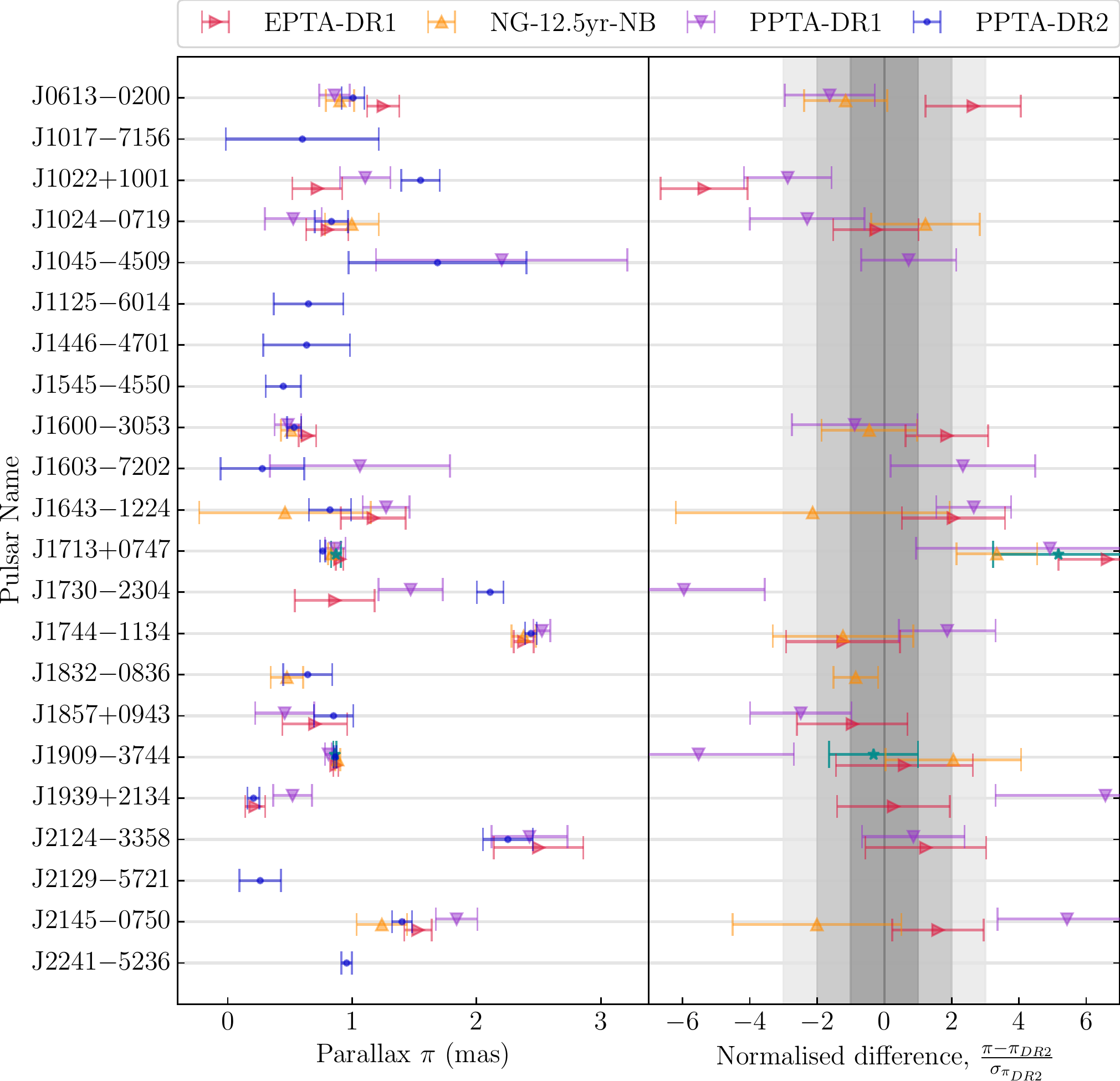}
\caption{Left:  Parallax measurements for each of the PPTA-DR2e pulsars (blue) with measurements from other PTAs (EPTA-DR1 in red, NANOGrav 12.5\,yr in orange, and PPTA-DR1 in purple).  Right:  Normalised difference between PPTA-DR2e and previous measurements, shown with respect to the PPTA-DR2e uncertainty. The cyan measurements with a star symbol for PSRs J1713+0747 and J1909$-$3744 are from the single-pulsar analyses of \citet{Zhu+19} and \citet{Liu+20}, respectively.}
\label{fig:PX_comparison}
\end{figure*}

\subsection{Performance of the extended data set} 

For many pulsar parameters, timing precision greatly improves as the timing baseline increases. However, the extension of timing baselines with older data will result in the inclusion of data with lower precision. Over the history of the PPTA project and previous precision timing programs, instrumentation (receivers and importantly digital systems) have improved, resulting in increases in fidelity and precision of the observations. We have assessed the importance of this early data by comparing parameter values and uncertainties obtained with the DR2 and DR2e data sets. 

For most of the pulsars, we see a marginal improvement in pulsar parameter estimation using the longer PPTA-DR2e data set. We compared astrometric and binary parameters from the $16$ pulsars in the extended data sets. Of the  $152$ parameters compared, we found that $113$ of the parameters ($74\%$) had improvements in their formal uncertainty from the extended data.  Only 15 showed more than 20\% improvement (and four showed more than 50\% improvement), with the largest improvements being in the binary period of PSR J1603$-$7202 (uncertainties decreasing by a factor of $2.1$) and binary period derivatives of PSRs J1603$-$7202, J2129$-$5721, and J2145$-$0750, which respectively improved in uncertainty by factors of $1.9$, $ 1.7$, and $2.2$. Only three of the parameters showed discrepancies greater than $1.1\sigma$, with the binary period derivative of PSR~J1713$+$0747 differing by $1.4\sigma$ and the remaining two being inconsistent at $1.1\sigma$. 

\subsection{Performance of the frequency-averaged data set}

The frequency-averaged data set derived from the sub-banded PPTA-DR2e is more sensitive than the original frequency-averaged data set released with PPTA-DR2 \citep{Kerr+20} because of the additional profile-evolution model and the robust white noise correction applied to the sub-bands before averaging. We have found that the timing model parameters measured with this alternate data set are largely consistent with the parameters presented in this paper. Of the $309$ parameters compared, $273$  (88\%) in the frequency averaged data set were consistent. The largest discrepancies between the two datasets is the position of PSR~J1600$-$3053, which shows a $3.3\sigma$ discrepancy in $\lambda$, and in PSR~J1909$-$3744, which shows a $4.3\sigma$ discrepancy in $T_{\rm asc}$  and $2\sigma$ discrepancies in $\dot{P_b}$ and $\mu_\lambda$.

We note that no barycenteric correction (pulsar or solar system barycentres) is made to the frequencies prior to frequency-averaging, and that any correlation between the timing model parameters and the profile-evolution (FD parameters) is also not accounted for. Frequency-averaged datasets such as this can be used for accelerated compute time in high-dimensional problems like Bayesian model selection in searches for low-frequency gravitational waves. However, we caution against their use in general because of the assumptions made in forming the dataset and the timing model errors that can result. Future PPTA datasets will make use of wide-bandwidth timing techniques \citep[e.g.][]{Pennucci+14, Liu+14} to reduce the size of the final dataset, instead of averaging many sub-bands.

\subsection{Comparison to previous analyses}

For pulsars where we have derived a mass, we compare these values to other PTA datasets by employing the same Monte Carlo simulation technique that we use to derive our values. The results are shown in Figure \ref{fig:mass_comparison}, and demonstrate good agreement between the independent datasets. The pulsar and companion masses for PSR~J1909$-$3744 are too precise to compare on the scale in this figure. Our measurements for this pulsar are $M_c = 0.2081 \pm 0.0009$\,M$_\odot$ and $M_p = 1.486 \pm 0.011$\,M$_\odot$. The most precise independent measurements are from the analysis of $\sim15$\,yrs of data from the EPTA, with $M_c = 0.209 \pm 0.001$\,M$_\odot$ and $M_p = 1.492 \pm 0.014$\,M$_\odot$, and from the NANOGrav 12.5\,yr dataset, $M_c = 0.2099 \pm 0.0017$\,M$_\odot$ and $M_p = 1.5060 \pm 0.021$\,M$_\odot$.

\subsubsection{Mass of PSR~J1600$-$3053}

The derived mass of PSR~J1600$-$3053, $M_p = 2.06^{+0.44}_{-0.41}$\,M$_\odot$, is of interest because massive pulsars are valuable tools for constraining the neutron star equation of state \citep[][]{Lattimer+01}. Pulsars with measured masses $\gtrsim 2$\,M$_\odot$ are rare and are not expected by all dense matter models that predict an equation of state. Thus they can be used to refute models that predict a lower maximum neutron star mass. The measurement uncertainty is far too large to constrain these models, but following the measurement of $\dot{\omega}$, this uncertainty will decrease more rapidly with time than with the Shapiro delay alone.

Our measurement is more precise than previous estimates because of the detection of this relativistic $\dot{\omega}$. The derivation assumes a Gaussian distribution for our measurement of Shapiro delay parameters $h_3$ and $\varsigma$. If we instead use $M_c$ and $\sin{i}$ the inferred mass distribution does not change significantly.

Considering only the Shapiro delay (with $h_3$, $\varsigma$) without $\dot{\omega}$, we would derive $M_p = 1.97^{+1.0}_{-0.7}$\,M$_\odot$. We compare this measurement with the other PTAs, which have only measured the Shapiro delay: $M_p = 2.1^{+1.5}_{-1.4}$\,M$_\odot$ (PPTA-DR1), $M_p = 1.22^{+0.5}_{-0.4}$\,M$_\odot$ (EPTA-DR1), and $M_p = 2.0^{+1.1}_{-1.0}$\,M$_\odot$ (NANOGrav-12.5yr-NB). The measurement from EPTA-DR1 \citep[with $h_3=3.3 \pm 0.2$\,$\mu$s, and $\varsigma = 0.68 \pm 0.05$,][]{Desvignes+16} is the most in tension with mass constraints from other PTAs, and differs at the $\sim 1.3\sigma$ level from our mass constraint including $\dot{\omega}$.

Combining the measurements from each independent PTA (i.e. without the PPTA-DR1 constraint above), we derive $M_p = 1.77^{+0.36}_{-0.32}$\,M$_\odot$. A combined dataset of these three PTAs, as well as observations from MeerKAT and other observatories, should be able to significantly improve the measurement of both the Shapiro delay and $\dot{\omega}$, which would then determine whether this pulsar is indeed massive. A future analysis of such a combined dataset could also use existing Bayesian inference techniques \citep[e.g.,][]{Lentati+14} to measure the noise model and timing model simultaneously, which would then capture any correlations between the noise and timing model parameters of interest.

\subsubsection{Parallaxes}\label{sec:parallax}

Our parallax measurements are displayed and compared with previous measurements in Figure \ref{fig:PX_comparison}. In general our measurements are in agreement with these earlier measurements, with the largest discrepancies being for PSRs~J1022$+$1001, ~J1730$-$2304 and ~J1713$+$0747. 

The inconsistent measurements for both PSRs~J1022$+$1001 and ~J1730$-$2304 between PPTA-DR2 and EPTA-DR1 may be due to their proximity to the ecliptic plane, with $\beta \sim -0.064 ^{\circ}$ and $\beta \sim 0.189 ^{\circ}$, respectively. 

Our measurement of $\pi=2.11 \pm 0.11$\,mas with PPTA-DR2 for PSR~J1730-2304 corresponds to a distance of $D_\pi = 0.47 \pm 0.03$\,kpc, which is consistent with distance estimates from Galactic electron density models, $D=0.53$\,kpc from \citet{Cordes+02} and $D=0.512$\,kpc from \citet{Yao+17}. These pulsars were also analysed with a combined data set of PPTA-DR1 and EPTA-DR1, under the International Pulsar Timing Array (IPTA) second data release \citep{Perera+19}. This combined analysis found $\pi=1.67 \pm 0.22$\,mas for PSR~J1730-2304, which is closer to our measurement than the values from PPTA-DR1 ($\pi=1.47 \pm 0.26$\, mas) or EPTA-DR1 ($\pi=0.86 \pm 0.32$\, mas) individually.

Our measured parallax for PSR~J1713+0747 is $\pi = 0.763 \pm 0.021$\,mas, while the most precise independent measurement is $\pi = 0.830 \pm 0.024$\,mas from \citet{NG12Narrowband}, which differs by 2.1$\sigma$. An earlier measurement of  $\pi = 0.90 \pm 0.03$\,mas from EPTA-DR1 is inconsistent at the 3.7$\sigma$ level \citep{Desvignes+16}. The data used for this measurement were included in a combined EPTA and NANOGrav analysis that measured $\pi = 0.87 \pm 0.04$\,mas \citep{Zhu+19}, which differs by 2.4$\sigma$ from our measurement.

The noise model for PSR~J1713$+$0747 is one of the most complex among this sample of pulsars, with a system-dependent noise processes, and two ``chromatic dip" events with exponential-like recoveries. These events were were initially observed through dispersion measure variations \citep{Coles+15, NG11}, thus they were thought to be associated with the interstellar medium. However, \citet{PPTADR2noise} found strong evidence for a chromaticity that was inconsistent with DM variations for the second of these events, and additionally identified a change in profile shape coincident with the event. This chromaticity, profile shape change, and the exponential recovery can be understood by a disturbance to the pulsar's magnetosphere similar to an event observed in PSR~J1643$-$1224 \citep{Shannon+16}. If this second event is removed from our timing model and the event is modelled only using DM variations, our parallax measurement becomes $\pi = 0.805 \pm 0.026$\,mas. If the first chromatic event is also removed from our model such that it too is modelled using only DM variations, the recovered parallax becomes $\pi = 0.822 \pm 0.027$\,mas, showing that the parallax measurement can be affected by sub-optimal noise models.

There are potential weaknesses in the noise models used to obtain previous measurements for PSR~J1713$+$0747, including: \citet{Reardon+16} modelled the first event using linearly-interpolated DM variations; \citet{Desvignes+16} modelled the first event with shapelet basis functions for DM instead of a step-function with exponential recovery; and \citet{NG12Narrowband} modelled the first and second events with DM variations. Recently, a third, much larger amplitude chromatic event has been observed (this event is not included in our data set), which is associated with a profile shape change \citep{Xu+21}. While observations are ongoing, the event may also have a chromatic dependence that is different from the $f^{-2}$ relationship of DM variations and an exponential-like recovery similar to the magnetospheric event in PSR~J1643$-$1224 \citep{Shannon+16}. This highlights the need for future analyses to characterise similar events with an appropriate model to avoid biases in timing model parameters and gravitational wave searches using these data sets.

We have verified that our noise model for PSR~J1713+0747 is sufficient to whiten and produce Gaussian-distributed normalised residuals with unit variance in each of our three frequency bands separately \citep[][see also Appendix \ref{sec:residuals}]{PPTADR2noise}.

\section{Conclusions}\label{sec:conclusions}

We have described the updated timing models for 25 MSPs using the second data release of the PPTA, which we extended to include up to 10\,yrs of legacy observations. New parameters were included in the timing models and a detailed noise analysis for each pulsar was conducted using the resulting set of parameters \citep[described in ][]{PPTADR2noise}. These noise models were determined using the sub-banded PPTA-DR2 dataset of \citet{Kerr+20}, and we have extrapolated them to describe the extended dataset by modifying the frequency coverage of the power-law noise processes. After determining the optimal set of timing model parameters with this PPTA-DR2e, including the best description of the pulsar's binary orbit, we formed whitened (i.e. with the low-frequency noise model subtracted) and normalised post-fit residuals to verify that the timing model parameter measurements and uncertainties from a generalised least-squares fit are reliable. The pulsar ephemerides with post-fit timing model solutions, noise models, and original and extended sets of TOAs, are available for download at \url{https://doi.org/10.25919/cx59-a798}.

Using the new ephemerides we derived additional physical parameters of interest, including pulsar and companion masses where available, limits or constraints on the orbital inclination, pulsar distances from parallax and the Shklovskii effect, and transverse space velocities. The measured and derived parameters were compared with earlier measurements. In general the timing models show good agreement, but we note that noise models can vary substantially between data sets. While some parameter discrepancies are expected to arise by chance when comparing hundreds of parameters across a sample of 24 pulsars, differences in noise models may be responsible for the largest differences.

We found new Shapiro delay signatures in four pulsars, including a high-significance measurement in PSR~J1125$-$6014. We derived pulsar masses with constraints from Shapiro delay measurements, kinematics ($\dot{x}$), and relativity ($\dot{\omega}$). We found $M_p = 1.5 \pm 0.2$\,M$_\odot$ from the Shapiro delay for PSR~J1125$-$6014, and $M_p = 2.06^{+0.44}_{-0.41}$\,M$_\odot$ from the Shapiro delay and new $\dot{\omega}$ measurement for PSR~J1600$-$3053.

We reported on one measurement of annual-orbital parallax for PSR~J1713+0747, which had previously been measured \citep{Zhu+19}, as well as weak evidence for an annual-orbital parallax in PSRs~J1643$-$1224 and J1909$-$3744, each giving a likely solution for $i$ and $\Omega$.

It is also possible to use measurements of scintillation to constrain these orbital properties of the pulsars. The relative transverse velocity of the pulsar and Earth change the temporal modulation in the interference pattern produced by diffractive scintillation, which provides complementary information to the radial effects measured through pulsar timing. Several scintillating pulsars show variability in their diffractive scintillation timescale, which can be modelled as in \citet{Rickett+14} and \citet{Reardon+19} to provide more precise measurements of $i$ and improve the mass constraints. Modulation in the curvature of scintillation arcs could also be detected in some pulsars and provide a more precise measurement of $i$ and $\Omega$ than the scintillation timescale \citep{Reardon+20}.

The Parkes Ultra-wideband low receiver \cite[][]{Hobbs+20} has been adopted for all PPTA observations after the conclusion of PPTA-DR2. The wide bandwidth coverage means that there is an increased chance of observing bright scintles for low-DM pulsars, which can have scintillation bandwidths of $10-10^3$~MHz. This should result in higher mean timing precision for the TOAs in the next PPTA data release, but is also important for the scintillation studies of binary pulsars.

The MeerKAT telescope and the MeerTime project \cite{Bailes+20} will provide the opportunity to study the binary and kinematic properties of a larger number of southern hemisphere millisecond pulsars. The MeerTime relativistic binary program \citep{Kramer+21} will include observing campaigns that target the Shapiro delay signatures of several pulsars included in this PPTA sample, which will lead to improved pulsar masses, particularly when combined with these PPTA data.

\section*{Data Availability}
The data presented in this work are available from \url{https://doi.org/10.25919/cx59-a798}. Included are the TOAs (original sub-banded data and frequency-averaged), \textsc{tempo2}-compatible pulsar ephemerides, and auxiliary files including the observatory clock correction and scripts used for frequency-averaging and parameter derivations.

\section*{Acknowledgements}

The Parkes radio telescope (Murriyang) is part of the Australia Telescope, which is funded by the Commonwealth Government for operation as a National Facility managed by CSIRO. This research was funded partially by the Australian Government through the Australian Research Council (ARC), grants CE170100004 (OzGrav) and FL150100148. This work was performed on the OzSTAR national facility at Swinburne University of Technology. The OzSTAR program receives funding in part from the Astronomy National Collaborative Research Infrastructure Strategy (NCRIS) allocation provided by the Australian Government. Work at NRL is supported by NASA. R.M.S. acknowledges funding support through Australian Research Council Future Fellowship FT190100155. S.D. is the recipient of an Australian Research Council Discovery Early Career Award (DE210101738) funded by the Australian Government. J.B.W. is supported by the National Natural Science Foundation of China (No.12041304) and the National SKA Program of China (No.2020SKA0120100).




\bibliographystyle{mnras}
\bibliography{ppta_ephem} 




\appendix

\section{Pulsar ephemerides}\label{sec:ephemerides}

Tables of pulsar ephemerides and derived parameters. Parameters fitted as in the timing model are highlighted in bold font. Values in parentheses are the standard error on the last quoted digit.

\begin{landscape}
\clearpage

        \begin{table}
        \footnotesize
        \begin{tabular}{llllllll}
        \hline\hline \noalign{\vskip 1.5mm}
        Pulsar Name 	 & 	 J0711$-$6830	 & 	 J1024$-$0719	 & 	 J1730$-$2304	 & 	 J1744$-$1134 
 \\ \hline \noalign{\vskip 1.5mm} 
Number of TOAs\dotfill	 & 	 $5628$	 & 	 $2735$	 & 	 $4624$	 & 	 $6860$\\ 
Number of observations\dotfill	 & 	 $1258$	 & 	 $776$	 & 	 $806$	 & 	 $1150$\\ 
MJD range\dotfill	 & 	 $49373$ -- $58232$	 & 	 $50117$ -- $58187$	 & 	 $49421$ -- $58232$	 & 	 $49729$ -- $58232$\\ 
Right ascension (RA), $\alpha$ (hh:mm:ss)\dotfill	 & 	 $7$:$11$:$54.180362(7)$	 & 	 $10$:$24$:$38.671419(4)$	 & 	 $17$:$30$:$21.67100(3)$	 & 	 $17$:$44$:$29.409783(1)$\\ 
Declination (DEC), $\delta$ (dd:mm:ss)\dotfill	 & 	 $-68$:$30$:$47.37034(4)$	 & 	 $-7$:$19$:$19.5146(1)$	 & 	 $-23$:$4$:$31.147(9)$	 & 	 $-11$:$34$:$54.71069(7)$\\ 
Proper motion in RA, $\mu_\alpha \cos\delta$ (${\rm mas}\,{\rm yr}^{-1}$)\dotfill	 & 	 $-15.568(10)$	 & 	 $-35.270(17)$	 & 	 $20.06(12)$	 & 	 $18.803(4)$\\ 
Proper motion in DEC, $\mu_\delta$ (${\rm mas}\,{\rm yr}^{-1}$)\dotfill	 & 	 $14.175(11)$	 & 	 $-48.22(3)$	 & 	 $-4(2)$	 & 	 $-9.390(18)$\\ 

 \noalign{\vskip 1.5mm} 
Spin frequency, $f$ (${\rm s}^{-1}$)\dotfill	 & 	 $\mathbf{ 182.1172346200484(6) }$	 & 	 $\mathbf{ 193.7156834116254(7) }$	 & 	 $\mathbf{ 123.1102871305625(3) }$	 & 	 $\mathbf{ 245.4261196602377(2) }$\\ 
First spin frequency derivative, ${\dot{f}}$ (${\rm s}^{-2}$)\dotfill	 & 	 $\mathbf{ -4.94426(6)\times 10^{-16} }$	 & 	 $\mathbf{ -6.96131(7)\times 10^{-16} }$	 & 	 $\mathbf{ -3.05912(3)\times 10^{-16} }$	 & 	 $\mathbf{ -5.38155(2)\times 10^{-16} }$\\ 
Dispersion measure, DM (${\rm cm}^{-3}\,{\rm pc}$)\dotfill	 & 	 $18.408$	 & 	 $6.4788$	 & 	 $9.6268$	 & 	 $3.13967$\\ 
Ecliptic latitude $\beta$ (deg.)\dotfill	 & 	 $\mathbf{ -82.888631513(11) }$	 & 	 $\mathbf{ -16.04469990(3) }$	 & 	 $\mathbf{ 0.188866(3) }$	 & 	 $\mathbf{ 11.805198877(19) }$\\ 
Ecliptic longitude $\lambda$ (deg.)\dotfill	 & 	 $\mathbf{ 204.06115141(9) }$	 & 	 $\mathbf{ 160.73435836(1) }$	 & 	 $\mathbf{ 263.186041168(8) }$	 & 	 $\mathbf{ 266.119406515(4) }$\\ 

 \noalign{\vskip 1.5mm} 
Proper motion in ecliptic latitude, $\mu_\beta$ (${\rm mas}\,{\rm yr}^{-1}$)\dotfill	 & 	 $\mathbf{ -17.267(9) }$	 & 	 $\mathbf{ -58.00(3) }$	 & 	 $\mathbf{ -2(2) }$	 & 	 $\mathbf{ -8.895(15) }$\\ 
Proper motion in ecliptic longitude, $\mu_\lambda \cos\beta$ (${\rm mas}\,{\rm yr}^{-1}$)\dotfill	 & 	 $\mathbf{ -12.049(9) }$	 & 	 $\mathbf{ -14.391(9) }$	 & 	 $\mathbf{ 20.251(7) }$	 & 	 $\mathbf{ 19.054(3) }$\\ 
Parallax, $\pi$ (${\rm mas}$)\dotfill	 & 	 $-$	 & 	 $\mathbf{ 0.83(13) }$	 & 	 $\mathbf{ 2.11(11) }$	 & 	 $\mathbf{ 2.44(5) }$\\ 
Parallax distance, $D_\pi$ (kpc)\dotfill	 & 	 $-$	 & 	 ${ 1.20 } ^{ +0.3 }_{ -0.17 }$	 & 	 $0.47(3)$	 & 	 ${ 0.410 } ^{ +0.009 }_{ -0.008 }$\\ 
Transverse velocity, $v_t$ (km\,s$^{-1}$)\dotfill	 & 	 $-$	 & 	 $340^{ +70 }_{ -50 }$	 & 	 $46(3)$	 & 	 ${ 40.9 } ^{ +0.9 }_{ -0.8 }$\\ 

        \noalign{\vskip 1.5mm}
        \hline\hline
        \end{tabular}\hfill\
        \caption{\label{tab:XXXXX}
        Measured and derived timing model parameters for solitary pulsars J0711$-$6830, J1024$-$0719 (the model for this pulsar also includes $\ddot{f} = -3.89(8)\times 10^{-27}$\,s$^{-3}$), J1730$-$2304, and J1744$-$1134. 
        }
        \end{table}
        
\clearpage

        \begin{table}
        \footnotesize
        \begin{tabular}{llllllll}
        \hline\hline \noalign{\vskip 1.5mm}
        Pulsar Name 	 & 	 J1824$-$2452A	 & 	 J1832$-$0836	 & 	 J1939+2134	 & 	 J2124$-$3358 
 \\ \hline \noalign{\vskip 1.5mm} 
Number of TOAs\dotfill	 & 	 $2626$	 & 	 $326$	 & 	 $5008$	 & 	 $5176$\\ 
Number of observations\dotfill	 & 	 $470$	 & 	 $111$	 & 	 $728$	 & 	 $1227$\\ 
MJD range\dotfill	 & 	 $53163$ -- $58202$	 & 	 $56260$ -- $58231$	 & 	 $49956$ -- $58229$	 & 	 $49489$ -- $58230$\\ 
Right ascension (RA), $\alpha$ (hh:mm:ss)\dotfill	 & 	 $18$:$24$:$32.007884(9)$	 & 	 $18$:$32$:$27.592338(5)$	 & 	 $19$:$39$:$38.5612568(6)$	 & 	 $21$:$24$:$43.845870(6)$\\ 
Declination (DEC), $\delta$ (dd:mm:ss)\dotfill	 & 	 $-24$:$52$:$10.872(3)$	 & 	 $-8$:$36$:$55.0339(3)$	 & 	 $21$:$34$:$59.12487(1)$	 & 	 $-33$:$58$:$45.0065(2)$\\ 
Proper motion in RA, $\mu_\alpha \cos\delta$ (${\rm mas}\,{\rm yr}^{-1}$)\dotfill	 & 	 $-0.25(4)$	 & 	 $-8.06(5)$	 & 	 $0.070(3)$	 & 	 $-14.109(19)$\\ 
Proper motion in DEC, $\mu_\delta$ (${\rm mas}\,{\rm yr}^{-1}$)\dotfill	 & 	 $-8.6(8)$	 & 	 $-21.01(19)$	 & 	 $-0.406(4)$	 & 	 $-50.36(4)$\\ 

 \noalign{\vskip 1.5mm} 
Spin frequency, $f$ (${\rm s}^{-1}$)\dotfill	 & 	 $\mathbf{ 327.405572743031(7) }$	 & 	 $\mathbf{ 367.767115417227(2) }$	 & 	 $\mathbf{ 641.928222127829(8) }$	 & 	 $\mathbf{ 202.793893699620(1) }$\\ 
First spin frequency derivative, ${\dot{f}}$ (${\rm s}^{-2}$)\dotfill	 & 	 $\mathbf{ -1.735323(3)\times 10^{-13} }$	 & 	 $\mathbf{ -1.11935(12)\times 10^{-15} }$	 & 	 $\mathbf{ -4.330987(4)\times 10^{-14} }$	 & 	 $\mathbf{ -8.45929(8)\times 10^{-16} }$\\ 
Dispersion measure, DM (${\rm cm}^{-3}\,{\rm pc}$)\dotfill	 & 	 $119.9069$	 & 	 $28.1906$	 & 	 $71.01662$	 & 	 $4.5957$\\ 
Ecliptic latitude $\beta$ (deg.)\dotfill	 & 	 $\mathbf{ -1.5488074(7) }$	 & 	 $\mathbf{ 14.59071381(8) }$	 & 	 $\mathbf{ 42.296752229(4) }$	 & 	 $\mathbf{ -17.81882907(4) }$\\ 
Ecliptic longitude $\lambda$ (deg.)\dotfill	 & 	 $\mathbf{ 275.564751368(18) }$	 & 	 $\mathbf{ 278.292004343(19) }$	 & 	 $\mathbf{ 301.973244511(3) }$	 & 	 $\mathbf{ 312.738850328(14) }$\\ 

 \noalign{\vskip 1.5mm} 
Proper motion in ecliptic latitude, $\mu_\beta$ (${\rm mas}\,{\rm yr}^{-1}$)\dotfill	 & 	 $\mathbf{ -8.6(8) }$	 & 	 $\mathbf{ -20.50(19) }$	 & 	 $\mathbf{ -0.413(3) }$	 & 	 $\mathbf{ -43.01(4) }$\\ 
Proper motion in ecliptic longitude, $\mu_\lambda \cos\beta$ (${\rm mas}\,{\rm yr}^{-1}$)\dotfill	 & 	 $\mathbf{ -0.613(20) }$	 & 	 $\mathbf{ -9.27(5) }$	 & 	 $\mathbf{ -0.022(2) }$	 & 	 $\mathbf{ -29.728(11) }$\\ 
Parallax, $\pi$ (${\rm mas}$)\dotfill	 & 	 $-$	 & 	 $\mathbf{ 0.64(20) }$	 & 	 $\mathbf{ 0.21(5) }$	 & 	 $\mathbf{ 2.3(2) }$\\ 
Parallax distance, $D_\pi$ (kpc)\dotfill	 & 	 $-$	 & 	 ${ 1.6 } ^{ +0.7 }_{ -0.4 }$	 & 	 ${ 4.8 } ^{ +1.5 }_{ -1 }$	 & 	 ${ 0.44 } ^{ +0.05 }_{ -0.04 }$\\ 
Transverse velocity, $v_t$ (km\,s$^{-1}$)\dotfill	 & 	 $-$	 & 	 $170^{ +80 }_{ -40 }$	 & 	 ${ 9.5 } ^{ +3 }_{ -1.8 }$	 & 	 $110^{ +11 }_{ -9 }$\\ 

        \noalign{\vskip 1.5mm}
        \hline\hline
        \end{tabular}\hfill\
        \caption{\label{tab:XXXXX}
        Measured and derived timing model parameters for solitary pulsars J1824$-$2452A, J1832$-$0836, J1939+2134, and J2124$-$3358.
        }
        \end{table}
        
\clearpage

        \begin{table}
        \footnotesize
        \begin{tabular}{llllllll}
        \hline\hline \noalign{\vskip 1.5mm}
        Pulsar Name 	 & 	 J0613$-$0200	 & 	 J1017$-$7156	 & 	 J1022+1001	 & 	 J1045$-$4509 
 \\ \hline \noalign{\vskip 1.5mm} 
Number of TOAs\dotfill	 & 	 $6044$	 & 	 $4053$	 & 	 $7612$	 & 	 $5807$\\ 
Number of observations\dotfill	 & 	 $1154$	 & 	 $736$	 & 	 $1166$	 & 	 $1100$\\ 
MJD range\dotfill	 & 	 $51526$ -- $58230$	 & 	 $55393$ -- $58232$	 & 	 $52649$ -- $58230$	 & 	 $49405$ -- $58212$\\ 
Right ascension (RA), $\alpha$ (hh:mm:ss)\dotfill	 & 	 $6$:$13$:$43.975902(2)$	 & 	 $10$:$17$:$51.321896(6)$	 & 	 $10$:$22$:$57.995(1)$	 & 	 $10$:$45$:$50.18519(1)$\\ 
Declination (DEC), $\delta$ (dd:mm:ss)\dotfill	 & 	 $-2$:$0$:$47.24352(8)$	 & 	 $-71$:$56$:$41.61771(3)$	 & 	 $10$:$1$:$52.69(4)$	 & 	 $-45$:$9$:$54.1062(1)$\\ 
Proper motion in RA, $\mu_\alpha \cos\delta$ (${\rm mas}\,{\rm yr}^{-1}$)\dotfill	 & 	 $1.836(5)$	 & 	 $-7.411(12)$	 & 	 $-19(3)$	 & 	 $-6.07(3)$\\ 
Proper motion in DEC, $\mu_\delta$ (${\rm mas}\,{\rm yr}^{-1}$)\dotfill	 & 	 $-10.349(13)$	 & 	 $6.870(11)$	 & 	 $-5(9)$	 & 	 $5.19(4)$\\ 

 \noalign{\vskip 1.5mm} 
Spin frequency, $f$ (${\rm s}^{-1}$)\dotfill	 & 	 $\mathbf{ 326.600561967271(1) }$	 & 	 $\mathbf{ 427.6219050534547(8) }$	 & 	 $\mathbf{ 60.7794479478993(2) }$	 & 	 $\mathbf{ 133.7931495240541(7) }$\\ 
First spin frequency derivative, ${\dot{f}}$ (${\rm s}^{-2}$)\dotfill	 & 	 $\mathbf{ -1.022991(19)\times 10^{-15} }$	 & 	 $\mathbf{ -4.0483(1)\times 10^{-16} }$	 & 	 $\mathbf{ -1.601009(15)\times 10^{-16} }$	 & 	 $\mathbf{ -3.16197(8)\times 10^{-16} }$\\ 
Dispersion measure, DM (${\rm cm}^{-3}\,{\rm pc}$)\dotfill	 & 	 $38.7595$	 & 	 $94.21829$	 & 	 $10.2442$	 & 	 $58.1238$\\ 
Ecliptic latitude $\beta$ (deg.)\dotfill	 & 	 $\mathbf{ -25.40713853(2) }$	 & 	 $\mathbf{ -67.736229381(7) }$	 & 	 $\mathbf{ -0.063965(10) }$	 & 	 $\mathbf{ -47.71478033(4) }$\\ 
Ecliptic longitude $\lambda$ (deg.)\dotfill	 & 	 $\mathbf{ 93.799007973(10) }$	 & 	 $\mathbf{ 222.427790415(20) }$	 & 	 $\mathbf{ 153.865859249(11) }$	 & 	 $\mathbf{ 186.51853722(4) }$\\ 

 \noalign{\vskip 1.5mm} 
Proper motion in ecliptic latitude, $\mu_\beta$ (${\rm mas}\,{\rm yr}^{-1}$)\dotfill	 & 	 $\mathbf{ -10.301(13) }$	 & 	 $\mathbf{ -4.820(12) }$	 & 	 $\mathbf{ -4(8) }$	 & 	 $\mathbf{ 0.88(3) }$\\ 
Proper motion in ecliptic longitude, $\mu_\lambda \cos\beta$ (${\rm mas}\,{\rm yr}^{-1}$)\dotfill	 & 	 $\mathbf{ 2.106(5) }$	 & 	 $\mathbf{ -8.882(11) }$	 & 	 $\mathbf{ -15.92(1) }$	 & 	 $\mathbf{ -7.95(2) }$\\ 
Parallax, $\pi$ (${\rm mas}$)\dotfill	 & 	 $\mathbf{ 1.01(9) }$	 & 	 $\mathbf{ 0.6(6) }$	 & 	 $\mathbf{ 1.55(16) }$	 & 	 $\mathbf{ 1.7(7) }$\\ 
Obital period, $P_{\mathrm{b}}$ ($\mathrm{d}$)\dotfill	 & 	 $\mathbf{ 1.19851257517(4) }$	 & 	 $\mathbf{ 6.5119041(9) }$	 & 	 $\mathbf{ 7.8051353(6) }$	 & 	 $\mathbf{ 4.08352925449(8) }$\\ 
Projected semimajor axis, $x$ (lt-s)\dotfill	 & 	 $\mathbf{ 1.09144421(4) }$	 & 	 $\mathbf{ 4.83003(7) }$	 & 	 $\mathbf{ 16.7654054(19) }$	 & 	 $\mathbf{ 3.0151324(5) }$\\ 

 \noalign{\vskip 1.5mm} 
Epoch of periastron, $T_0$ (MJD)\dotfill	 & 	 $50324.47(15)$	 & 	 $\mathbf{ 55335.061(7) }$	 & 	 $\mathbf{ 49778.4079(5) }$	 & 	 $50196.88(14)$\\ 
Longitude of periastron, $\omega$ (deg)\dotfill	 & 	 $48.3(8)$	 & 	 $\mathbf{ 329.5(4) }$	 & 	 $\mathbf{ 97.64(2) }$	 & 	 $-117.9(2)$\\ 
Eccentricity of orbit, $e$\dotfill	 & 	 $5.24(7)\times 10^{-6}$	 & 	 $\mathbf{ 0.0001424(6) }$	 & 	 $\mathbf{ 9.702(4)\times 10^{-5} }$	 & 	 $2.349(9)\times 10^{-5}$\\ 
Epoch of ascending node, $T_{\mathrm{ASC}}$ (MJD)\dotfill	 & 	 $\mathbf{ 50315.26949110(9) }$	 & 	 $-$	 & 	 $-$	 & 	 $\mathbf{ 50273.50700484(12) }$\\ 
First Laplace parameter, $\epsilon_1 = e \sin \omega$\dotfill	 & 	 $\mathbf{ 3.91(7)\times 10^{-6} }$	 & 	 $-$	 & 	 $-$	 & 	 $\mathbf{ -2.076(9)\times 10^{-5} }$\\ 

 \noalign{\vskip 1.5mm} 
Second Laplace parameter, $\epsilon_2 = e \cos \omega$\dotfill	 & 	 $\mathbf{ 3.49(7)\times 10^{-6} }$	 & 	 $-$	 & 	 $-$	 & 	 $\mathbf{ -1.099(9)\times 10^{-5} }$\\ 
First time derivative of orbital period, ${\dot P}_{\mathrm{b}}$ \dotfill	 & 	 $\mathbf{ 2.0(8)\times 10^{-14} }$	 & 	 $\mathbf{ 4(2)\times 10^{-13} }$	 & 	 $-$	 & 	 $-$\\ 
Rate of advance of periastron, ${\dot \omega}$ (deg\,yr$^{-1}$)\dotfill	 & 	 $-$	 & 	 $\mathbf{ 0.016(3) }$	 & 	 $\mathbf{ 0.0109(13) }$	 & 	 $-$\\ 
Rate of change of projected semi-major axis, ${\dot x}$ \dotfill	 & 	 $-$	 & 	 $\mathbf{ 7.7(6)\times 10^{-15} }$	 & 	 $\mathbf{ 1.32(6)\times 10^{-14} }$	 & 	 $\mathbf{ -2.9(1.1)\times 10^{-15} }$\\ 
Sine of inclination angle, $\sin i$\dotfill	 & 	 $-$	 & 	 $-$	 & 	 $-$	 & 	 $-$\\ 

 \noalign{\vskip 1.5mm} 
Companion mass, $M_{\mathrm{c}}$ ($M_{\odot}$)\dotfill	 & 	 ${ 0.12 } ^{ +0.16 }_{ -0.07 }$	 & 	 ${ 0.27 } ^{ +0.6 }_{ -0.19 }$	 & 	 ${ 0.52 } ^{ +0.3 }_{ -0.16 }$	 & 	 $-$\\ 
Orthometric Shapiro delay parameter, $h_3$ ($\mu\,$s)\dotfill	 & 	 $\mathbf{ 2.4(3)\times 10^{-7} }$	 & 	 $\mathbf{ 1.1(3)\times 10^{-7} }$	 & 	 $\mathbf{ 6.3(6)\times 10^{-7} }$	 & 	 $-$\\ 
Orthometric Shapiro delay parameter, $h_4$ ($\mu\,$s)\dotfill	 & 	 $\mathbf{ 1.8(4)\times 10^{-7} }$	 & 	 $-$	 & 	 $-$	 & 	 $-$\\ 
Orthometric Shapiro delay parameter, $\varsigma = h_4 / h_3$\dotfill	 & 	 $-$	 & 	 $\mathbf{ 0.2(3) }$	 & 	 $\mathbf{ 0.62(8) }$	 & 	 $-$\\ 
Longitude of ascending node, $\Omega$ (deg.)\dotfill	 & 	 $-$	 & 	 $-$	 & 	 $-$	 & 	 $-$\\ 

 \noalign{\vskip 1.5mm} 
Inclination angle, $i$ (deg.)\dotfill	 & 	 $73^{ +12 }_{ -14 }$	 & 	 $41^{ +4 }_{ -7 }$	 & 	 $68^{ +4 }_{ -7 }$	 & 	 $-$\\ 
Upper limit on $i$ (deg.)\dotfill	 & 	 $-$	 & 	 $<44$	 & 	 $<73$	 & 	 $<52$\\ 
Parallax distance, $D_\pi$ (kpc)\dotfill	 & 	 ${ 0.99 } ^{ +0.1 }_{ -0.09 }$	 & 	 ${ 1.4 } ^{ +3 }_{ -0.6 }$	 & 	 ${ 0.64 } ^{ +0.08 }_{ -0.06 }$	 & 	 ${ 0.59 } ^{ +0.5 }_{ -0.18 }$\\ 
Shklovskii distance $D_{\rm shk}$ (kpc)\dotfill	 & 	 $0.6(3)$	 & 	 $3.5(1.3)$	 & 	 $-$	 & 	 $-$\\ 
Pulsar mass, $M_{\mathrm{P}}$ ($M_{\odot}$) \dotfill	 & 	 ${ 1.2 } ^{ +3 }_{ -0.8 }$	 & 	 $2.0(8)$	 & 	 $1.4(5)$	 & 	 $-$\\ 

 \noalign{\vskip 1.5mm} 
Total system mass $M_{\rm tot}$ ($M_{\odot}$)\dotfill	 & 	 ${ 1.3 } ^{ +3 }_{ -0.9 }$	 & 	 $2.4(7)$	 & 	 $2.0(4)$	 & 	 $-$\\ 
$M_{\rm tot}$ derived from periastron advance ${\dot \omega}$ ($M_{\odot}$)\dotfill	 & 	 $-$	 & 	 $-$	 & 	 $-$	 & 	 $-$\\ 
Transverse velocity, $v_t$ (km\,s$^{-1}$)\dotfill	 & 	 $49(5)$	 & 	 $65^{ +110 }_{ -30 }$	 & 	 $54^{ +12 }_{ -7 }$	 & 	 $22^{ +16 }_{ -7 }$\\ 

        \noalign{\vskip 1.5mm}
        \hline\hline
        \end{tabular}\hfill\
        \caption{\label{tab:XXXXX}
        Measured and derived timing model parameters for binary pulsars J0613$-$0200, J1017$-$7156, J1022+1001, and J1045$-$4509.
        }
        \end{table}
        
\clearpage

        \begin{table}
        \footnotesize
        \begin{tabular}{llllllll}
        \hline\hline \noalign{\vskip 1.5mm}
        Pulsar Name 	 & 	 J1125$-$6014	 & 	 J1446$-$4701	 & 	 J1545$-$4550	 & 	 J1600$-$3053 
 \\ \hline \noalign{\vskip 1.5mm} 
Number of TOAs\dotfill	 & 	 $1407$	 & 	 $508$	 & 	 $1634$	 & 	 $7135$\\ 
Number of observations\dotfill	 & 	 $348$	 & 	 $231$	 & 	 $337$	 & 	 $1159$\\ 
MJD range\dotfill	 & 	 $53722$ -- $58229$	 & 	 $55520$ -- $58210$	 & 	 $55685$ -- $58232$	 & 	 $52301$ -- $58232$\\ 
Right ascension (RA), $\alpha$ (hh:mm:ss)\dotfill	 & 	 $11$:$25$:$55.230337(8)$	 & 	 $14$:$46$:$35.712384(8)$	 & 	 $15$:$45$:$55.945691(4)$	 & 	 $16$:$0$:$51.903223(2)$\\ 
Declination (DEC), $\delta$ (dd:mm:ss)\dotfill	 & 	 $-60$:$14$:$6.69635(6)$	 & 	 $-47$:$1$:$26.7780(1)$	 & 	 $-45$:$50$:$37.52246(8)$	 & 	 $-30$:$53$:$49.3873(1)$\\ 
Proper motion in RA, $\mu_\alpha \cos\delta$ (${\rm mas}\,{\rm yr}^{-1}$)\dotfill	 & 	 $11.106(13)$	 & 	 $-4.36(4)$	 & 	 $-0.48(2)$	 & 	 $-0.960(7)$\\ 
Proper motion in DEC, $\mu_\delta$ (${\rm mas}\,{\rm yr}^{-1}$)\dotfill	 & 	 $-13.037(14)$	 & 	 $-3.00(7)$	 & 	 $2.37(4)$	 & 	 $-6.96(3)$\\ 

 \noalign{\vskip 1.5mm} 
Spin frequency, $f$ (${\rm s}^{-1}$)\dotfill	 & 	 $\mathbf{ 380.173093964712(2) }$	 & 	 $\mathbf{ 455.644016228211(3) }$	 & 	 $\mathbf{ 279.6976986512927(8) }$	 & 	 $\mathbf{ 277.9377069492822(7) }$\\ 
First spin frequency derivative, ${\dot{f}}$ (${\rm s}^{-2}$)\dotfill	 & 	 $\mathbf{ -5.39294(18)\times 10^{-16} }$	 & 	 $\mathbf{ -2.03614(6)\times 10^{-15} }$	 & 	 $\mathbf{ -4.10355(2)\times 10^{-15} }$	 & 	 $\mathbf{ -7.33846(13)\times 10^{-16} }$\\ 
Dispersion measure, DM (${\rm cm}^{-3}\,{\rm pc}$)\dotfill	 & 	 $52.9303$	 & 	 $55.8278$	 & 	 $68.4162$	 & 	 $52.36494$\\ 
Ecliptic latitude $\beta$ (deg.)\dotfill	 & 	 $\mathbf{ -55.659047536(17) }$	 & 	 $\mathbf{ -29.41052111(4) }$	 & 	 $\mathbf{ -25.29111651(2) }$	 & 	 $\mathbf{ -10.07183776(3) }$\\ 
Ecliptic longitude $\lambda$ (deg.)\dotfill	 & 	 $\mathbf{ 209.50402091(2) }$	 & 	 $\mathbf{ 234.21534222(2) }$	 & 	 $\mathbf{ 244.821546010(11) }$	 & 	 $\mathbf{ 244.347677741(5) }$\\ 

 \noalign{\vskip 1.5mm} 
Proper motion in ecliptic latitude, $\mu_\beta$ (${\rm mas}\,{\rm yr}^{-1}$)\dotfill	 & 	 $\mathbf{ -1.600(15) }$	 & 	 $\mathbf{ -4.33(7) }$	 & 	 $\mathbf{ 2.18(5) }$	 & 	 $\mathbf{ -7.02(3) }$\\ 
Proper motion in ecliptic longitude, $\mu_\lambda \cos\beta$ (${\rm mas}\,{\rm yr}^{-1}$)\dotfill	 & 	 $\mathbf{ 17.052(12) }$	 & 	 $\mathbf{ -3.08(4) }$	 & 	 $\mathbf{ -1.043(20) }$	 & 	 $\mathbf{ 0.457(4) }$\\ 
Parallax, $\pi$ (${\rm mas}$)\dotfill	 & 	 $\mathbf{ 0.6(3) }$	 & 	 $\mathbf{ 0.6(3) }$	 & 	 $\mathbf{ 0.45(14) }$	 & 	 $\mathbf{ 0.53(6) }$\\ 
Obital period, $P_{\mathrm{b}}$ ($\mathrm{d}$)\dotfill	 & 	 $\mathbf{ 8.7526036468(4) }$	 & 	 $\mathbf{ 0.27766607699(15) }$	 & 	 $\mathbf{ 6.20306492579(17) }$	 & 	 $\mathbf{ 14.3484645(9) }$\\ 
Projected semimajor axis, $x$ (lt-s)\dotfill	 & 	 $\mathbf{ 8.3391939(4) }$	 & 	 $\mathbf{ 0.06401226(15) }$	 & 	 $\mathbf{ 3.84690472(19) }$	 & 	 $\mathbf{ 8.8016538(7) }$\\ 

 \noalign{\vskip 1.5mm} 
Epoch of periastron, $T_0$ (MJD)\dotfill	 & 	 $53418(4)$	 & 	 $55657.7(15)$	 & 	 $55826.44(15)$	 & 	 $\mathbf{ 53281.1899(4) }$\\ 
Longitude of periastron, $\omega$ (deg)\dotfill	 & 	 $180(130)$	 & 	 $-130(80)$	 & 	 $-138.32(15)$	 & 	 $\mathbf{ 181.817(11) }$\\ 
Eccentricity of orbit, $e$\dotfill	 & 	 $6.15(11)\times 10^{-7}$	 & 	 $1.1(4)\times 10^{-5}$	 & 	 $1.304(3)\times 10^{-5}$	 & 	 $\mathbf{ 0.000173719(7) }$\\ 
Epoch of ascending node, $T_{\mathrm{ASC}}$ (MJD)\dotfill	 & 	 $\mathbf{ 53171.58579575(8) }$	 & 	 $\mathbf{ 55647.8044393(3) }$	 & 	 $\mathbf{ 55607.59239512(5) }$	 & 	 $-$\\ 
First Laplace parameter, $\epsilon_1 = e \sin \omega$\dotfill	 & 	 $\mathbf{ 3(3)\times 10^{-8} }$	 & 	 $\mathbf{ -7(5)\times 10^{-6} }$	 & 	 $\mathbf{ -8.67(4)\times 10^{-6} }$	 & 	 $-$\\ 

 \noalign{\vskip 1.5mm} 
Second Laplace parameter, $\epsilon_2 = e \cos \omega$\dotfill	 & 	 $\mathbf{ -6.14(11)\times 10^{-7} }$	 & 	 $\mathbf{ -7(5)\times 10^{-6} }$	 & 	 $\mathbf{ -9.74(3)\times 10^{-6} }$	 & 	 $-$\\ 
First time derivative of orbital period, ${\dot P}_{\mathrm{b}}$ \dotfill	 & 	 $\mathbf{ 7(1)\times 10^{-13} }$	 & 	 $\mathbf{ 2(1)\times 10^{-13} }$	 & 	 $-$	 & 	 $\mathbf{ 5(1)\times 10^{-13} }$\\ 
Rate of advance of periastron, ${\dot \omega}$ (deg\,yr$^{-1}$)\dotfill	 & 	 $-$	 & 	 $-$	 & 	 $-$	 & 	 $\mathbf{ 0.0043(6) }$\\ 
Rate of change of projected semi-major axis, ${\dot x}$ \dotfill	 & 	 $-$	 & 	 $-$	 & 	 $\mathbf{ -2.0(1.3)\times 10^{-15} }$	 & 	 $\mathbf{ -3.5(2)\times 10^{-15} }$\\ 
Sine of inclination angle, $\sin i$\dotfill	 & 	 $\mathbf{ 0.978(5) }$	 & 	 $-$	 & 	 $-$	 & 	 $-$\\ 

 \noalign{\vskip 1.5mm} 
Companion mass, $M_{\mathrm{c}}$ ($M_{\odot}$)\dotfill	 & 	 $\mathbf{ 0.31(3) }$	 & 	 $-$	 & 	 $-$	 & 	 ${ 0.29 } ^{ +0.08 }_{ -0.06 }$\\ 
Orthometric Shapiro delay parameter, $h_3$ ($\mu\,$s)\dotfill	 & 	 $-$	 & 	 $-$	 & 	 $\mathbf{ 9(5)\times 10^{-8} }$	 & 	 $\mathbf{ 3.37(20)\times 10^{-7} }$\\ 
Orthometric Shapiro delay parameter, $h_4$ ($\mu\,$s)\dotfill	 & 	 $-$	 & 	 $-$	 & 	 $-$	 & 	 $-$\\ 
Orthometric Shapiro delay parameter, $\varsigma = h_4 / h_3$\dotfill	 & 	 $-$	 & 	 $-$	 & 	 $\mathbf{ 0.3(7) }$	 & 	 $\mathbf{ 0.62(5) }$\\ 
Longitude of ascending node, $\Omega$ (deg.)\dotfill	 & 	 $-$	 & 	 $-$	 & 	 $-$	 & 	 $-$\\ 

 \noalign{\vskip 1.5mm} 
Inclination angle, $i$ (deg.)\dotfill	 & 	 ${ 77.9 } ^{ +1.4 }_{ -1.3 }$	 & 	 $-$	 & 	 $-$	 & 	 $66^{ +3 }_{ -5 }$\\ 
Upper limit on $i$ (deg.)\dotfill	 & 	 $-$	 & 	 $-$	 & 	 $<35$	 & 	 $<70$\\ 
Parallax distance, $D_\pi$ (kpc)\dotfill	 & 	 ${ 1.5 } ^{ +1.1 }_{ -0.5 }$	 & 	 ${ 1.5 } ^{ +1.6 }_{ -0.6 }$	 & 	 ${ 2.2 } ^{ +1.1 }_{ -0.6 }$	 & 	 ${ 1.87 } ^{ +0.3 }_{ -0.18 }$\\ 
Shklovskii distance $D_{\rm shk}$ (kpc)\dotfill	 & 	 $1.4(3)$	 & 	 $150(70)$	 & 	 $-$	 & 	 $3.0(8)$\\ 
Pulsar mass, $M_{\mathrm{P}}$ ($M_{\odot}$) \dotfill	 & 	 $1.5(2)$	 & 	 $-$	 & 	 $-$	 & 	 $2.1(5)$\\ 

 \noalign{\vskip 1.5mm} 
Total system mass $M_{\rm tot}$ ($M_{\odot}$)\dotfill	 & 	 $1.8(3)$	 & 	 $-$	 & 	 $-$	 & 	 $2.4(5)$\\ 
$M_{\rm tot}$ derived from periastron advance ${\dot \omega}$ ($M_{\odot}$)\dotfill	 & 	 $-$	 & 	 $-$	 & 	 $-$	 & 	 $-$\\ 
Transverse velocity, $v_t$ (km\,s$^{-1}$)\dotfill	 & 	 $124^{ +90 }_{ -40 }$	 & 	 $39^{ +40 }_{ -14 }$	 & 	 $26^{ +12 }_{ -7 }$	 & 	 $62^{ +8 }_{ -7 }$\\ 

        \noalign{\vskip 1.5mm}
        \hline\hline
        \end{tabular}\hfill\
        \caption{\label{tab:XXXXX}
        Measured and derived timing model parameters for binary pulsars J1125$-$6014, J1446$-$4701, J1545$-$4550, and J1600$-$3053.
        }
        \end{table}
        
\clearpage

        \begin{table}
        \footnotesize
        \begin{tabular}{llllllll}
        \hline\hline \noalign{\vskip 1.5mm}
        Pulsar Name 	 & 	 J1603$-$7202	 & 	 J1643$-$1224	 & 	 J1713+0747	 & 	 J1732$-$5049 
 \\ \hline \noalign{\vskip 1.5mm} 
Number of TOAs\dotfill	 & 	 $5464$	 & 	 $6070$	 & 	 $7967$	 & 	 $817$\\ 
Number of observations\dotfill	 & 	 $988$	 & 	 $937$	 & 	 $1197$	 & 	 $149$\\ 
MJD range\dotfill	 & 	 $50026$ -- $58232$	 & 	 $49421$ -- $58232$	 & 	 $49421$ -- $58232$	 & 	 $52679$ -- $55724$\\ 
Right ascension (RA), $\alpha$ (hh:mm:ss)\dotfill	 & 	 $16$:$3$:$35.675848(9)$	 & 	 $16$:$43$:$38.162227(5)$	 & 	 $17$:$13$:$49.5337733(4)$	 & 	 $17$:$32$:$47.76675(3)$\\ 
Declination (DEC), $\delta$ (dd:mm:ss)\dotfill	 & 	 $-72$:$2$:$32.75278(5)$	 & 	 $-12$:$24$:$58.6664(3)$	 & 	 $7$:$47$:$37.48575(1)$	 & 	 $-50$:$49$:$0.1891(4)$\\ 
Proper motion in RA, $\mu_\alpha \cos\delta$ (${\rm mas}\,{\rm yr}^{-1}$)\dotfill	 & 	 $-2.447(11)$	 & 	 $5.970(18)$	 & 	 $4.9254(14)$	 & 	 $-0.43(14)$\\ 
Proper motion in DEC, $\mu_\delta$ (${\rm mas}\,{\rm yr}^{-1}$)\dotfill	 & 	 $-7.356(13)$	 & 	 $3.77(8)$	 & 	 $-3.917(3)$	 & 	 $-9.9(2)$\\ 

 \noalign{\vskip 1.5mm} 
Spin frequency, $f$ (${\rm s}^{-1}$)\dotfill	 & 	 $\mathbf{ 67.3765811248755(2) }$	 & 	 $\mathbf{ 216.3733370950632(9) }$	 & 	 $\mathbf{ 218.8118403947170(3) }$	 & 	 $\mathbf{ 188.23351221745(1) }$\\ 
First spin frequency derivative, ${\dot{f}}$ (${\rm s}^{-2}$)\dotfill	 & 	 $\mathbf{ -7.09502(15)\times 10^{-17} }$	 & 	 $\mathbf{ -8.64392(9)\times 10^{-16} }$	 & 	 $\mathbf{ -4.08387(2)\times 10^{-16} }$	 & 	 $\mathbf{ -5.0282(18)\times 10^{-16} }$\\ 
Dispersion measure, DM (${\rm cm}^{-3}\,{\rm pc}$)\dotfill	 & 	 $38.0398$	 & 	 $62.418$	 & 	 $15.98477$	 & 	 $56.823061257620836$\\ 
Ecliptic latitude $\beta$ (deg.)\dotfill	 & 	 $\mathbf{ -49.963017351(15) }$	 & 	 $\mathbf{ 9.77833276(9) }$	 & 	 $\mathbf{ 30.700359822(3) }$	 & 	 $\mathbf{ -27.49159975(10) }$\\ 
Ecliptic longitude $\lambda$ (deg.)\dotfill	 & 	 $\mathbf{ 256.520289582(17) }$	 & 	 $\mathbf{ 251.087221466(15) }$	 & 	 $\mathbf{ 256.6686963898(17) }$	 & 	 $\mathbf{ 265.16176861(7) }$\\ 

 \noalign{\vskip 1.5mm} 
Proper motion in ecliptic latitude, $\mu_\beta$ (${\rm mas}\,{\rm yr}^{-1}$)\dotfill	 & 	 $\mathbf{ -7.754(12) }$	 & 	 $\mathbf{ 4.63(7) }$	 & 	 $\mathbf{ -3.442(2) }$	 & 	 $\mathbf{ -9.88(20) }$\\ 
Proper motion in ecliptic longitude, $\mu_\lambda \cos\beta$ (${\rm mas}\,{\rm yr}^{-1}$)\dotfill	 & 	 $\mathbf{ -0.125(9) }$	 & 	 $\mathbf{ 5.402(12) }$	 & 	 $\mathbf{ 5.2651(13) }$	 & 	 $\mathbf{ 0.02(13) }$\\ 
Parallax, $\pi$ (${\rm mas}$)\dotfill	 & 	 $\mathbf{ 0.3(3) }$	 & 	 $\mathbf{ 0.82(17) }$	 & 	 $\mathbf{ 0.763(21) }$	 & 	 $-$\\ 
Obital period, $P_{\mathrm{b}}$ ($\mathrm{d}$)\dotfill	 & 	 $\mathbf{ 6.30862966949(19) }$	 & 	 $\mathbf{ 147.017397803(8) }$	 & 	 $\mathbf{ 67.825139(3) }$	 & 	 $\mathbf{ 5.2629972176(5) }$\\ 
Projected semimajor axis, $x$ (lt-s)\dotfill	 & 	 $\mathbf{ 6.8806581(2) }$	 & 	 $\mathbf{ 25.0726139(4) }$	 & 	 $\mathbf{ 32.34242273(10) }$	 & 	 $\mathbf{ 3.9828698(3) }$\\ 

 \noalign{\vskip 1.5mm} 
Epoch of periastron, $T_0$ (MJD)\dotfill	 & 	 $50596.87(12)$	 & 	 $\mathbf{ 49577.9682(3) }$	 & 	 $\mathbf{ 54303.6340(2) }$	 & 	 $51535.7(8)$\\ 
Longitude of periastron, $\omega$ (deg)\dotfill	 & 	 $169.89(12)$	 & 	 $\mathbf{ 321.8481(7) }$	 & 	 $\mathbf{ 176.1966(12) }$	 & 	 $166(1)$\\ 
Eccentricity of orbit, $e$\dotfill	 & 	 $9.33(2)\times 10^{-6}$	 & 	 $\mathbf{ 0.000505740(6) }$	 & 	 $\mathbf{ 7.49408(5)\times 10^{-5} }$	 & 	 $8.43(14)\times 10^{-6}$\\ 
Epoch of ascending node, $T_{\mathrm{ASC}}$ (MJD)\dotfill	 & 	 $\mathbf{ 50426.28702395(7) }$	 & 	 $-$	 & 	 $-$	 & 	 $\mathbf{ 51396.3661229(3) }$\\ 
First Laplace parameter, $\epsilon_1 = e \sin \omega$\dotfill	 & 	 $\mathbf{ 1.64(2)\times 10^{-6} }$	 & 	 $-$	 & 	 $-$	 & 	 $\mathbf{ 1.99(15)\times 10^{-6} }$\\ 

 \noalign{\vskip 1.5mm} 
Second Laplace parameter, $\epsilon_2 = e \cos \omega$\dotfill	 & 	 $\mathbf{ -9.18(2)\times 10^{-6} }$	 & 	 $-$	 & 	 $-$	 & 	 $\mathbf{ -8.19(14)\times 10^{-6} }$\\ 
First time derivative of orbital period, ${\dot P}_{\mathrm{b}}$ \dotfill	 & 	 $\mathbf{ 1.9(4)\times 10^{-13} }$	 & 	 $-$	 & 	 $\mathbf{ 2(1)\times 10^{-13} }$	 & 	 $-$\\ 
Rate of advance of periastron, ${\dot \omega}$ (deg\,yr$^{-1}$)\dotfill	 & 	 $-$	 & 	 $-$	 & 	 $\mathbf{ 0.00023(8) }$	 & 	 $-$\\ 
Rate of change of projected semi-major axis, ${\dot x}$ \dotfill	 & 	 $\mathbf{ 1.30(5)\times 10^{-14} }$	 & 	 $\mathbf{ -4.97(7)\times 10^{-14} }$	 & 	 $-$	 & 	 $-$\\ 
Sine of inclination angle, $\sin i$\dotfill	 & 	 $-$	 & 	 $-$	 & 	 $-$	 & 	 $-$\\ 

 \noalign{\vskip 1.5mm} 
Companion mass, $M_{\mathrm{c}}$ ($M_{\odot}$)\dotfill	 & 	 $-$	 & 	 $-$	 & 	 $\mathbf{ 0.283(9) }$	 & 	 ${ 0.12 } ^{ +0.5 }_{ -0.11 }$\\ 
Orthometric Shapiro delay parameter, $h_3$ ($\mu\,$s)\dotfill	 & 	 $-$	 & 	 $-$	 & 	 $-$	 & 	 $\mathbf{ 5(2)\times 10^{-7} }$\\ 
Orthometric Shapiro delay parameter, $h_4$ ($\mu\,$s)\dotfill	 & 	 $-$	 & 	 $-$	 & 	 $-$	 & 	 $\mathbf{ 3(3)\times 10^{-7} }$\\ 
Orthometric Shapiro delay parameter, $\varsigma = h_4 / h_3$\dotfill	 & 	 $-$	 & 	 $-$	 & 	 $-$	 & 	 $-$\\ 
Longitude of ascending node, $\Omega$ (deg.)\dotfill	 & 	 $-$	 & 	 $-$	 & 	 $\mathbf{ 89.2(1.4) }$	 & 	 $-$\\ 

 \noalign{\vskip 1.5mm} 
Inclination angle, $i$ (deg.)\dotfill	 & 	 $-$	 & 	 $-$	 & 	 $\mathbf{ 72.0(5) }$	 & 	 $70^{ +14 }_{ -16 }$\\ 
Upper limit on $i$ (deg.)\dotfill	 & 	 $<32$	 & 	 $<29$	 & 	 $-$	 & 	 $-$\\ 
Parallax distance, $D_\pi$ (kpc)\dotfill	 & 	 ${ 2.7 } ^{ +5 }_{ -1.3 }$	 & 	 ${ 1.2 } ^{ +0.4 }_{ -0.3 }$	 & 	 $1.31(4)$	 & 	 $-$\\ 
Shklovskii distance $D_{\rm shk}$ (kpc)\dotfill	 & 	 $3.4(5)$	 & 	 $-$	 & 	 $1.0(3)$	 & 	 $-$\\ 
Pulsar mass, $M_{\mathrm{P}}$ ($M_{\odot}$) \dotfill	 & 	 $-$	 & 	 $-$	 & 	 $1.28(8)$	 & 	 ${ 0.7 } ^{ +6 }_{ -0.8 }$\\ 

 \noalign{\vskip 1.5mm} 
Total system mass $M_{\rm tot}$ ($M_{\odot}$)\dotfill	 & 	 $-$	 & 	 $-$	 & 	 ${ 1.57 } ^{ +0.09 }_{ -0.08 }$	 & 	 ${ 0.9 } ^{ +7 }_{ -0.9 }$\\ 
$M_{\rm tot}$ derived from periastron advance ${\dot \omega}$ ($M_{\odot}$)\dotfill	 & 	 $-$	 & 	 $-$	 & 	 $-$	 & 	 $-$\\ 
Transverse velocity, $v_t$ (km\,s$^{-1}$)\dotfill	 & 	 $100^{ +170 }_{ -50 }$	 & 	 $41^{ +11 }_{ -8 }$	 & 	 $39.1(11)$	 & 	 $-$\\ 

        \noalign{\vskip 1.5mm}
        \hline\hline
        \end{tabular}\hfill\
        \caption{\label{tab:XXXXX}
        Measured and derived timing model parameters for binary pulsars J1603$-$7202, J1643$-$1224, J1713+0747, and J1732$-$5049.
        }
        \end{table}
        
\clearpage

        \begin{table}
        \footnotesize
        \begin{tabular}{llllllll}
        \hline\hline \noalign{\vskip 1.5mm}
        Pulsar Name 	 & 	 J1857+0943	 & 	 J1909$-$3744	 & 	 J2129$-$5721	 & 	 J2145$-$0750	 & 	 J2241$-$5236 
 \\ \hline \noalign{\vskip 1.5mm} 
Number of TOAs\dotfill	 & 	 $3840$	 & 	 $14774$	 & 	 $3019$	 & 	 $7065$	 & 	 $5224$\\ 
Number of observations\dotfill	 & 	 $656$	 & 	 $2383$	 & 	 $739$	 & 	 $1175$	 & 	 $823$\\ 
MJD range\dotfill	 & 	 $53042$ -- $58232$	 & 	 $52618$ -- $58229$	 & 	 $49987$ -- $58231$	 & 	 $49517$ -- $58229$	 & 	 $55235$ -- $58230$\\ 
Right ascension (RA), $\alpha$ (hh:mm:ss)\dotfill	 & 	 $18$:$57$:$36.390309(2)$	 & 	 $19$:$9$:$47.4321890(3)$	 & 	 $21$:$29$:$22.770723(4)$	 & 	 $21$:$45$:$50.459478(6)$	 & 	 $22$:$41$:$42.026483(1)$\\ 
Declination (DEC), $\delta$ (dd:mm:ss)\dotfill	 & 	 $9$:$43$:$17.19768(6)$	 & 	 $-37$:$44$:$14.57793(1)$	 & 	 $-57$:$21$:$14.24329(5)$	 & 	 $-7$:$50$:$18.5048(2)$	 & 	 $-52$:$36$:$36.23755(1)$\\ 
Proper motion in RA, $\mu_\alpha \cos\delta$ (${\rm mas}\,{\rm yr}^{-1}$)\dotfill	 & 	 $-2.665(9)$	 & 	 $-9.5146(8)$	 & 	 $9.30(1)$	 & 	 $-9.48(2)$	 & 	 $18.881(4)$\\ 
Proper motion in DEC, $\mu_\delta$ (${\rm mas}\,{\rm yr}^{-1}$)\dotfill	 & 	 $-5.413(17)$	 & 	 $-35.776(3)$	 & 	 $-9.576(13)$	 & 	 $-9.11(7)$	 & 	 $-5.294(5)$\\ 

 \noalign{\vskip 1.5mm} 
Spin frequency, $f$ (${\rm s}^{-1}$)\dotfill	 & 	 $\mathbf{ 186.4940783438289(5) }$	 & 	 $\mathbf{ 339.3156871298895(1) }$	 & 	 $\mathbf{ 268.3592272034145(7) }$	 & 	 $\mathbf{ 62.29588783082522(9) }$	 & 	 $\mathbf{ 457.3101495463383(2) }$\\ 
First spin frequency derivative, ${\dot{f}}$ (${\rm s}^{-2}$)\dotfill	 & 	 $\mathbf{ -6.20476(6)\times 10^{-16} }$	 & 	 $\mathbf{ -1.614814(5)\times 10^{-15} }$	 & 	 $\mathbf{ -1.501807(7)\times 10^{-15} }$	 & 	 $\mathbf{ -1.156186(9)\times 10^{-16} }$	 & 	 $\mathbf{ -1.442296(6)\times 10^{-15} }$\\ 
Dispersion measure, DM (${\rm cm}^{-3}\,{\rm pc}$)\dotfill	 & 	 $13.2995$	 & 	 $10.39103$	 & 	 $31.8642$	 & 	 $9.00425$	 & 	 $11.40964$\\ 
Ecliptic latitude $\beta$ (deg.)\dotfill	 & 	 $\mathbf{ 32.321485158(18) }$	 & 	 $\mathbf{ -15.155520529(3) }$	 & 	 $\mathbf{ -39.899967603(13) }$	 & 	 $\mathbf{ 5.31305359(6) }$	 & 	 $\mathbf{ -40.393412229(3) }$\\ 
Ecliptic longitude $\lambda$ (deg.)\dotfill	 & 	 $\mathbf{ 286.863487475(11) }$	 & 	 $\mathbf{ 284.2208512143(9) }$	 & 	 $\mathbf{ 303.827964329(11) }$	 & 	 $\mathbf{ 326.024614276(6) }$	 & 	 $\mathbf{ 318.696383267(3) }$\\ 

 \noalign{\vskip 1.5mm} 
Proper motion in ecliptic latitude, $\mu_\beta$ (${\rm mas}\,{\rm yr}^{-1}$)\dotfill	 & 	 $\mathbf{ -5.064(17) }$	 & 	 $\mathbf{ -34.327(3) }$	 & 	 $\mathbf{ -12.532(11) }$	 & 	 $\mathbf{ -5.41(6) }$	 & 	 $\mathbf{ -13.900(5) }$\\ 
Proper motion in ecliptic longitude, $\mu_\lambda \cos\beta$ (${\rm mas}\,{\rm yr}^{-1}$)\dotfill	 & 	 $\mathbf{ -3.280(9) }$	 & 	 $\mathbf{ -13.8619(7) }$	 & 	 $\mathbf{ 4.551(7) }$	 & 	 $\mathbf{ -11.973(5) }$	 & 	 $\mathbf{ 13.832(4) }$\\ 
Parallax, $\pi$ (${\rm mas}$)\dotfill	 & 	 $\mathbf{ 0.85(16) }$	 & 	 $\mathbf{ 0.86(1) }$	 & 	 $\mathbf{ 0.26(17) }$	 & 	 $\mathbf{ 1.40(8) }$	 & 	 $\mathbf{ 0.96(4) }$\\ 
Obital period, $P_{\mathrm{b}}$ ($\mathrm{d}$)\dotfill	 & 	 $\mathbf{ 12.32717138280(14) }$	 & 	 $\mathbf{ 1.533449474571(1) }$	 & 	 $\mathbf{ 6.6254930897(5) }$	 & 	 $\mathbf{ 6.83890261495(9) }$	 & 	 $\mathbf{ 0.14567224025(2) }$\\ 
Projected semimajor axis, $x$ (lt-s)\dotfill	 & 	 $\mathbf{ 9.2307802(2) }$	 & 	 $\mathbf{ 1.897991160(17) }$	 & 	 $\mathbf{ 3.5005655(2) }$	 & 	 $\mathbf{ 10.16410673(18) }$	 & 	 $\mathbf{ 0.025795324(11) }$\\ 

 \noalign{\vskip 1.5mm} 
Epoch of periastron, $T_0$ (MJD)\dotfill	 & 	 $48062.90(8)$	 & 	 $53669.4(9)$	 & 	 $50650.1(7)$	 & 	 $51020.74(11)$	 & 	 $56729.7(5)$\\ 
Longitude of periastron, $\omega$ (deg)\dotfill	 & 	 $-83.50(4)$	 & 	 $158(4)$	 & 	 $-163.2(6)$	 & 	 $-159.3(1)$	 & 	 $120(30)$\\ 
Eccentricity of orbit, $e$\dotfill	 & 	 $2.173(2)\times 10^{-5}$	 & 	 $1.07(5)\times 10^{-7}$	 & 	 $1.214(13)\times 10^{-5}$	 & 	 $1.934(3)\times 10^{-5}$	 & 	 $2.8(8)\times 10^{-6}$\\ 
Epoch of ascending node, $T_{\mathrm{ASC}}$ (MJD)\dotfill	 & 	 $\mathbf{ 47520.43234506(9) }$	 & 	 $\mathbf{ 53630.7232148960(9) }$	 & 	 $\mathbf{ 50442.64312451(18) }$	 & 	 $\mathbf{ 50802.29822952(3) }$	 & 	 $\mathbf{ 56726.96380406(2) }$\\ 
First Laplace parameter, $\epsilon_1 = e \sin \omega$\dotfill	 & 	 $\mathbf{ -2.159(2)\times 10^{-5} }$	 & 	 $\mathbf{ 3.9(8)\times 10^{-8} }$	 & 	 $\mathbf{ -3.50(13)\times 10^{-6} }$	 & 	 $\mathbf{ -6.83(3)\times 10^{-6} }$	 & 	 $\mathbf{ 2.3(9)\times 10^{-6} }$\\ 

 \noalign{\vskip 1.5mm} 
Second Laplace parameter, $\epsilon_2 = e \cos \omega$\dotfill	 & 	 $\mathbf{ 2.458(15)\times 10^{-6} }$	 & 	 $\mathbf{ -9.9(4)\times 10^{-8} }$	 & 	 $\mathbf{ -1.162(13)\times 10^{-5} }$	 & 	 $\mathbf{ -1.809(3)\times 10^{-5} }$	 & 	 $\mathbf{ -1.3(8)\times 10^{-6} }$\\ 
First time derivative of orbital period, ${\dot P}_{\mathrm{b}}$ \dotfill	 & 	 $-$	 & 	 $\mathbf{ 5.093(7)\times 10^{-13} }$	 & 	 $\mathbf{ 1.51(9)\times 10^{-12} }$	 & 	 $\mathbf{ 1.3(2)\times 10^{-13} }$	 & 	 $-$\\ 
Rate of advance of periastron, ${\dot \omega}$ (deg\,yr$^{-1}$)\dotfill	 & 	 $-$	 & 	 $-$	 & 	 $-$	 & 	 $-$	 & 	 $-$\\ 
Rate of change of projected semi-major axis, ${\dot x}$ \dotfill	 & 	 $-$	 & 	 $\mathbf{ -3.2(3)\times 10^{-16} }$	 & 	 $\mathbf{ 4.3(4)\times 10^{-15} }$	 & 	 $\mathbf{ 6.1(4)\times 10^{-15} }$	 & 	 $\mathbf{ 1.3(1.6)\times 10^{-16} }$\\ 
Sine of inclination angle, $\sin i$\dotfill	 & 	 $\mathbf{ 0.9989(4) }$	 & 	 $\mathbf{ 0.99809(4) }$	 & 	 $-$	 & 	 $-$	 & 	 $-$\\ 

 \noalign{\vskip 1.5mm} 
Companion mass, $M_{\mathrm{c}}$ ($M_{\odot}$)\dotfill	 & 	 $\mathbf{ 0.263(14) }$	 & 	 $\mathbf{ 0.2081(9) }$	 & 	 $-$	 & 	 $-$	 & 	 $-$\\ 
Orthometric Shapiro delay parameter, $h_3$ ($\mu\,$s)\dotfill	 & 	 $-$	 & 	 $-$	 & 	 $-$	 & 	 $-$	 & 	 $-$\\ 
Orthometric Shapiro delay parameter, $h_4$ ($\mu\,$s)\dotfill	 & 	 $-$	 & 	 $-$	 & 	 $-$	 & 	 $-$	 & 	 $-$\\ 
Orthometric Shapiro delay parameter, $\varsigma = h_4 / h_3$\dotfill	 & 	 $-$	 & 	 $-$	 & 	 $-$	 & 	 $-$	 & 	 $-$\\ 
Longitude of ascending node, $\Omega$ (deg.)\dotfill	 & 	 $-$	 & 	 $-$	 & 	 $-$	 & 	 $-$	 & 	 $-$\\ 

 \noalign{\vskip 1.5mm} 
Inclination angle, $i$ (deg.)\dotfill	 & 	 ${ 87.4 } ^{ +0.7 }_{ -0.5 }$	 & 	 $86.46(5)$	 & 	 $-$	 & 	 $-$	 & 	 $-$\\ 
Upper limit on $i$ (deg.)\dotfill	 & 	 $-$	 & 	 $<88$	 & 	 $<59$	 & 	 $<73$	 & 	 $<28$\\ 
Parallax distance, $D_\pi$ (kpc)\dotfill	 & 	 ${ 1.18 } ^{ +0.3 }_{ -0.19 }$	 & 	 ${ 1.157 } ^{ +0.014 }_{ -0.013 }$	 & 	 ${ 3.6 } ^{ +5 }_{ -1.4 }$	 & 	 ${ 0.71 } ^{ +0.05 }_{ -0.04 }$	 & 	 $1.05(5)$\\ 
Shklovskii distance $D_{\rm shk}$ (kpc)\dotfill	 & 	 $-$	 & 	 $1.152(3)$	 & 	 $7.0(4)$	 & 	 $0.83(9)$	 & 	 $-$\\ 
Pulsar mass, $M_{\mathrm{P}}$ ($M_{\odot}$) \dotfill	 & 	 $1.54(13)$	 & 	 $1.486(11)$	 & 	 $-$	 & 	 $-$	 & 	 $-$\\ 

 \noalign{\vskip 1.5mm} 
Total system mass $M_{\rm tot}$ ($M_{\odot}$)\dotfill	 & 	 ${ 1.80 } ^{ +0.15 }_{ -0.14 }$	 & 	 $1.694(12)$	 & 	 $-$	 & 	 $-$	 & 	 $-$\\ 
$M_{\rm tot}$ derived from periastron advance ${\dot \omega}$ ($M_{\odot}$)\dotfill	 & 	 $-$	 & 	 $-$	 & 	 $29^{ +30 }_{ -17 }$	 & 	 ${ 2.3 } ^{ +1.8 }_{ -1.4 }$	 & 	 $-$\\ 
Transverse velocity, $v_t$ (km\,s$^{-1}$)\dotfill	 & 	 $34^{ +8 }_{ -6 }$	 & 	 $203(3)$	 & 	 $230^{ +260 }_{ -90 }$	 & 	 $44(3)$	 & 	 $97(5)$\\ 

        \noalign{\vskip 1.5mm}
        \hline\hline
        \end{tabular}\hfill\
        \caption{\label{tab:XXXXX}
        Measured and derived timing model parameters for binary pulsars J1857+0943, J1909$-$3744, J2129$-$5721, J2145$-$0750 and J2241$-$5236 (10 orbital frequency derivative parameters not shown).
        }
        \end{table}
        
\end{landscape}

\section{Timing residuals for each pulsar}\label{sec:residuals}

In the figures below we show, for each pulsar, the frequency-averaged timing residuals from the extended DR2 data set. Each figure has four panels; the top panel shows the post-fit timing residuals including all sources of noise, the second panel shows the residuals with DM variations and chromatic noise (CN) subtracted (but achromatic red noise, system noise, and band noise remains), the third panel shows the fully-whitened residuals, and the fourth panel shows the whitened and normalised residuals. The pulsar name is shown in the title of each figure, and the legends in the top three panels show the weighted root mean square residual, $\sigma_{\rm RMS}$ (to two significant digits) for each of the three observing frequency bands; 700\,MHz in red, 1400\,MHz in teal, and 3100\,MHz in blue. The legend for the fourth panel gives the Anderson-Darling statistic (ADS) for each band, which tests for consistency of the normalised residuals with a Gaussian distribution with zero mean and unit variance. 

\begin{figure*}
\centering
\includegraphics[width=0.9\textwidth]{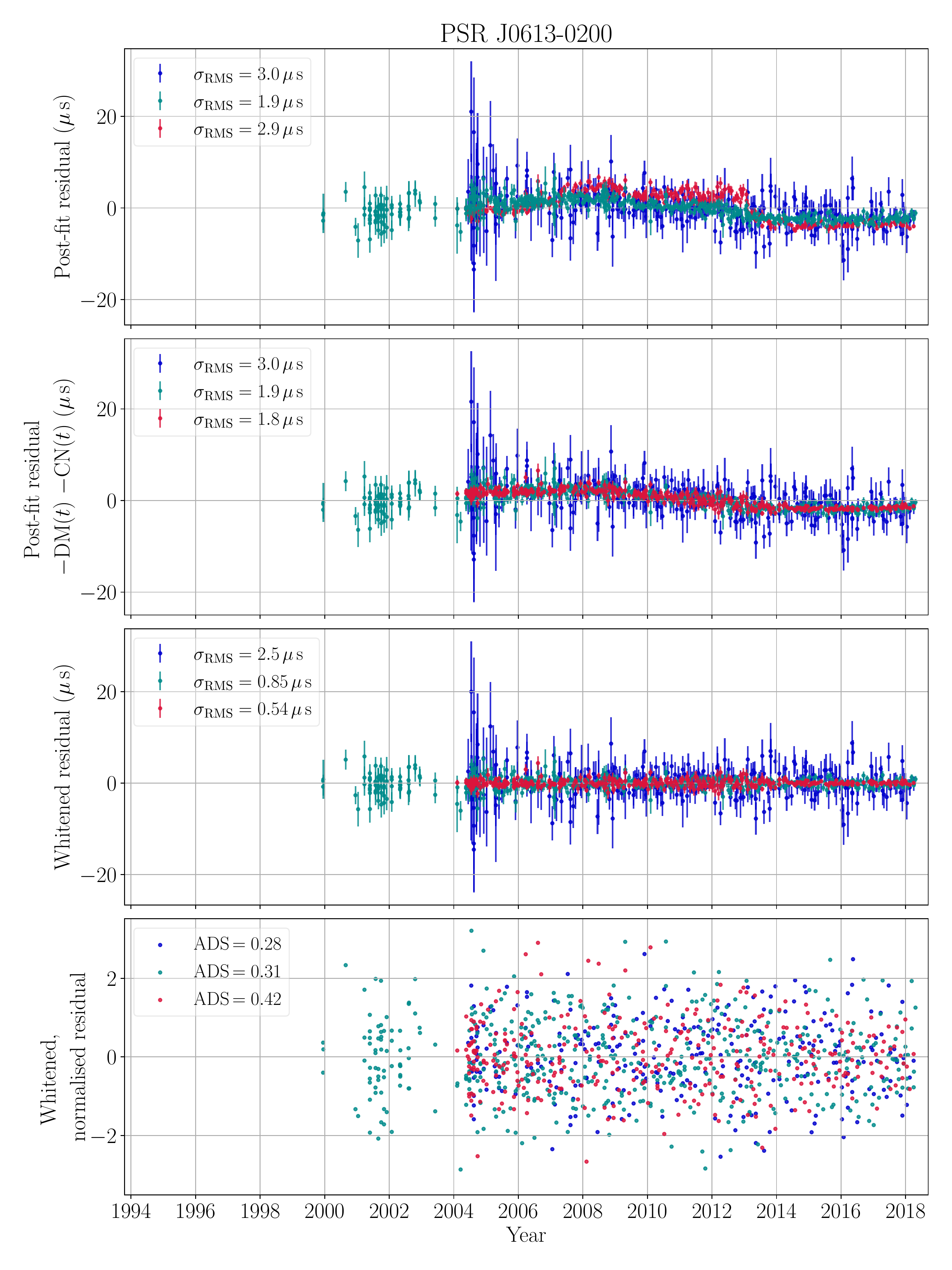}
\caption{Timing residuals for PSR~J0613$-$0200.}
\label{fig:0613_res}
\end{figure*}

\begin{figure*}
\centering
\includegraphics[width=0.95\textwidth]{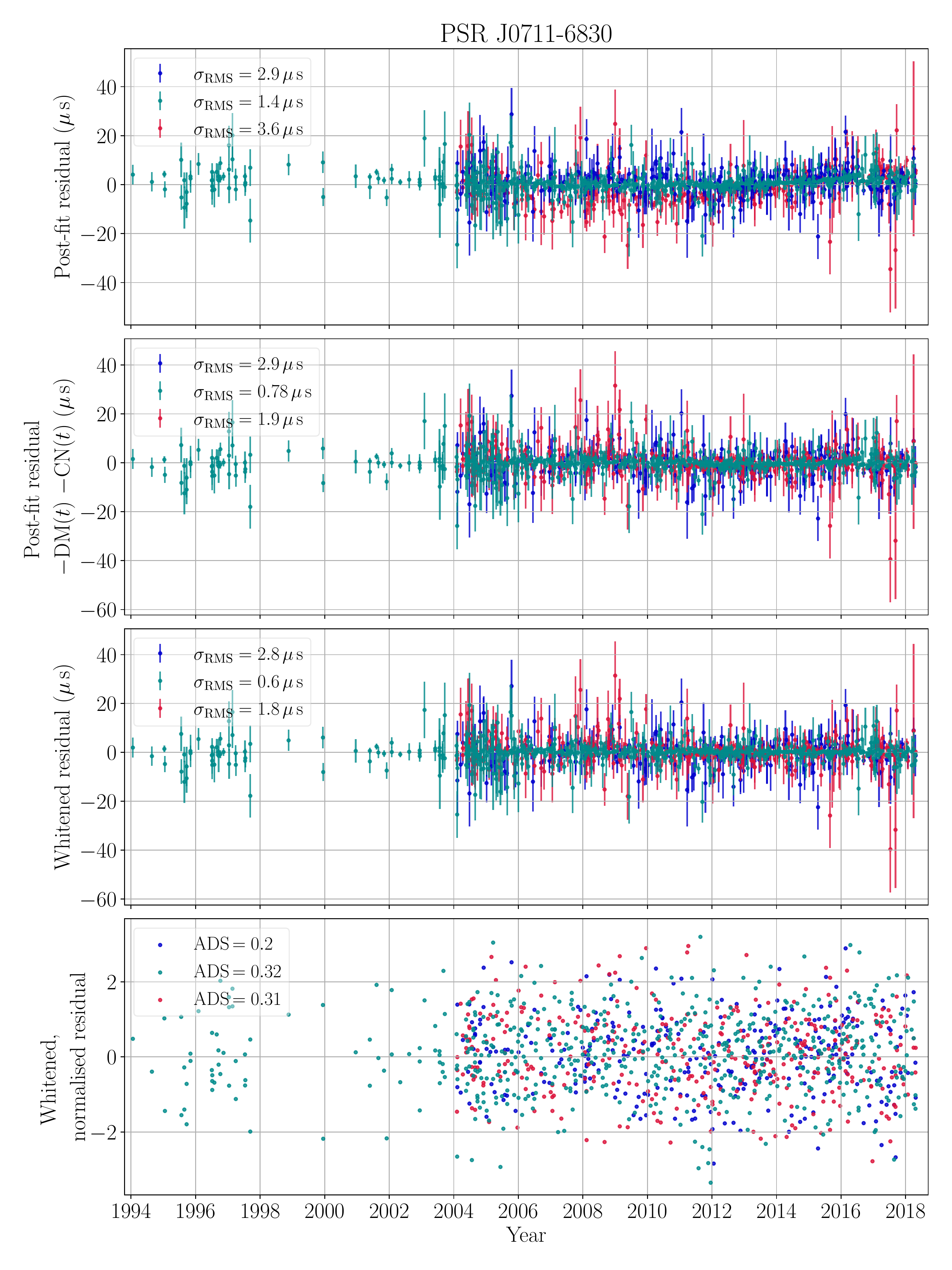}
\caption{Timing residuals for PSR~J0711$-$6830}
\label{fig:0711_res}
\end{figure*}

\begin{figure*}
\centering
\includegraphics[width=0.95\textwidth]{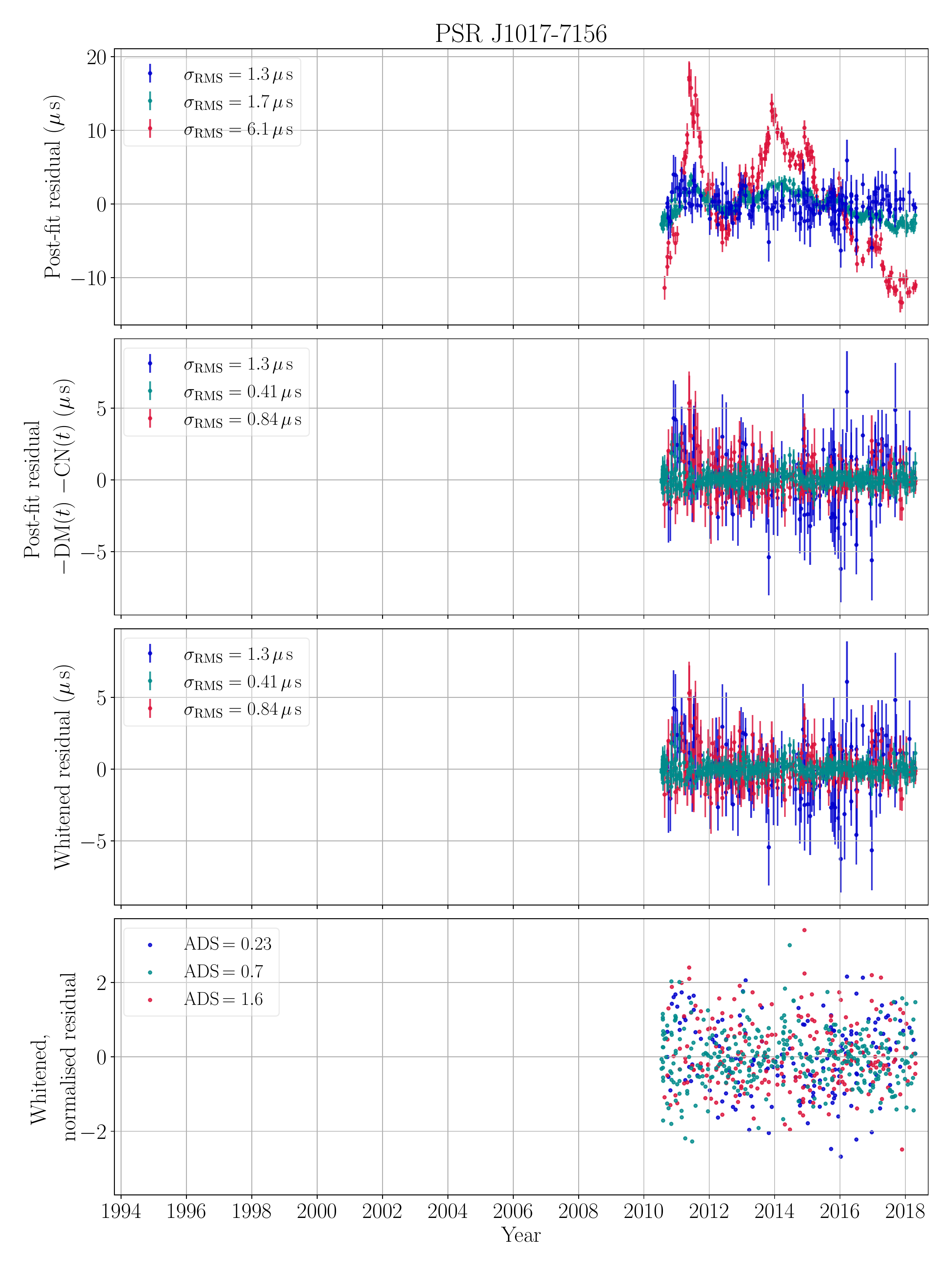}
\caption{Timing residuals for PSR~J1017$-$7156}
\label{fig:1017_res}
\end{figure*}

\begin{figure*}
\centering
\includegraphics[width=0.95\textwidth]{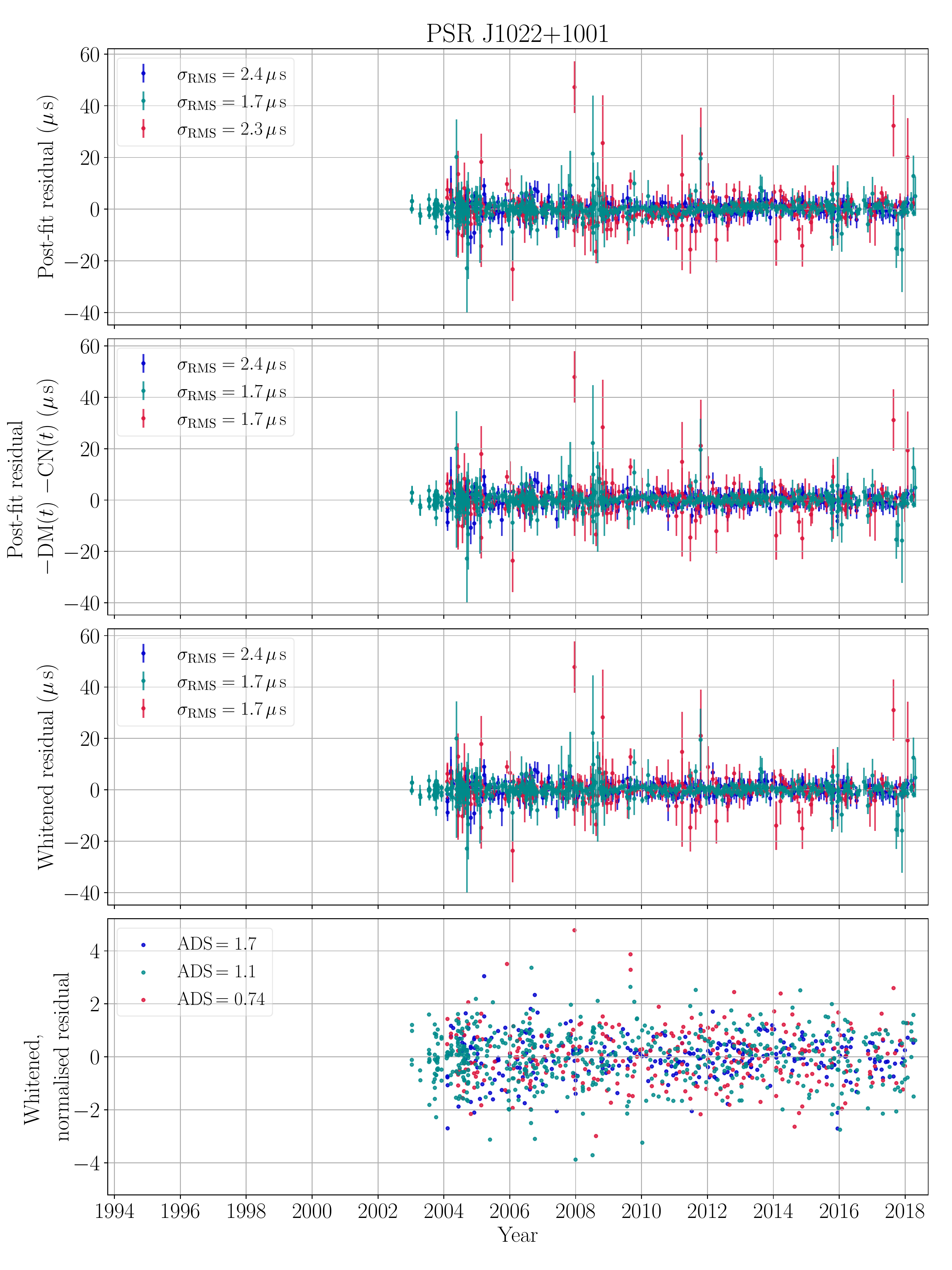}
\caption{Timing residuals for PSR~J1022+1001}
\label{fig:1022_res}
\end{figure*}

\begin{figure*}
\centering
\includegraphics[width=0.95\textwidth]{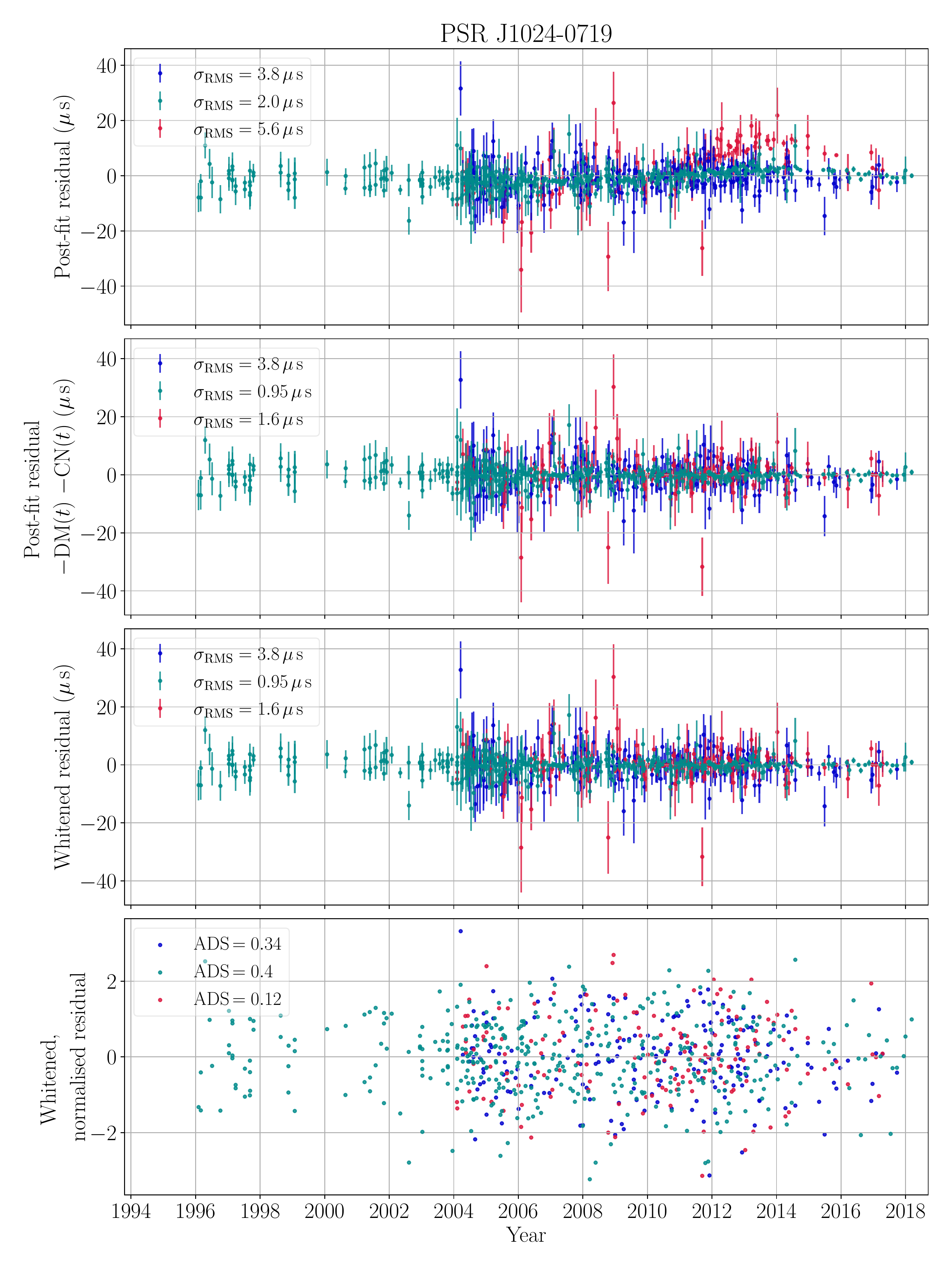}
\caption{Timing residuals for PSR~J1024$-$0719}
\label{fig:1024_res}
\end{figure*}

\begin{figure*}
\centering
\includegraphics[width=0.95\textwidth]{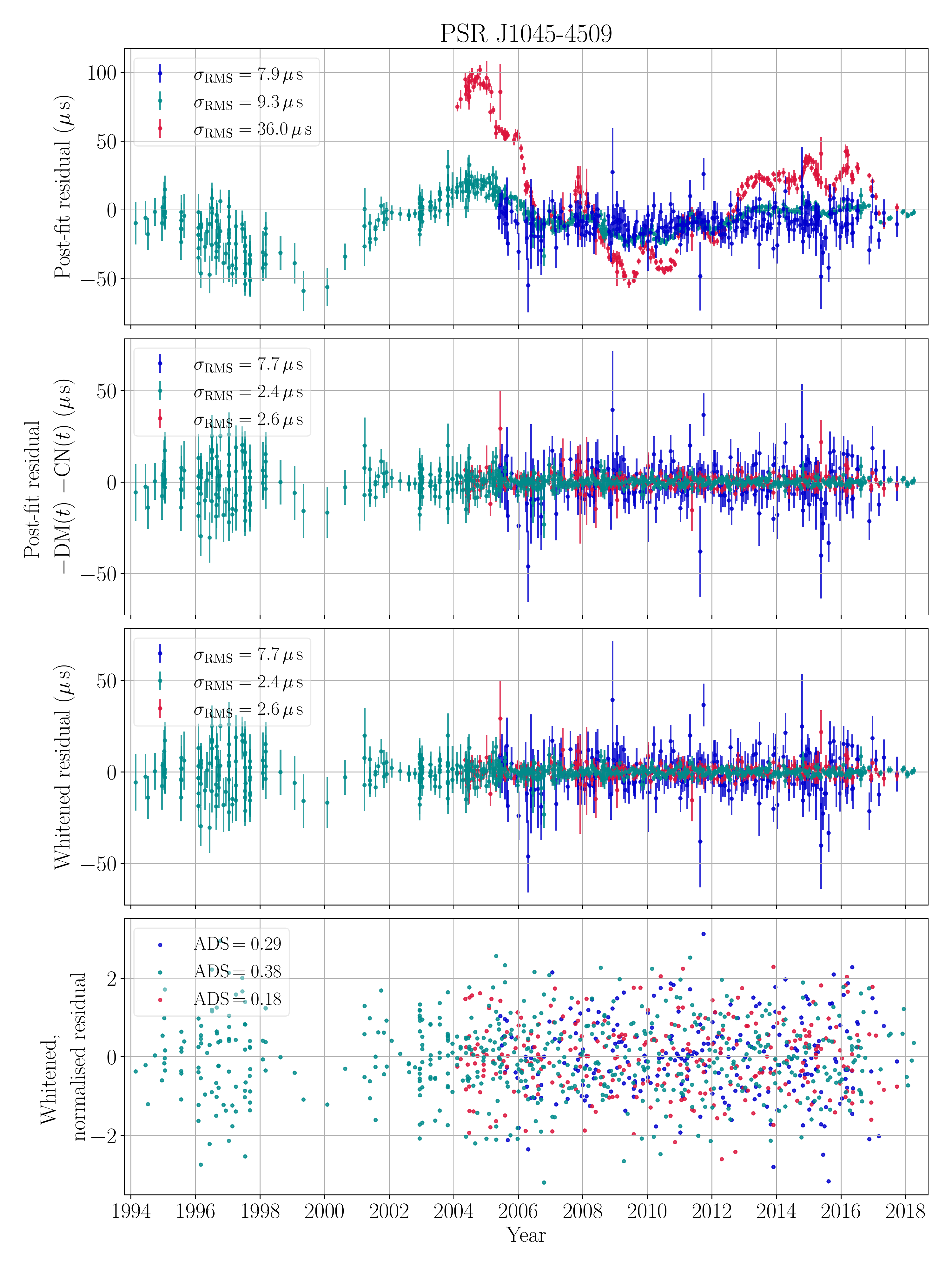}
\caption{Timing residuals for PSR~J1045$-$4509}
\label{fig:1045_res}
\end{figure*}

\begin{figure*}
\centering
\includegraphics[width=0.95\textwidth]{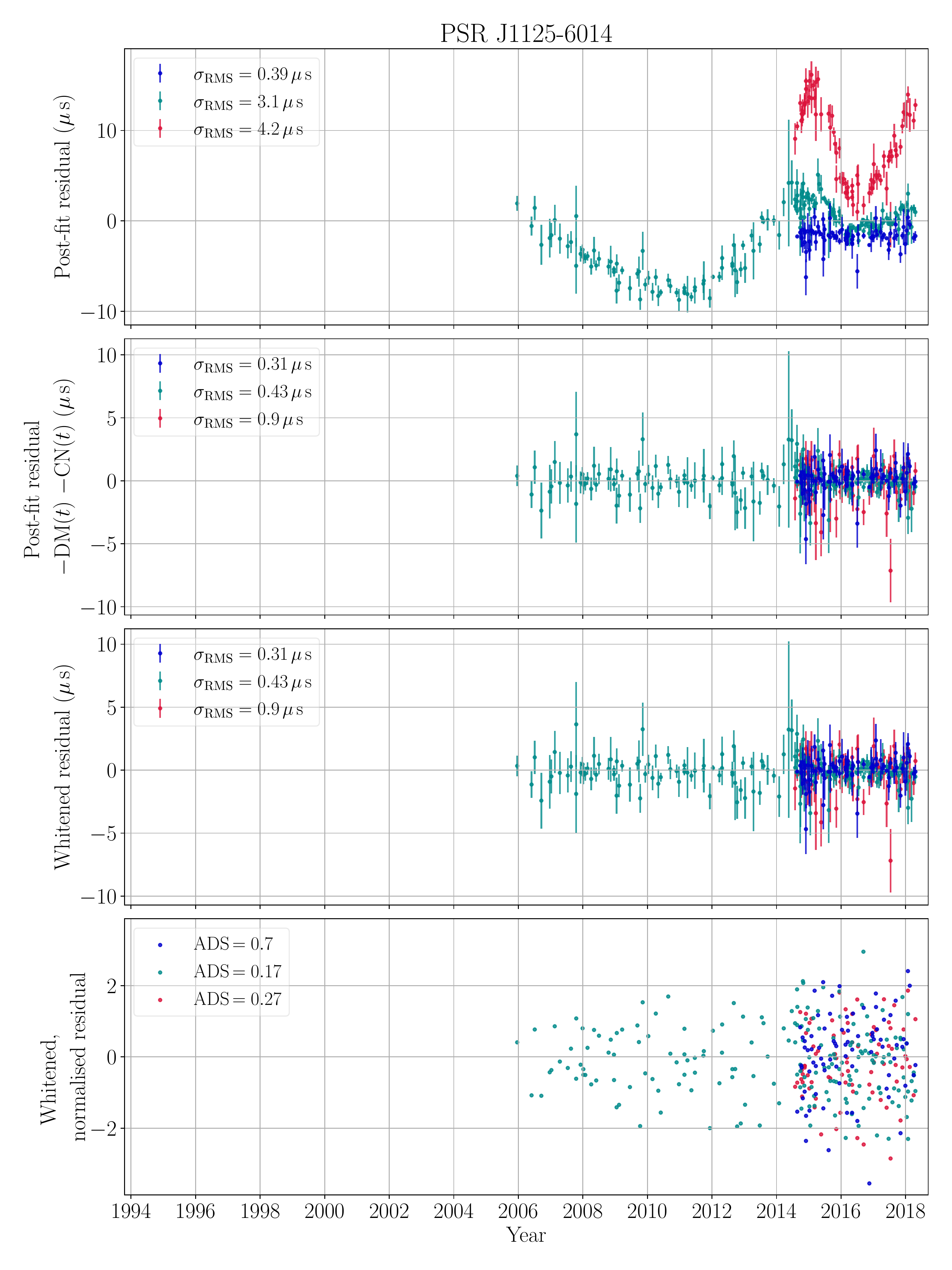}
\caption{Timing residuals for PSR~J1125$-$6014.}
\label{fig:1125_res}
\end{figure*}

\begin{figure*}
\centering
\includegraphics[width=0.95\textwidth]{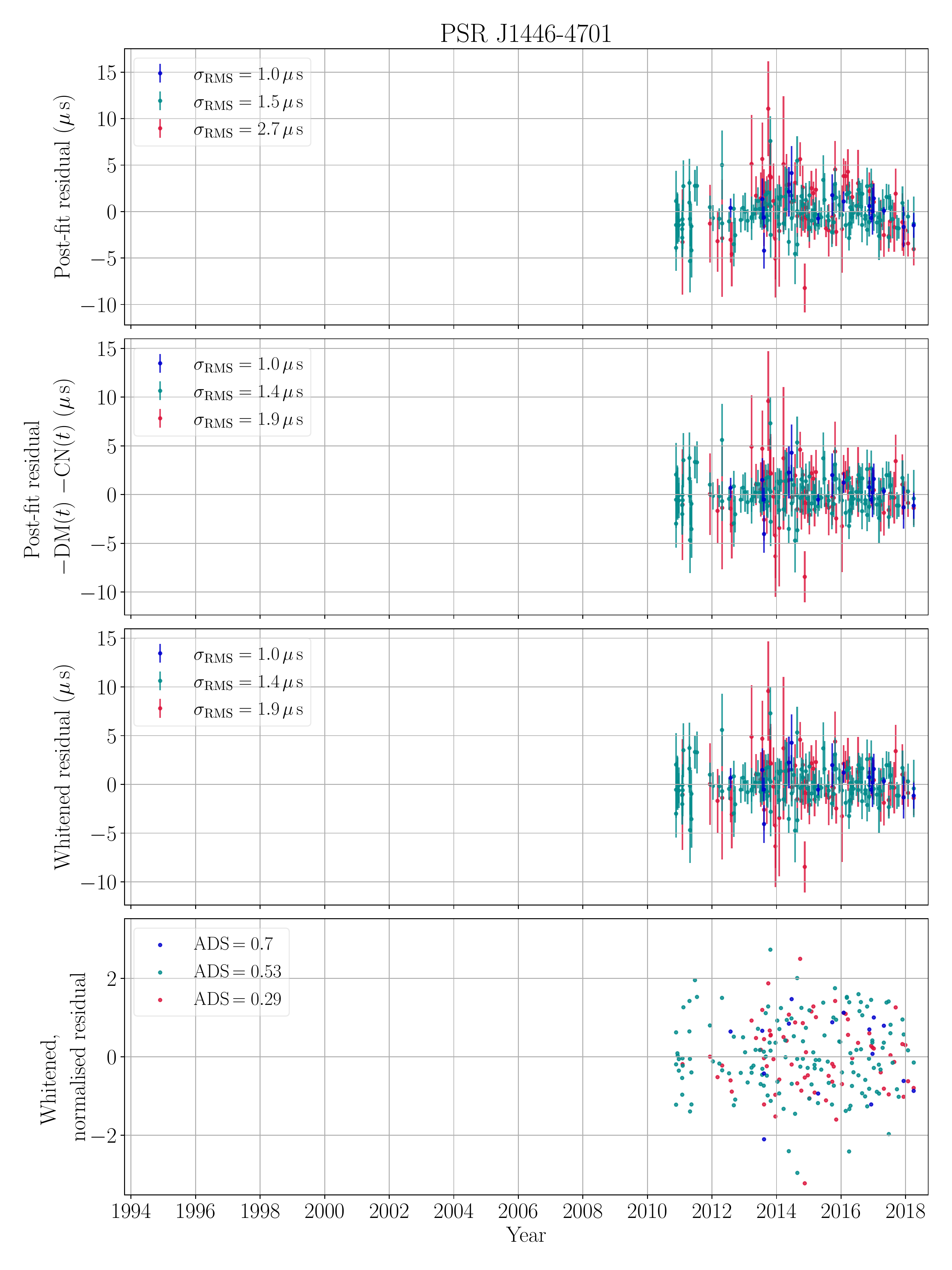}
\caption{Timing residuals for PSR~J1446$-$4701}
\label{fig:1446_res}
\end{figure*}

\begin{figure*}
\centering
\includegraphics[width=0.95\textwidth]{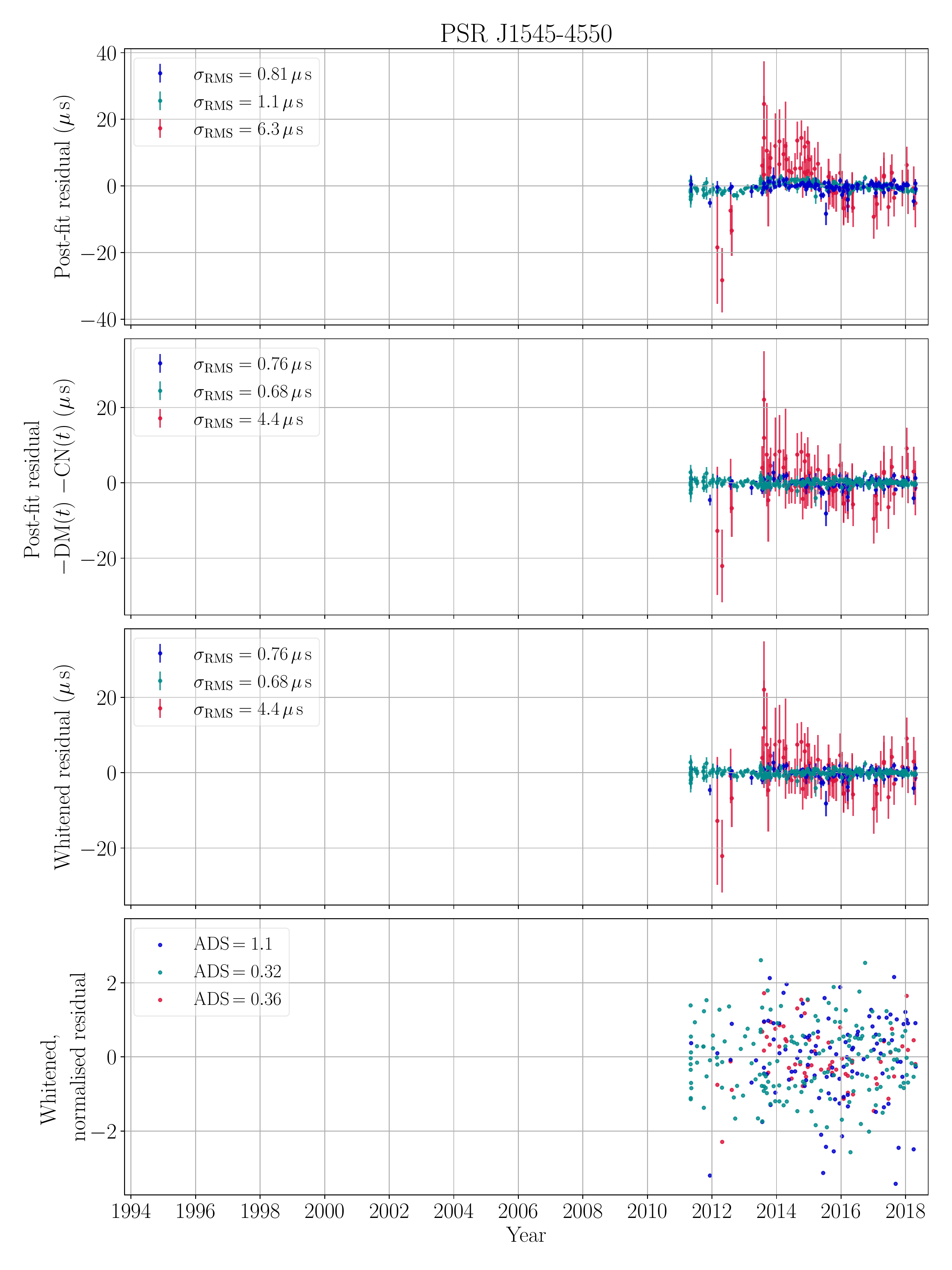}
\caption{Timing residuals for PSR~J1545$-$4550}
\label{fig:1545_res}
\end{figure*}

\begin{figure*}
\centering
\includegraphics[width=0.95\textwidth]{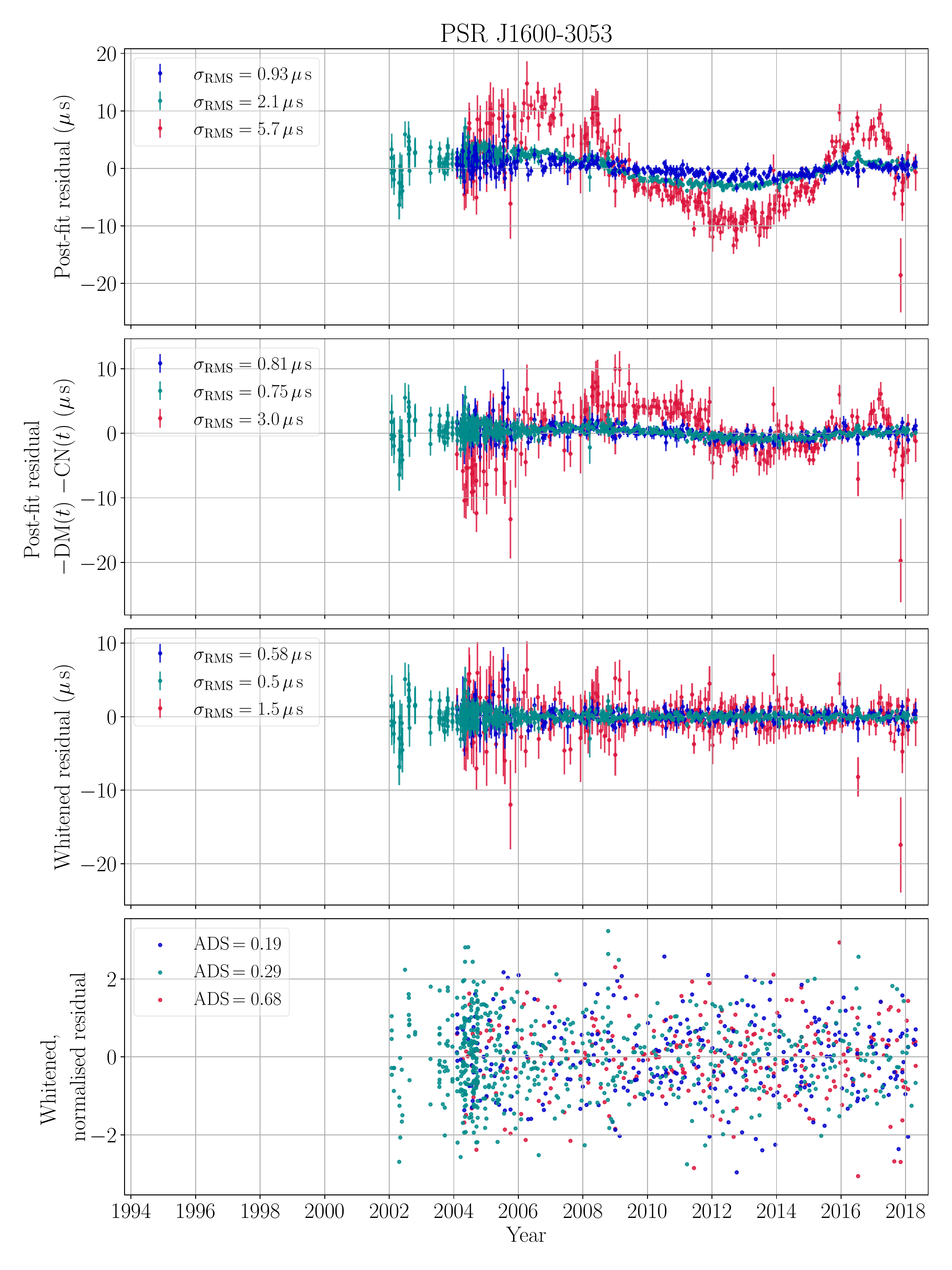}
\caption{Timing residuals for PSR~J1600$-$3053}
\label{fig:1600_res}
\end{figure*}

\begin{figure*}
\centering
\includegraphics[width=0.95\textwidth]{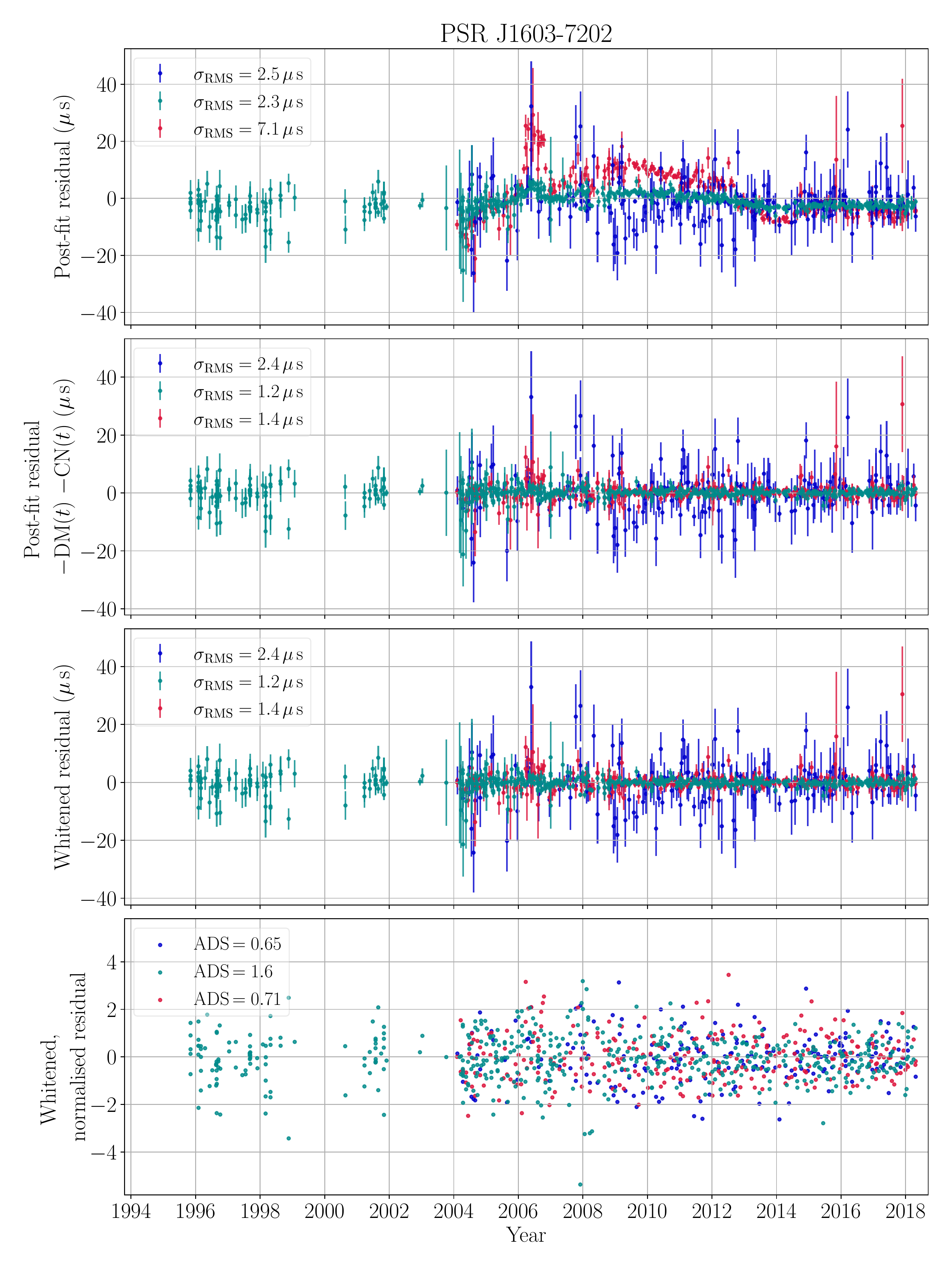}
\caption{Timing residuals for PSR~J1603$-$7202}
\label{fig:1603_res}
\end{figure*}

\begin{figure*}
\centering
\includegraphics[width=0.95\textwidth]{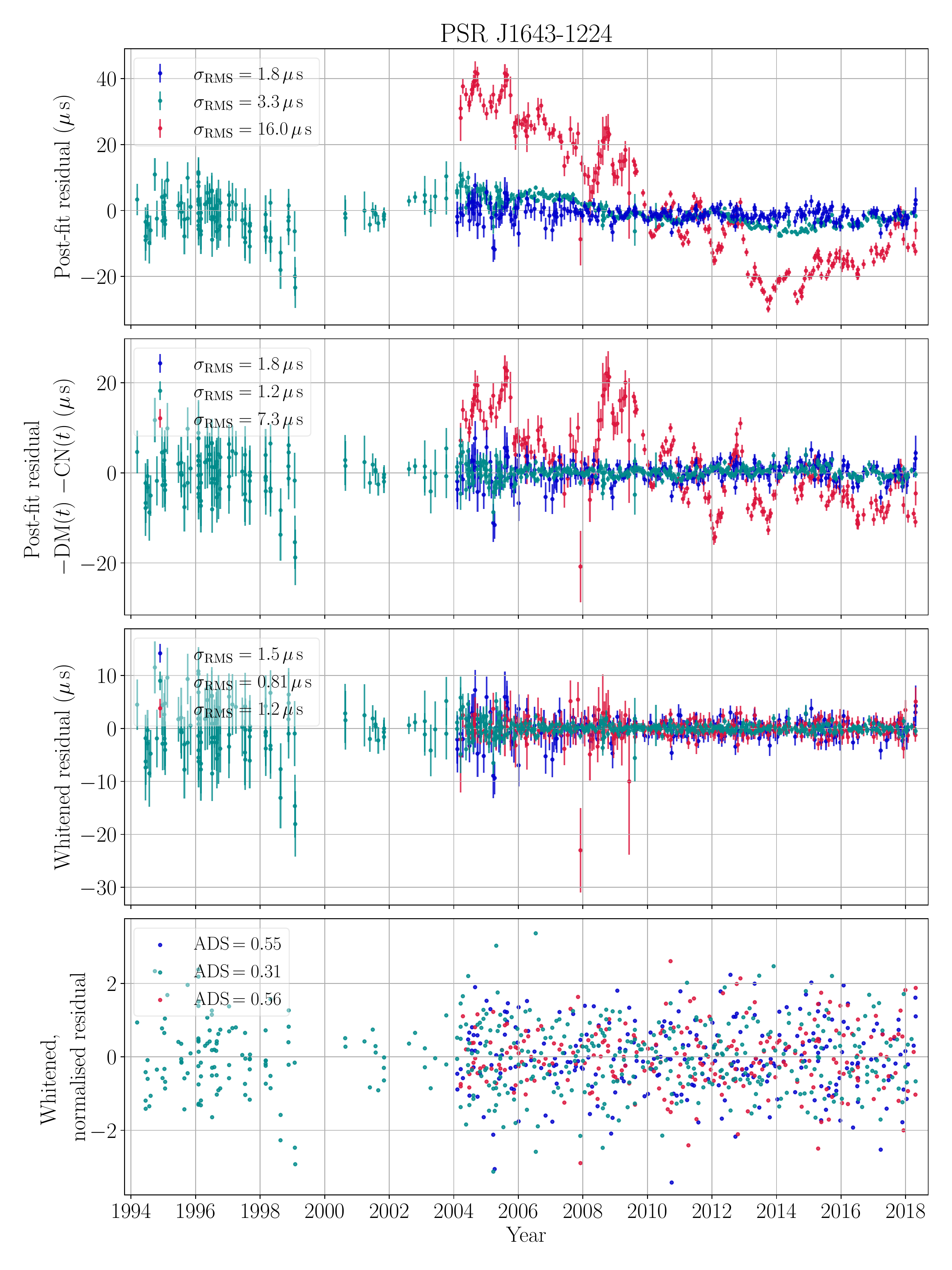}
\caption{Timing residuals for PSR~J1643$-$1224}
\label{fig:1643_res}
\end{figure*}

\begin{figure*}
\centering
\includegraphics[width=0.95\textwidth]{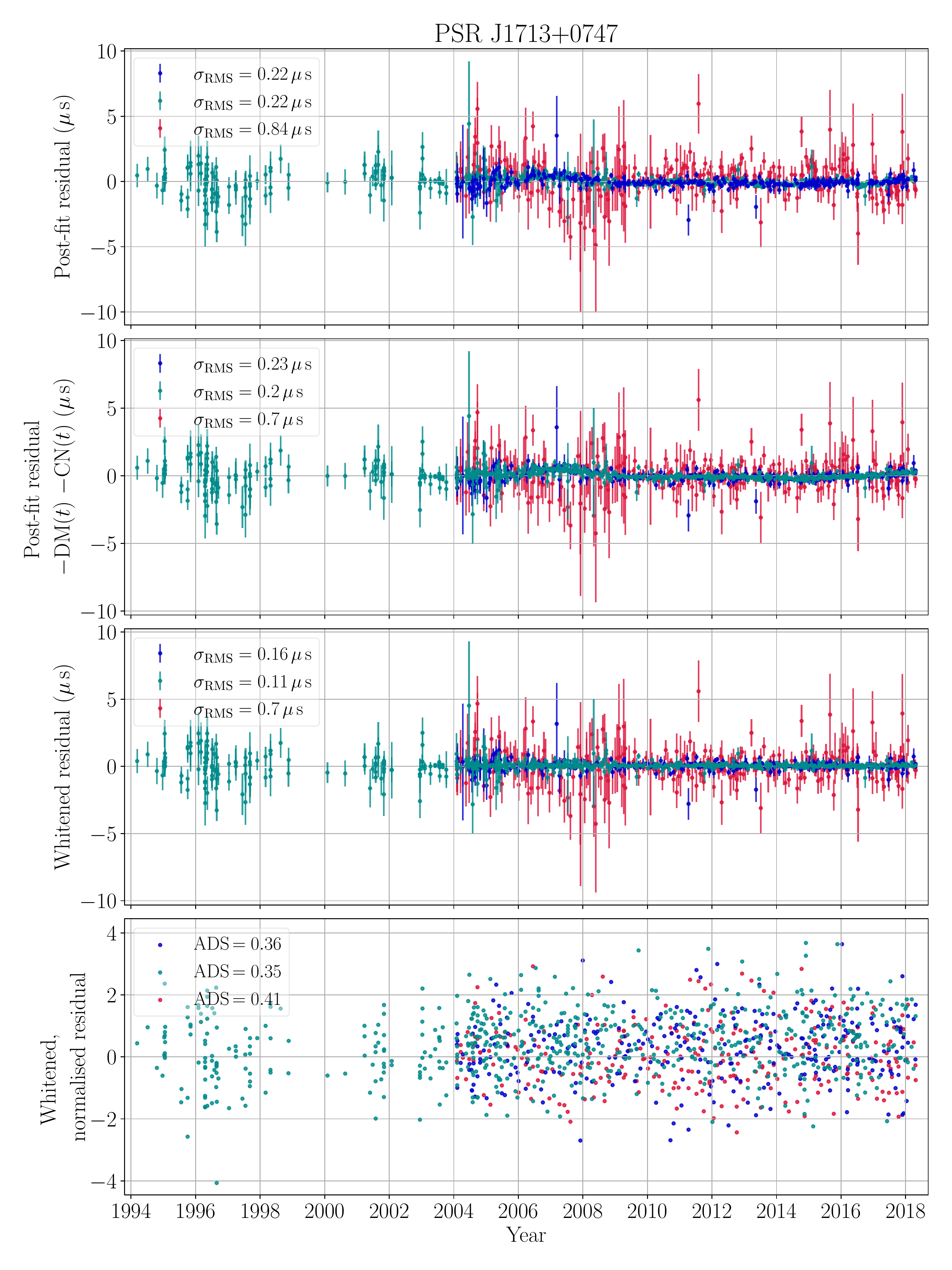}
\caption{Timing residuals for PSR~J1713+0747}
\label{fig:1713_res}
\end{figure*}

\begin{figure*}
\centering
\includegraphics[width=0.95\textwidth]{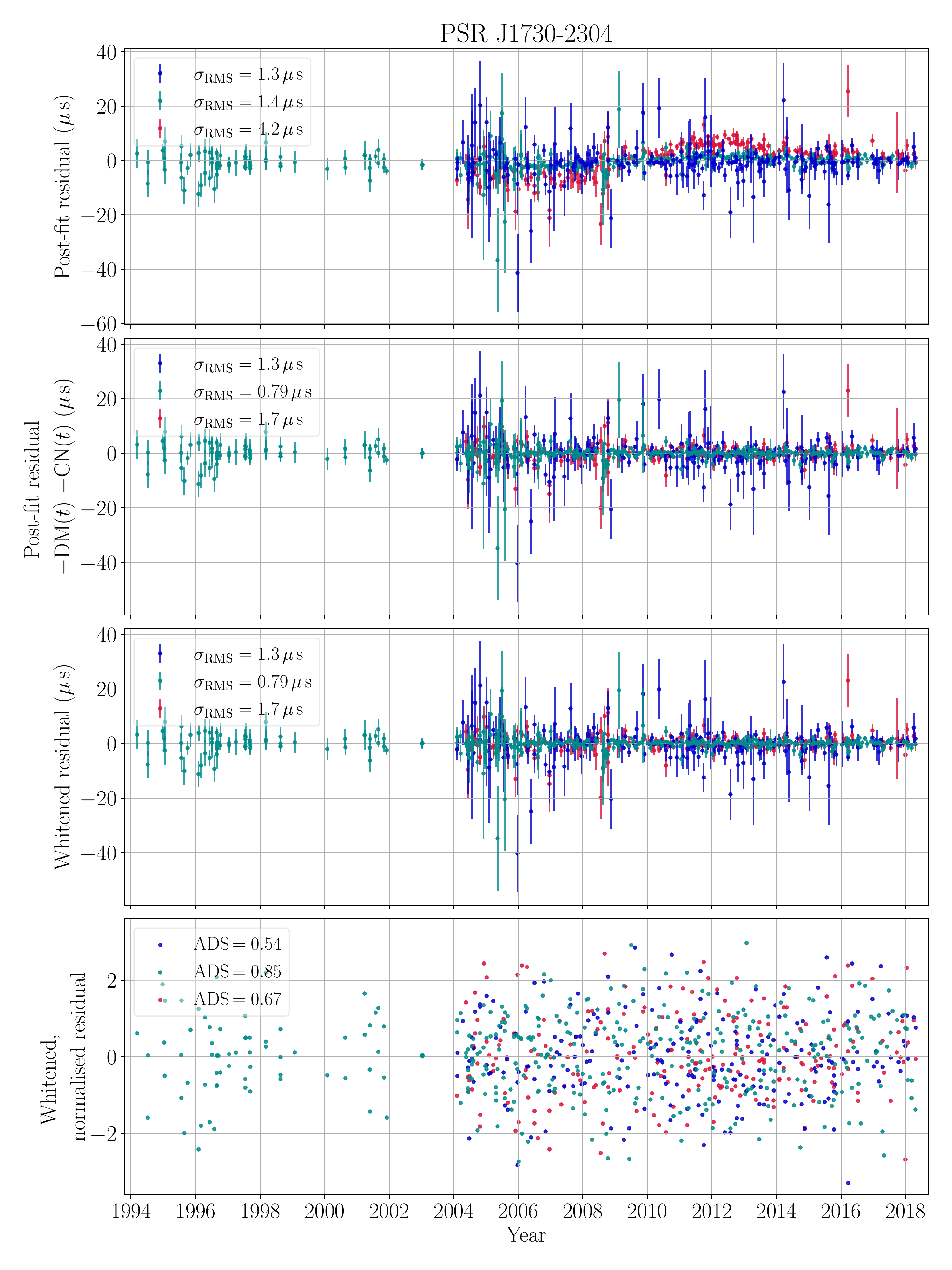}
\caption{Timing residuals for PSR~J1730$-$2304}
\label{fig:1730_res}
\end{figure*}

\begin{figure*}
\centering
\includegraphics[width=0.95\textwidth]{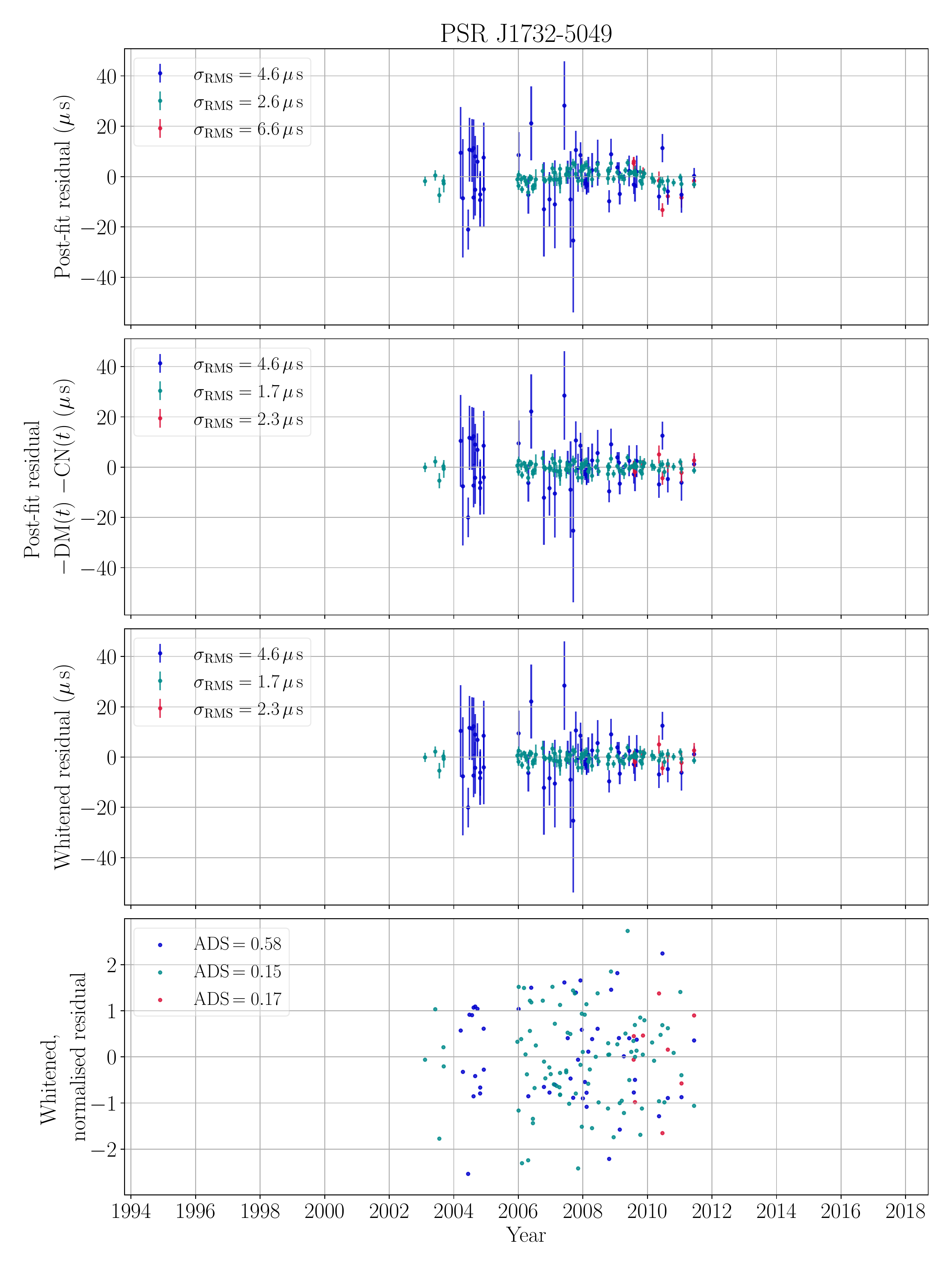}
\caption{Timing residuals for PSR~J1732$-$5049}
\label{fig:1732_res}
\end{figure*}

\begin{figure*}
\centering
\includegraphics[width=0.95\textwidth]{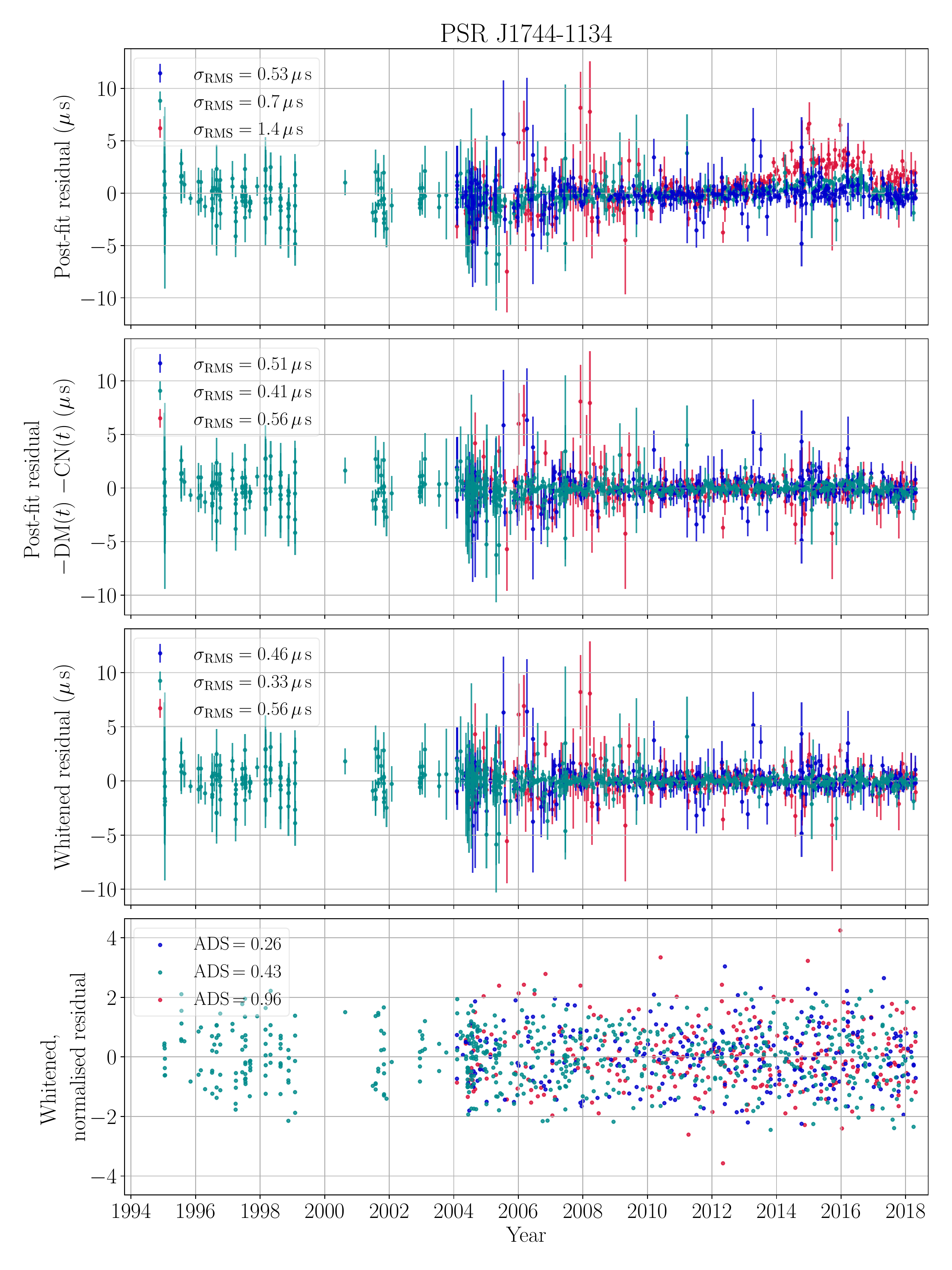}
\caption{Timing residuals for PSR~J1744$-$1134}
\label{fig:1744_res}
\end{figure*}

\begin{figure*}
\centering
\includegraphics[width=0.95\textwidth]{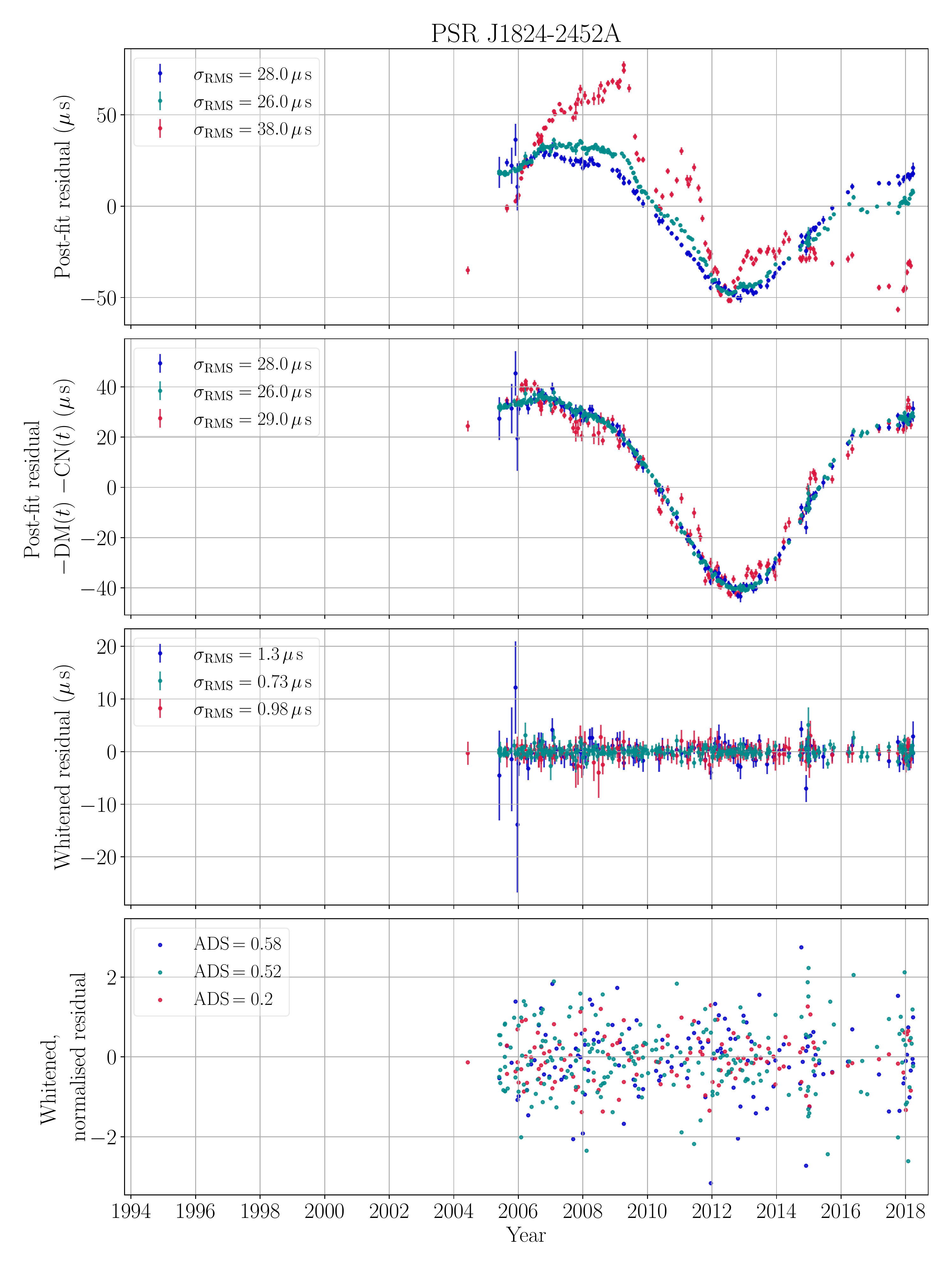}
\caption{Timing residuals for PSR~J1824$-$2452A}
\label{fig:1824_res}
\end{figure*}

\begin{figure*}
\centering
\includegraphics[width=0.95\textwidth]{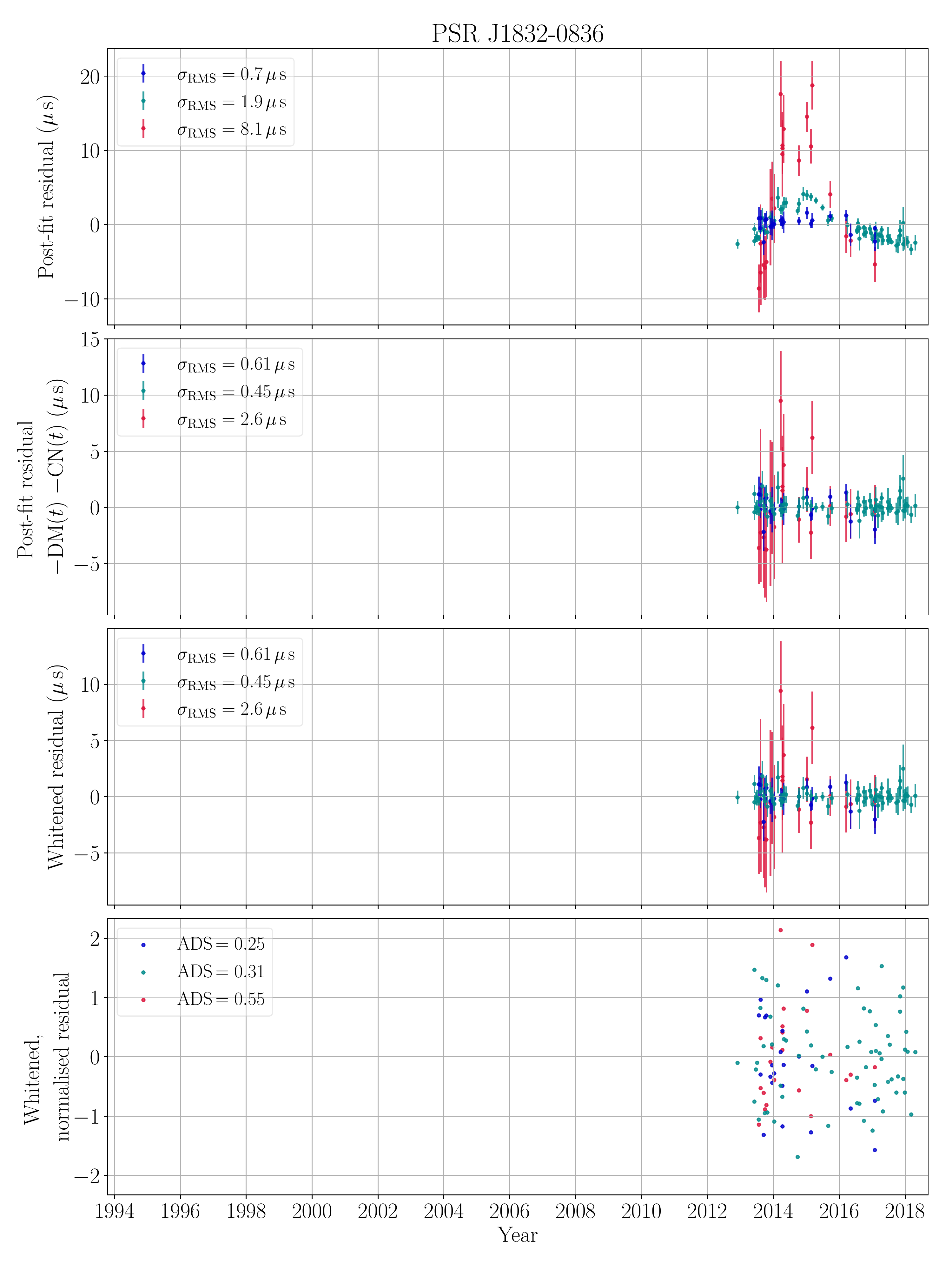}
\caption{Timing residuals for PSR~J1832$-$0836}
\label{fig:1832_res}
\end{figure*}

\begin{figure*}
\centering
\includegraphics[width=0.95\textwidth]{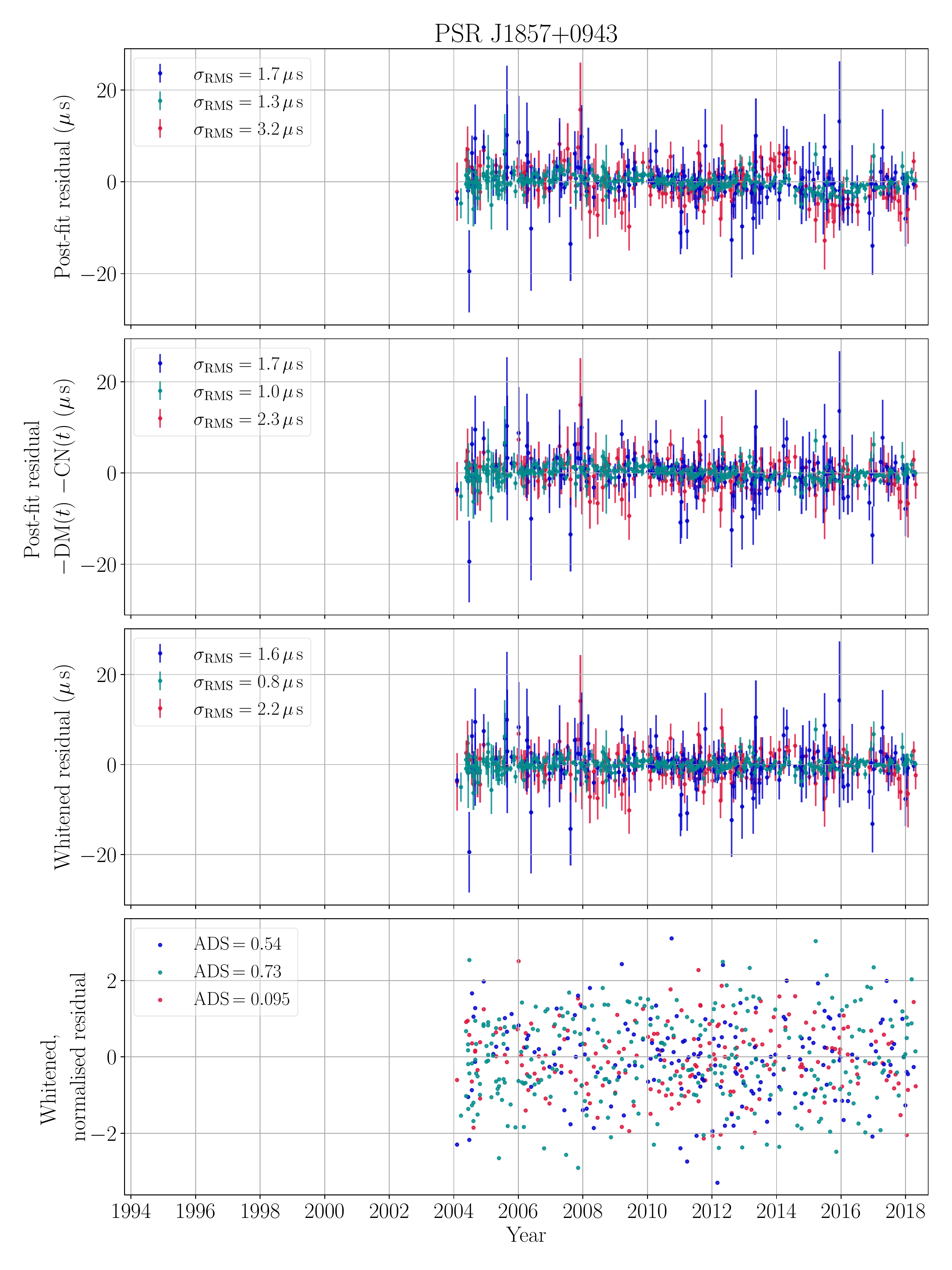}
\caption{Timing residuals for PSR~J1857+0943}
\label{fig:1857_res}
\end{figure*}

\begin{figure*}
\centering
\includegraphics[width=0.95\textwidth]{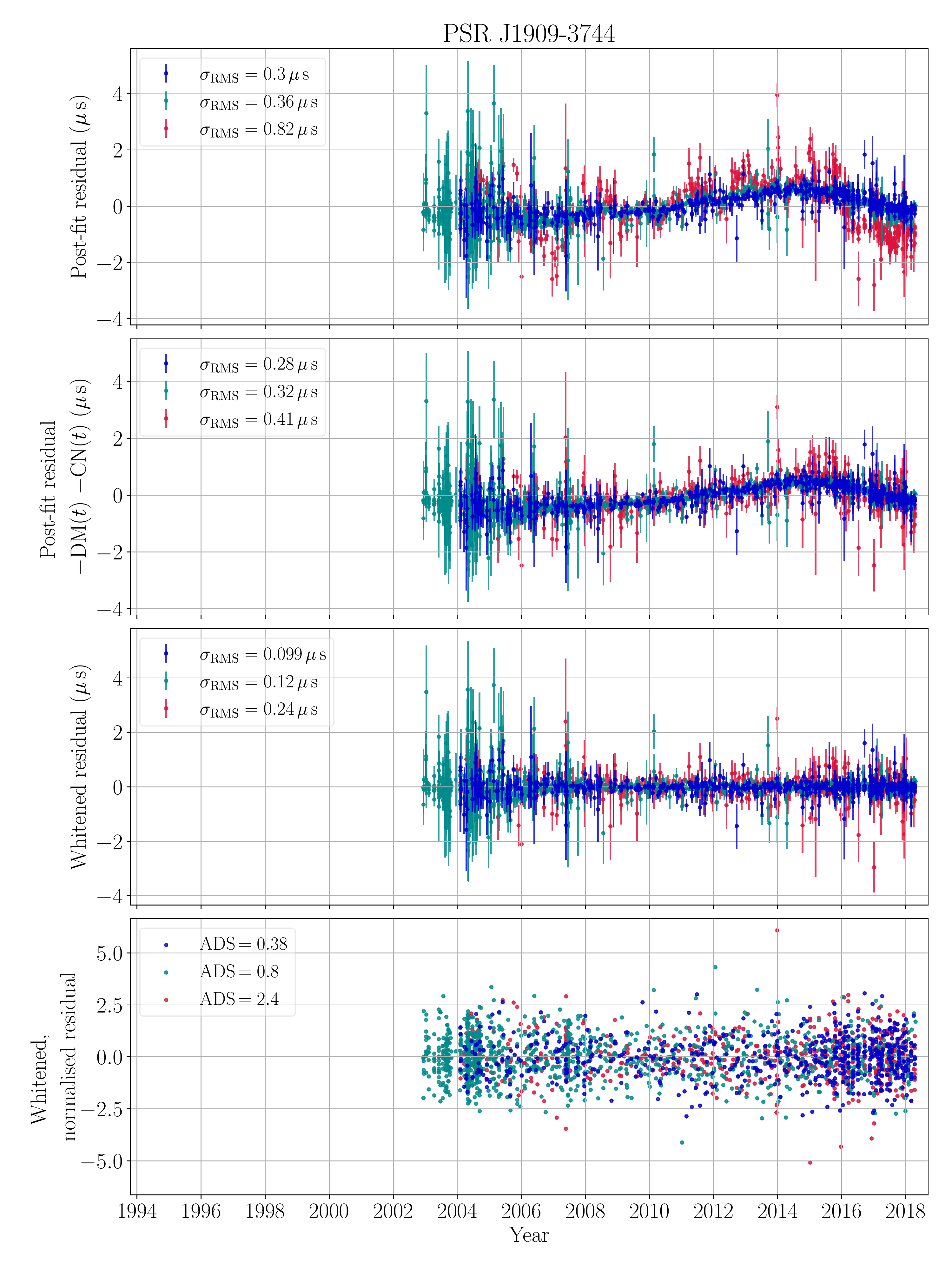}
\caption{Timing residuals for PSR~J1909$-$3744}
\label{fig:1909_res}
\end{figure*}

\begin{figure*}
\centering
\includegraphics[width=0.95\textwidth]{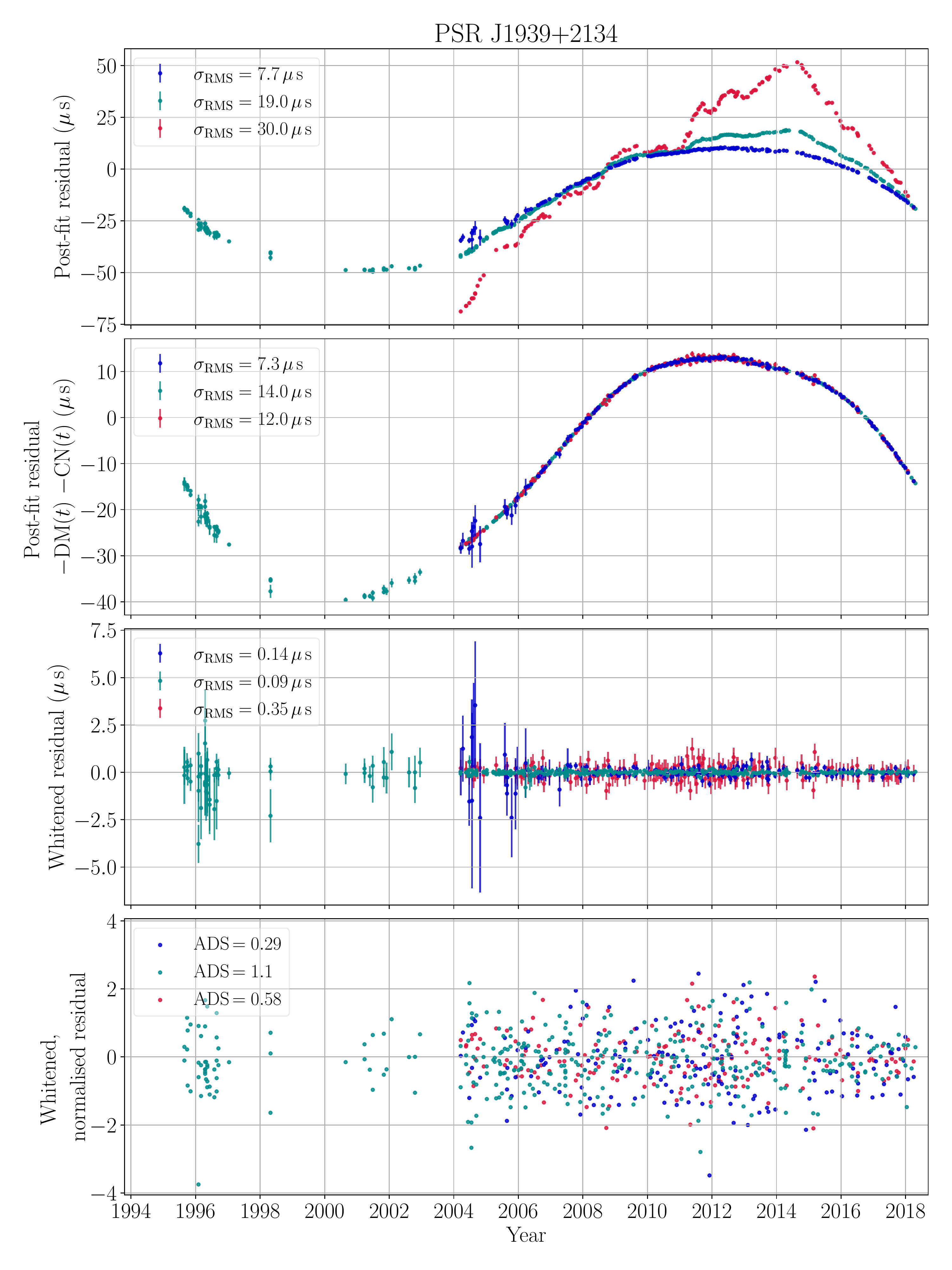}
\caption{Timing residuals for PSR~J1939+2134}
\label{fig:1939_res}
\end{figure*}

\begin{figure*}
\centering
\includegraphics[width=0.95\textwidth]{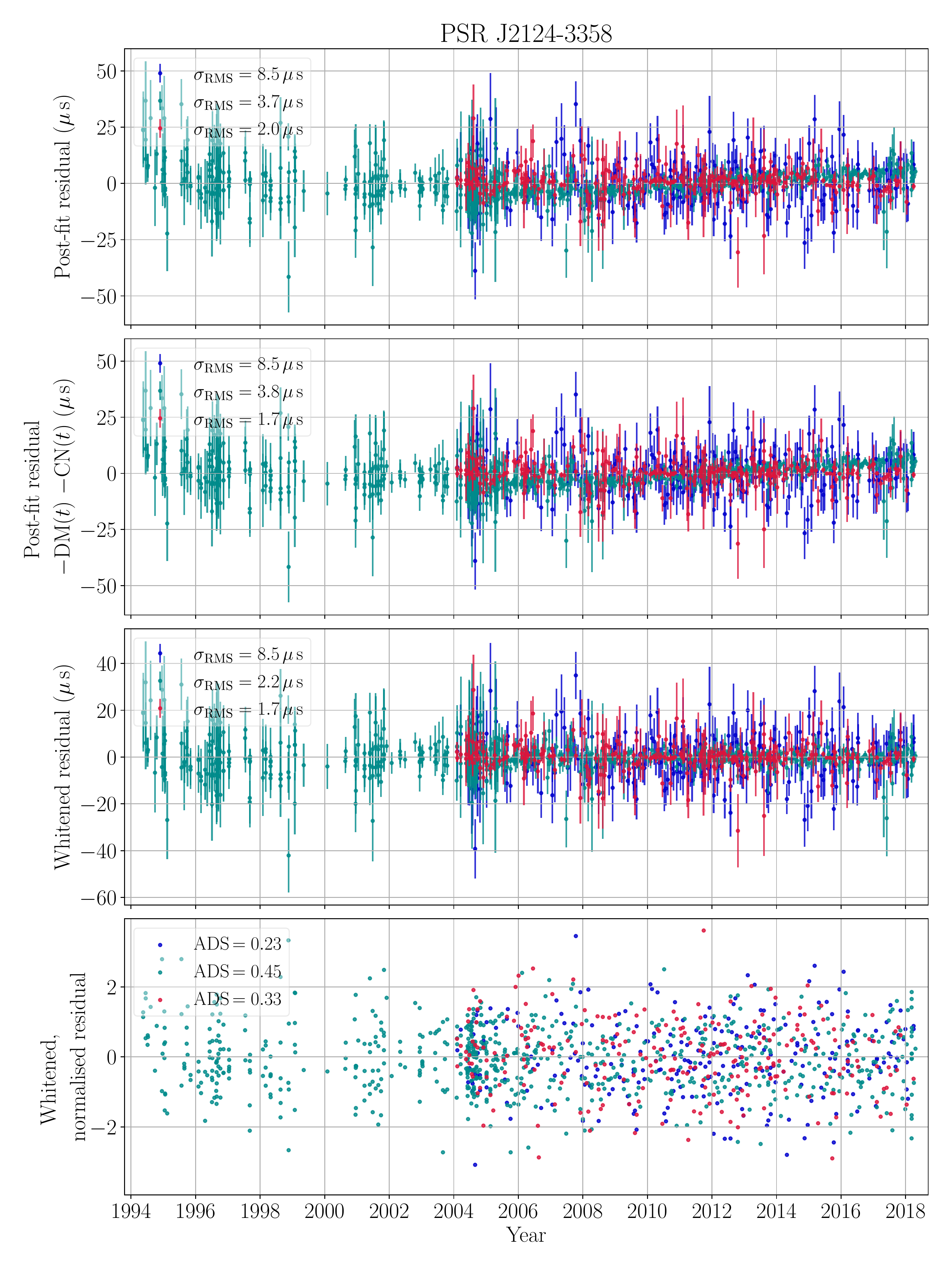}
\caption{Timing residuals for PSR~J2124$-$3358}
\label{fig:2124_res}
\end{figure*}

\begin{figure*}
\centering
\includegraphics[width=0.95\textwidth]{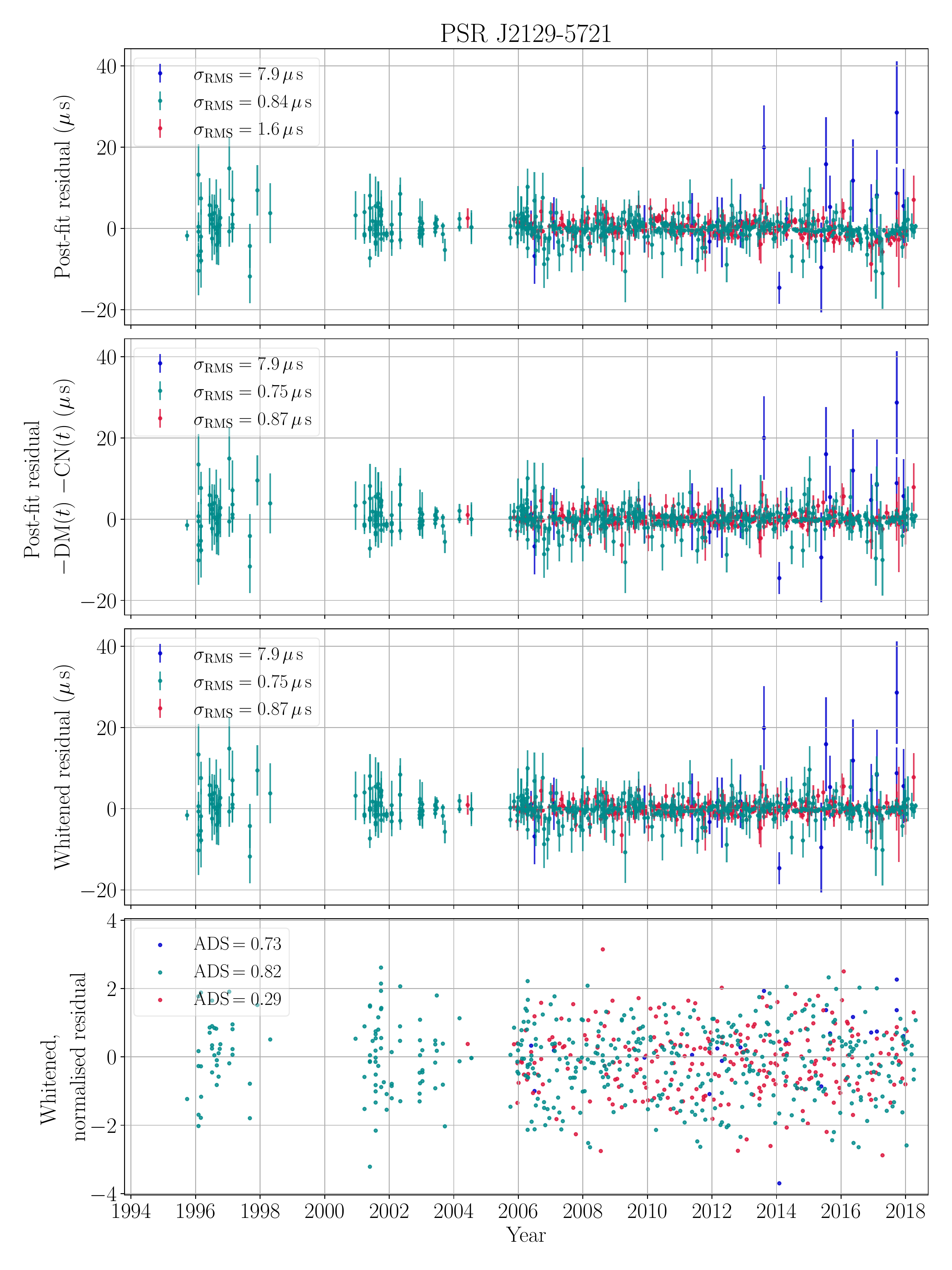}
\caption{Timing residuals for PSR~J2129$-$5721}
\label{fig:2129_res}
\end{figure*}

\begin{figure*}
\centering
\includegraphics[width=0.95\textwidth]{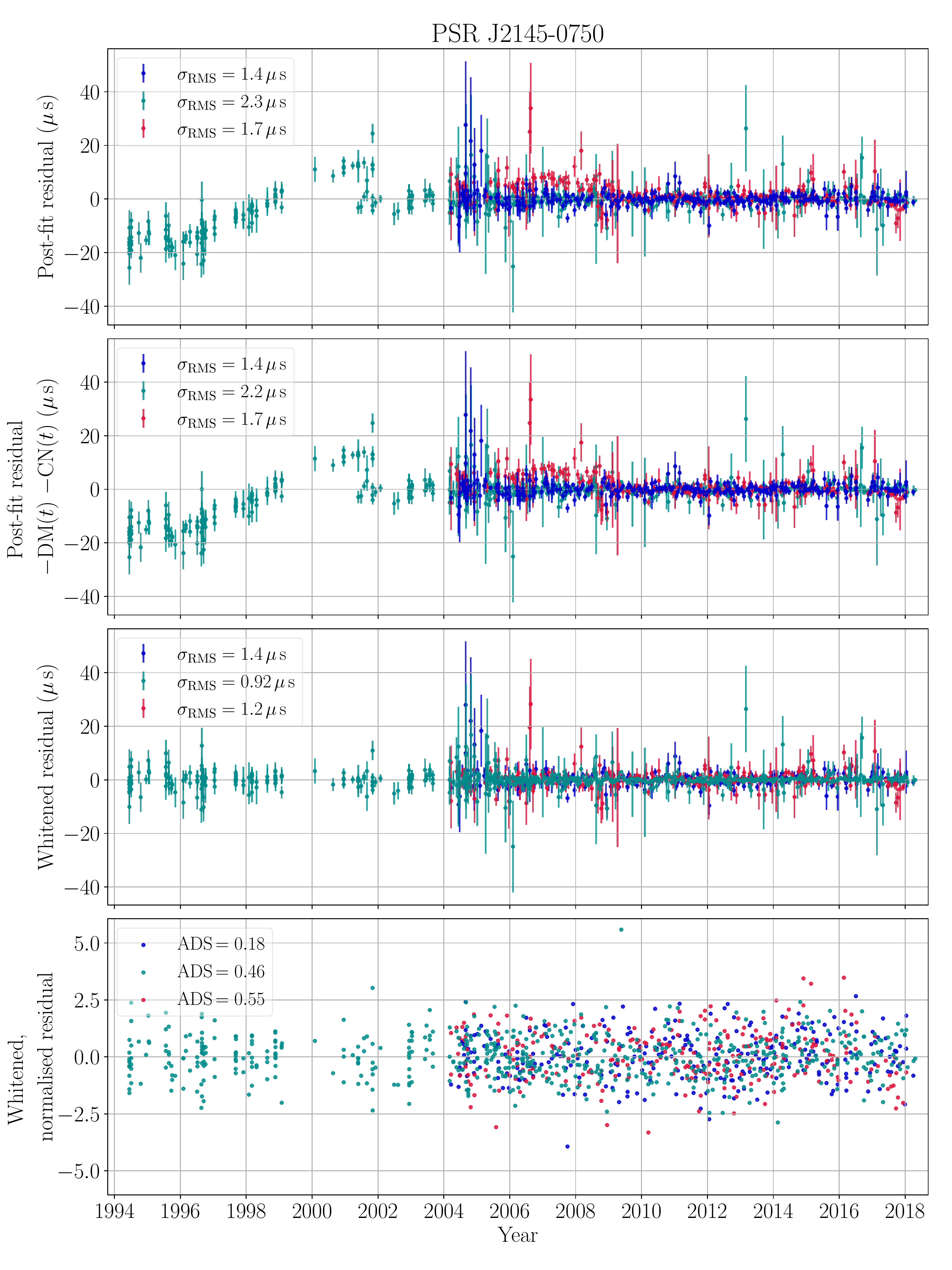}
\caption{Timing residuals for PSR~J2145$-$0750}
\label{fig:2145_res}
\end{figure*}

\begin{figure*}
\centering
\includegraphics[width=0.95\textwidth]{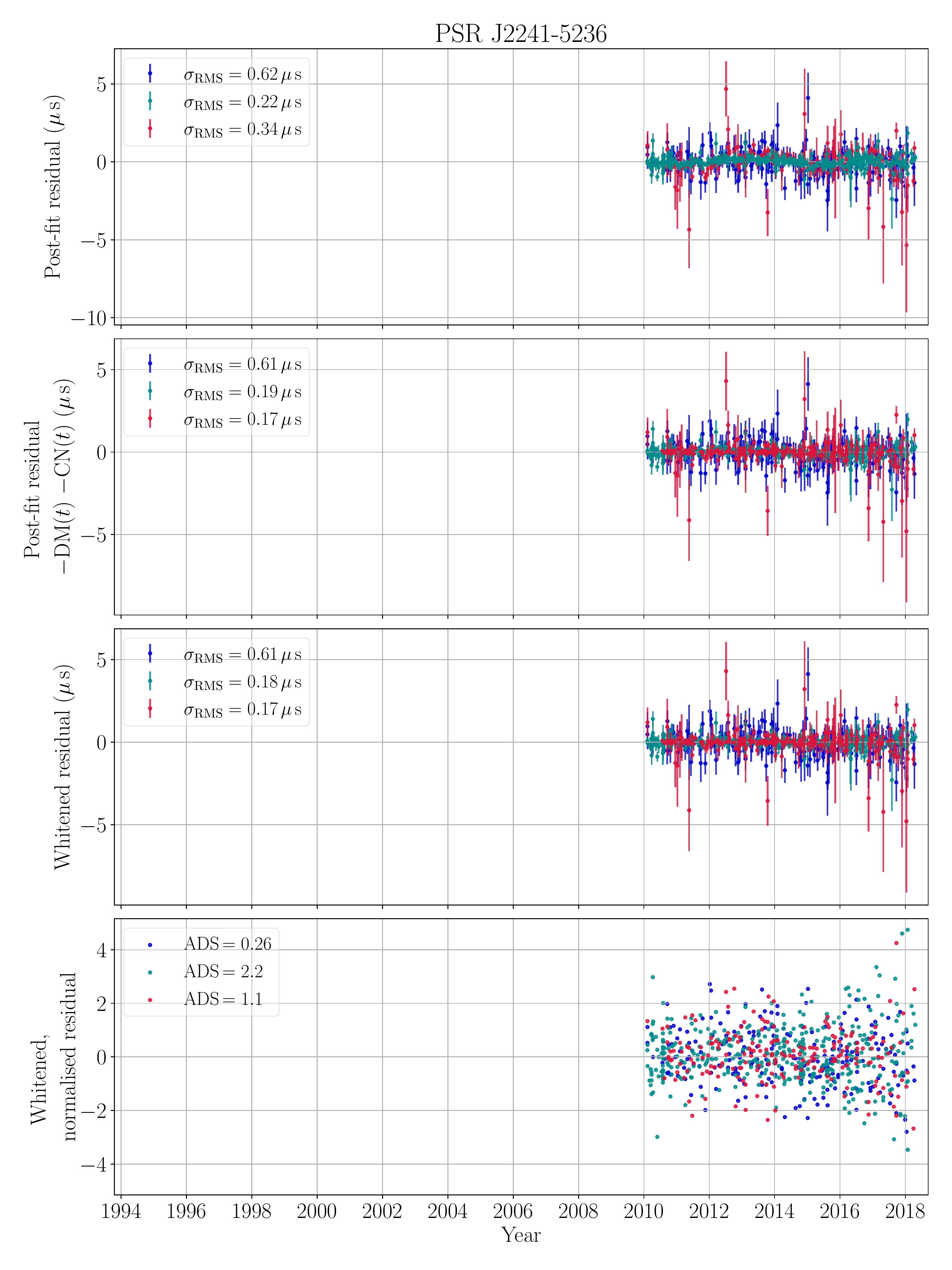}
\caption{Timing residuals for PSR~J2241$-$5236}
\label{fig:2241_res}
\end{figure*}


\bsp	
\label{lastpage}
\end{document}